\documentclass{amsart}

\usepackage{amsmath}
\usepackage{amssymb}
\usepackage{amsthm}
\usepackage{mathrsfs}
\usepackage{pdfpages}
\usepackage{dsfont}
\usepackage{lineno}
\usepackage{enumerate}
\usepackage{graphicx}
\usepackage{epstopdf}
\usepackage{color}
\usepackage{xspace}
\usepackage{subfig}
\usepackage{pdfsync}

\DeclareMathOperator*{\argmin}{argmin}

\DeclareGraphicsExtensions{.pdf}
\graphicspath{{./PDF/}}

\newcommand{\bq}{\begin{equation}}
\newcommand{\eq}{\end{equation}}
\newcommand{\R}{\mathbb{R}}
\newcommand{\E}{\mathbb{E}}
\newcommand{\abs}[1]{\left\vert#1\right\vert}

\newcommand{\bO}{\mathcal{O}}

\newcommand{\Dt}{\mathcal{D}}

\newcommand{\QW}{quadratic Wasserstein metric\xspace}
\newcommand{\MA}{Monge-Amp\`ere\xspace}

\newcommand{\diag}{\text{diag}}
\newcommand{\vp}{v^\perp}
\newcommand{\M}{\mathcal{M}}
\newcommand\numberthis{\addtocounter{equation}{1}\tag{\theequation}}
\newcommand{\addresseshere}{%
  \enddoc@text\let\enddoc@text\relax
}

\newtheorem{theorem}{Theorem}
\theoremstyle{lemma}

\newtheorem{definition}{Definition}

\theoremstyle{remark}

\begin{document}
\title[Application of Quadratic Wasserstein Metric to FWI]{Application of Optimal Transport and the Quadratic Wasserstein Metric to Full-Waveform Inversion}

\author{Yunan Yang}
\thanks{We thank Sergey Fomel and Zhiguang Xue for very helpful discussions, and thank the sponsors of the Texas Consortium for Computational Seismology (TCCS) for financial support. The third author was additionally supported by the Statoil Fellows Program at the University of Texas at Austin. The second author was partially supported by NSF DMS-1620396 and the fourth author was partially supported by NSF DMS-1619807.}. 
\address{Department of Mathematics, The University of Texas at Austin, 1 University Station C1200, Austin, TX 78712 USA}
\email{yunanyang@math.utexas.edu}

\author{Bj\"orn Engquist}
\address{Department of Mathematics and ICES, The University of Texas at Austin, 1 University Station C1200, Austin, TX 78712 USA}
\email{engquist@math.utexas.edu}

\author{Junzhe Sun}
\address{Bureau of Economic Geology, John A. and Katherine G. Jackson School of Geosciences, The University of Texas at Austin, University Station, Box X, Austin, TX 78713 USA}
\email{sunjzhe@gmail.com}

\author{Brittany D. Froese}
\address{Department of Mathematical Sciences, New Jersey Institute of Technology, University Heights, Newark, NJ 07102 USA}
\email{bdfroese@njit.edu}

\begin{abstract}
Conventional full-waveform inversion (FWI) using the least-squares norm ($L^2$) as a misfit function is known to suffer from cycle skipping. This increases the risk of computing a local rather than the global minimum of the misfit. In our previous work, we proposed the quadratic Wasserstein metric ($W_2$) as a new misfit function for FWI. The $W_2$ metric has been proved to have many ideal properties with regards to convexity and insensitivity to noise. When the observed and predicted seismic data are regarded as two density functions, the quadratic Wasserstein metric corresponds to the optimal cost of rearranging one density into the other, where the transportation cost is quadratic in distance. The difficulty of transforming seismic signals into nonnegative density functions is discussed. Unlike the $L^2$ norm, $W_2$ measures not only amplitude differences, but also global phase shifts, which helps to avoid cycle skipping issues. In this work, we build on our earlier method to cover more realistic high-resolution applications by embedding the $W_2$ technique into the framework of the adjoint-state method and applying it to seismic relevant 2D examples: the Camembert, the Marmousi, and the 2004 BP models. We propose a new way of using the $W_2$ metric trace-by-trace in FWI and compare it to global $W_2$ via the solution of the \MA equation. With corresponding adjoint source, the velocity model can be updated using the l-BFGS method. Numerical results show the effectiveness of $W_2$ for alleviating cycle skipping issues and sensitivity to noise. Both mathematical theory and numerical examples demonstrate that the quadratic Wasserstein metric is a good candidate for a misfit function in seismic inversion. 
\end{abstract}

\date{\today}
\maketitle

\section{Introduction}
Full-waveform inversion (FWI) was originally  proposed three decades ago in an attempt to obtain high resolution subsurface properties based on seismic waveforms~\cite{lailly1983seismic,TarantolaInversion}. Over the last decade, there have been many encouraging results employing FWI in seismic processing of marine and land data \cite{sirgue2010thematic, virieux2009overview}. FWI iteratively updates a subsurface model and computes corresponding synthetic data to reduce the difference (the data misfit) between the synthetic and recorded seismic data. 

The objective of FWI is to match the synthetic and recorded data in a comprehensive way such that all information in the waveforms is accounted for in the data misfit. If we denote the predicted data by $f$ and the observed data by $g$, then the unknown velocities are determined by minimizing the mismatch $d(f,g)$. 

FWI has the potential to generate high resolution subsurface models, but suffers from the ill-posedness of the inverse problem. This issue can be handled by considering multiple data components ranging from low to high frequency  \cite{bunks1995multiscale} or by adding regularization terms \cite{esser2016total, gholami2010regularization, regular_pgs}.

The least-squares norm ($L^2$) is the most widely used misfit function in FWI, but suffers from cycle skipping and sensitivity to noise. Other norms have been proposed in the literature including the $L_1$ norm, the Huber norm~\cite{ha2009waveform}, and hybrid $L_1/L^2$ norms~\cite{brossier2010data}. These misfit functions follow the same path of dealing with the predicted and observed data independently.

Differences between the predicted velocity model and true model produce a misfit, which is the information FWI uses to update the velocity model. This motivates us to take a different view of the predicted and observed data by considering a ``map'' connecting them~\cite{ma2013wave}. The idea of mapping synthetic data to observed data with stationary and non-stationary filters in the time domain has been promoted recently~\cite{AWI,zhu2016building}. Although the misfits in these two approaches are not critical metrics between two objects in mathematics, they demonstrate the advantages and feasibility of map-based ideas.

Optimal transport has become a well developed topic in mathematics since it was first proposed by~\cite{Monge}. 
Due to their ability to incorporate differences in both intensity and spatial information, optimal transport based metrics for modeling and signal processing have recently been adopted in a variety of applications including image retrieval, cancer detection, and machine learning \cite{kolouri2016transport}.

The idea of using optimal transport for seismic inversion was first proposed by~\cite{EFWass}. The Wasserstein metric is a concept based on optimal transportation~\cite{Villani}. Here, we treat our data sets of seismic signals as density functions of two probability distributions, which can be imagined as the distributions of two piles of sand with equal mass. Given a particular cost function, different plans of transporting one pile into the other lead to different costs.  The plan with the lowest cost is the optimal map and this lowest cost is the Wasserstein metric. In computer science the metric is often called the ``Earth Mover's Distance'' (EMD). Here we will focus on the quadratic cost functions. The corresponding misfit is the quadratic Wasserstein metric ($W_2$).

Following the idea that changes in velocity cause a shift or ``transport''  in the arrival time, \cite{engquist2016optimal} demonstrated the advantageous mathematical properties of the quadratic Wasserstein metric ($W_2$) and provided rigorous proofs that laid a solid theoretical foundation for this new misfit function. In this paper, we continue the study of the quadratic Wasserstein metric with more focus on its applications to FWI.  We also develop a fast and robust trace-by-trace technique.

After the paper of \cite{EFWass}, researchers in geophysics started to work on other optimal transport related misfit functions~\cite{metivier2016increasing,W1_2D,W1_3D}. The Kantorovich-Rubinstein (KR) norm in their papers is a relaxation of the 1-Wasserstein distance, which is another optimal transport metric with the absolute value cost function. The advantage of the KR norm is that it does not require data to satisfy nonnegativity or mass balance conditions.

The Wasserstein distance measures the difference between nonnegative measures or functions with equal mass. These are not natural constraints for seismic signals and thus they first have to be normalized. In our earlier work we separated the positive and negative part of the signals to achieve nonnegativity. The resulting signal was then divided by its integral. This worked well in our earlier test cases, but is less effective for the larger scale problems with the adjoint-state method studied here. Linear normalization, on the other hand, is effective in spite of the fact that it results in a measure that is not convex with respect to simple shifts.

In this paper, we briefly review the theory of optimal transport and revisit the mathematical properties of $W_2$ that were proved in \cite{engquist2016optimal}, including the convexity and insensitivity to noise. Next, we apply the quadratic Wasserstein metric ($W_2$) as misfit function in two different ways: trace-by-trace comparison and entire data set comparison. The trace-by-trace strategy and global strategy lead to different formulations of the misfit computation and the adjoint source \cite{Plessix}. The trace-by-trace technique is new and the results for inversion are very encouraging. The computational cost is low and similar to that of the classical $L^2$ method. Finally, after deriving the adjoint source expressions, we show the application of FWI using the $W_2$ metric on three synthetic models: the Camembert, the Marmousi and the 2004 BP models. Discussions and comparisons between the FWI results using $W_2$ and $L^2$ metrics illustrate that the $W_2$ metric is very promising for overcoming the cycle skipping issue in FWI.

\section{Theory}

\subsection{Formulation}
Conventional FWI defines a least squares waveform misfit as
\begin{equation}
d(f,g) = J_0(m)=\frac{1}{2}\sum_r\int\abs{f(x_r,t;m)-g(x_r,t)}^2dt,
\end{equation}
where $g$ is observed data, $f$ is simulated data, $x_r$ are receiver locations, and $m$ is the model parameter. This formulation can also be extended to the case with multiple shots. We get the modeled data $f(x,t;m)$ by solving a wave equation with a finite difference method (FDM) in both the space and time domain. 

In this paper, we propose using the quadratic Wasserstein metric ($W_2$) as an alternative misfit function to measure the difference between synthetic data $f$ and observed data $g$. There are two ways to apply this idea: trace-by-trace $W_2$ and global $W_2$.

We can compare the data trace by trace and use the quadratic Wasserstein metric ($W_2$) in 1D to measure the misfit. The overall misfit is then
\bq \label{eqn:W21D}
J_{1}(m) = \sum\limits_{r=1}^R W_2^2(f(x_r,t;m),g(x_r,t)), 
\eq
where $R$ is the total number of traces.

In the global case we compare the full data sets and consider the whole synthetic data $f$ and observed data $g$ as objects with the general quadratic Wasserstein metric ($W_2$):
\bq \label{eqn:W22D}
J_2(m) = W_2^2(f(x_r,t;m),g(x_r,t)).
\eq

We treat the misfit $J(m)$ as a function of the model parameter $m$. Our aim is to find the model parameter $m^{\ast}$ that minimizes the objective function, i.e. \(m^{\ast} = \argmin J(m) \). This is a PDE-constrained optimization problem, and we use a gradient-based iterative scheme to update the model $m$.

\subsection{Background}
Optimal transport originated in 1781 with the French mathematician Monge.  This problem seeks the minimum cost required to transport mass of one distribution into another given a cost function. More specifically, we consider two probability measures $\mu$ and $\nu$ defined on spaces $X$ and $Y$ respectively. For simplicity, we regard $X$ and $Y$ as subsets of $\mathbb{R}^d$. Measures $\mu$ and $\nu$ have density functions $f$ and $g$: $d\mu = f(x)dx$ and $d\nu = g(y)dy$. In applications, $f(x)$ can represent the height of a pile of sand at location $x$, the gray scale of one pixel $x$ for a image, or as here the amplitude of a seismic waveform at mesh grid point $x$.

Although they must share the same total mass, measures $\mu$ and $\nu$ are not the same, i.e. $f\neq g$. We want to redistribute ``sand'' from $\mu$ into $\nu$ and it requires effort. The cost function $c(x,y)$ maps pairs $(x,y) \in X\times Y$ to $\mathbb{R}\cup \{+\infty\}$, which denotes the cost of transporting one unit mass from location $x$ to $y$. The most common choices of $c(x,y)$ include $|x-y|$ and $|x-y|^2$. Once we find a transport plan $T: X\rightarrow Y$ such that for any measurable set $B \subset Y$, $\nu[B] = \mu[T^{-1}(B)]$, the cost corresponding to this plan $T$ is 
\bq
I(T,f,g,c) = \int\limits_Xc(x,T(x))f(x)\,dx. 
\eq

While there are many maps $T$ that can perform the relocation, we are interested in finding the optimal map that minimizes the total cost
\bq
I(f,g,c) = \inf\limits_{T\in\M}\int\limits_Xc(x,T(x))f(x)\,dx,
\eq
where $\M$ is the set of all maps that rearrange $f$ into $g$.

Thus we have informally defined the optimal transport problem, the optimal map as well as the optimal cost, which is also called the Wasserstein distance:
\begin{definition}[The Wasserstein distance]
  We denote by $\mathscr{P}_p(X)$ the set of probability measures with finite moments of order $p$. For all $p \in [1, \infty)$,   
\bq
W_p(\mu,\nu)=\left( \inf _{T\in \mathcal{M}}\int_{\mathbb{R}^n}\left|x-T(x)\right|^p d\mu(x)\right) ^{\frac{1}{p}},\quad \mu, \nu \in \mathscr{P}_p(X).
\eq
$\mathcal{M}$ is the set of all maps that rearrange the distribution $\mu$ into $\nu$.
\end{definition}

In this paper, we focus on the case of a quadratic cost function: $c(x,y) = |x-y|^2$. The mathematical definition of the distance between the distributions $f:X\to\R^+$ and $g:Y\to\R^+$ can then be formulated as
\bq\label{eq:W2}  W_2^2(f,g) = \inf\limits_{T\in\M} \int\limits_X\abs{x-T(x)}^2f(x)\,dx, \eq
where $\M$ is the set of all maps that rearrange the distribution $f$ into $g$. For details see~\cite{Villani}. The optimal transport formulation requires non-negative distributions and equal total masses, which are not natural for seismic signals. We will discuss this in the section on data normalization below.

\subsection{Optimal transport on the real line}
For $f$ and $g$ in one dimension, it is possible to exactly solve the optimal transportation problem~\cite{Villani} in terms of the cumulative distribution functions
\bq \label{eq:F&G}
F(x) = \int_{-\infty}^x f(t)\,dt, \quad G(y) = \int_{-\infty}^y g(t)\,dt.
\eq

\begin{figure}
	\centering
	\includegraphics[width=0.6\textwidth]{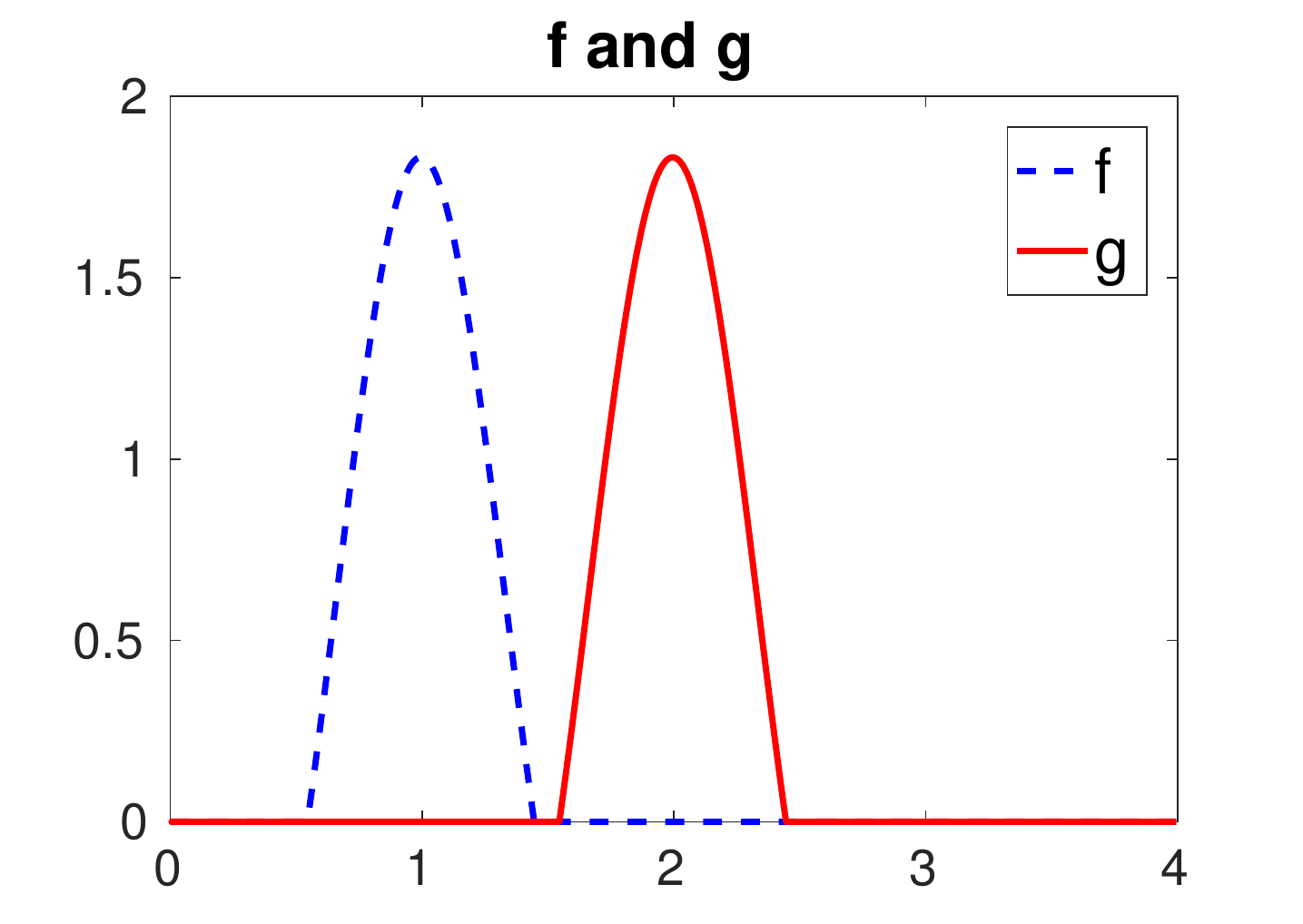}
  	\caption{1D densities $f$ and $g$}\label{fig:f&g}
\end{figure}

\begin{figure}
	\centering
  	\subfloat[]{\includegraphics[width=0.45\textwidth]{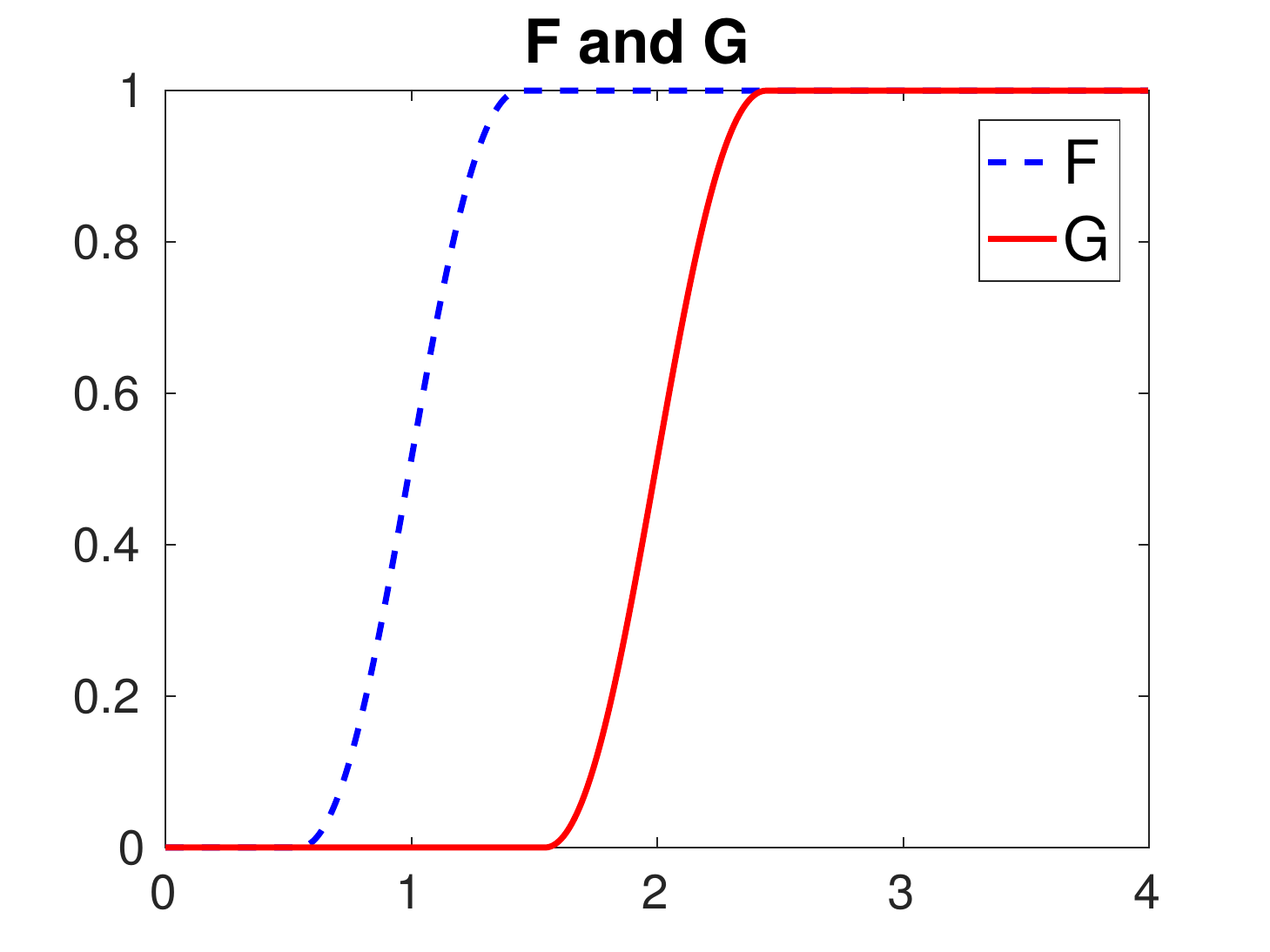}\label{fig:F&G}}
  	\subfloat[]{\includegraphics[width=0.45\textwidth]{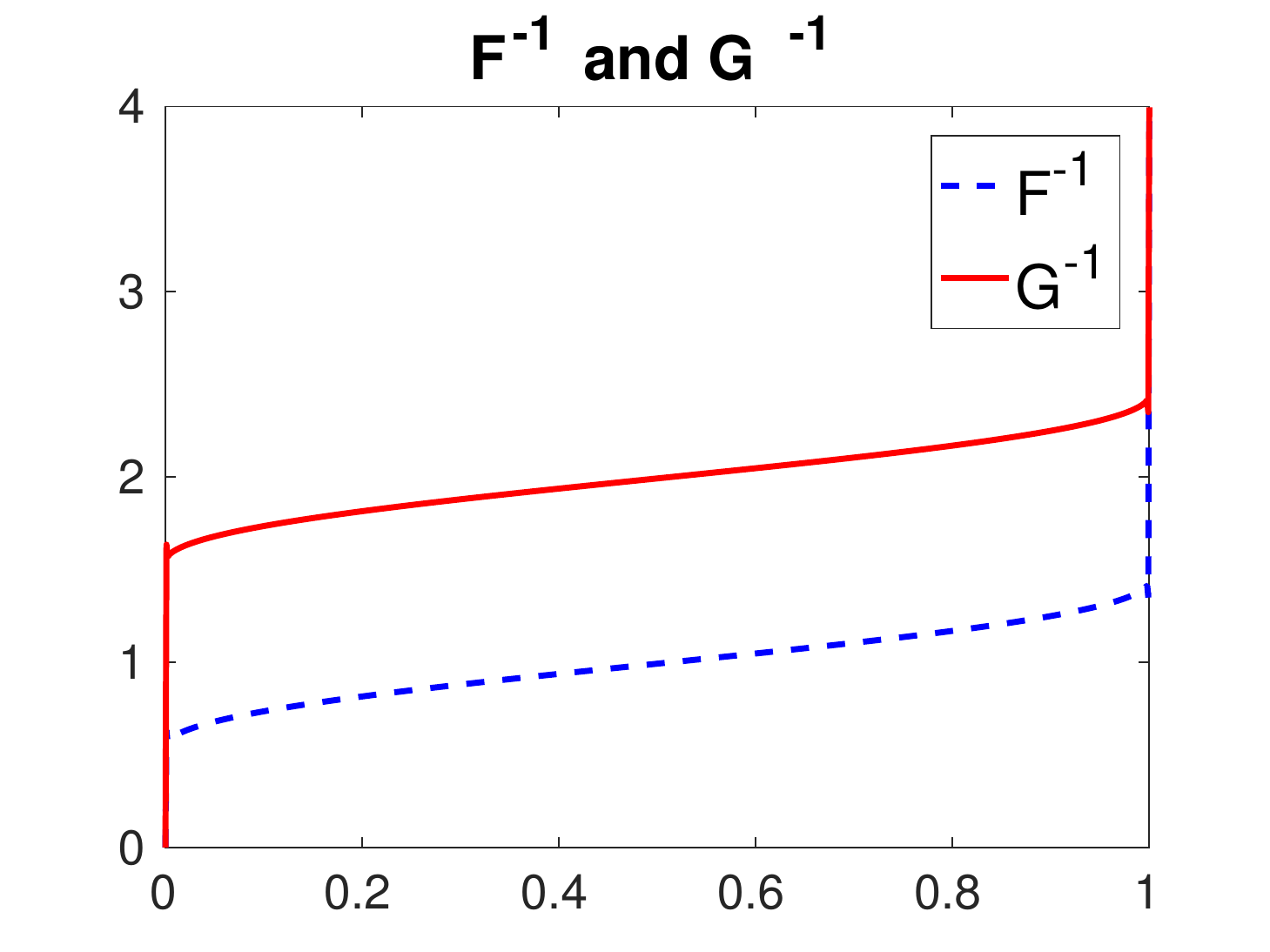}\label{fig:F1&G1}}
	\caption{(A)~Cumulative distribution functions $F$ and $G$ and (B)~the inverse distribution function $F^{-1}$ and $G^{-1}$ for probability density functions $f$ and $g$}
	\label{fig:F&G,F1&G1}
\end{figure}

In fact, the optimal map is just the unique monotone rearrangement of the density $f$ into $g$ (Figure~\ref{fig:f&g}). In order to compute the quadratic Wasserstein metric ($W_2$), we need the cumulative distribution functions $F$ and $G$ (Figure~\ref{fig:F&G}) and their inverses $F^{-1}$ and $G^{-1}$ (Figure~\ref{fig:F1&G1}) as the following theorem states.

\begin{theorem}[Optimal transportation for a quadratic cost on $\R$]\label{OT1D}
Let $0 < f, g < \infty$ be two probability density functions, each supported on a connected subset of $\R$.  Then the optimal map from $f$ to $g$ is $T = G^{-1}\circ F$.
\end{theorem}

If $f$ and $g$ are the datasets of synthetic data and observed data in a corresponding trace,  we can consider $f$ and $g$ continuous in time since there is no singularity in the waveform. With proper normalization, signals $f$ and $g$ can be rescaled to be positive, supported on $[0,1]$, and have total mass 1. From the theorem above, we derive another formulation for the 1D quadratic Wasserstein metric:
\bq\label{myOT1D}
W_2^2(f,g) = \int_0^1|x-G^{-1}(F(x))|^2 f(x)dx.
\eq

\subsection{Optimal transport in higher dimensions}
The simple exact formula for 1D optimal transportation does not extend to optimal transportation in higher dimensions.  Nevertheless, it can be computed by relying on two important properties of the optimal mapping~$T(x)$: conservation of mass and cyclical monotonicity.  From the definition of the problem, $T(x)$ maps $f$ into $g$.  The change of variables formula formally leads to the requirement
\bq\label{eq:massConserved}
f(x) = g(T(x))\det(\nabla T(x)).
\eq 

The optimal map takes on additional structure in the special case of a quadratic cost function: it is cyclically monotone~\cite{Brenier,KnottSmith}.
\begin{definition}[Cyclical monotonicity]
\label{cyclical}
We say that $T:X\to Y$ is cyclically monotone if for any $m\in\mathbb{N}^+$, \(x_i\in X, \,1\leq i \leq m\),
\bq\label{eq:cyclical}
\sum_{i=1}^{m}|x_i-T(x_i)|^2 \leq  \sum_{i=1}^{m}|x_i-T(x_{i-1})|^2
\eq 
or equivalently 
\bq
\sum_{i=1}^{m}\langle T(x_i),x_i-x_{i-1}\rangle \geq 0  
\eq 
where $x_0\equiv x_m$.
\newline
\end{definition}

Additionally, a cyclically monotone mapping is formally equivalent to the gradient of a convex function~\cite{Brenier,KnottSmith}.  Making the substitution $T(x) = \nabla u(x)$ into the constraint~\eqref{eq:massConserved} leads to the \MA equation
\bq\label{eq:MAA}
\det (D^2 u(x)) = \frac{f(x)}{g(\nabla u(x))}, \quad u \text{ is convex}.
\eq

In order to compute the misfit between distributions $f$ and $g$, we first compute the optimal map $T(x) = \nabla u(x)$ via the solution of this \MA equation coupled to the non-homogeneous Neumann boundary condition 
\bq\label{eq:BC}
\nabla u(x) \cdot n = x\cdot n, \,\, x \in \partial X.
\eq
The squared Wasserstein metric is then given by
\bq\label{eq:WassMA}
W_2^2(f,g) = \int_X f(x)\abs{x-\nabla u(x)}^2\,dx.
\eq

\subsection{Convexity}
As demonstrated in~\cite{engquist2016optimal}, the squared Wasserstein metric has several properties that make it attractive as a choice of misfit function.  One highly desirable feature is its convexity with respect to data shifts, dilation and partial amplitude change, which occur naturally in seismic waveform inversion.  

We recall the overall set-up for FWI, in which we have a fixed observation $g$ and a simulation $f(m)$ that depends on unknown model parameters $m$.  The model parameters are recovered via the minimization
\bq m^* = \argmin\limits_m \{W_2^2(f(m), g)\}. \eq
In order to perform this minimization effectively and efficiently, we desire the distance $W_2^2(f(m), g)$ to be convex in the model parameter $m$.

This is certainly not the case for all possible functions $f(m)$, but it is true for many settings that occur naturally in seismic inversion.  For example, variations in the  wave velocity lead to simulations $f(m)$ that are derived from shifts,
\bq\label{eq:shift}
f(x;s) = g(x+s\eta), \quad \eta \in \R^n,
\eq
 or dilations,
\bq\label{eq:dilation}
f(x;A) = g(Ax), \quad A^T = A, \, A > 0,
\eq
 applied to the observation $g$.
Variations in the strength of a reflecting surface or the focusing of seismic waves can also lead to local rescalings of the form
\bq\label{eq:rescaleLocal}
f(x;\beta) = \begin{cases}  \beta g(x), & x \in E\\ g(x), & x \in \R^n\backslash E.\end{cases}
\eq

Proving the convexity of $W_2^2$ follows nicely from the interpretation of the misfit as a transportation cost, with the underlying transportation cost exhibiting a great deal of structure.  In particular, the cyclical monotonicity of the transport map $T(x)$ leads readily to estimates of 
\bq W_2^2(f({\lambda m_1 + (1-\lambda)m_2}), g), \quad 0 < \lambda < 1,  \eq
which in turn yields the desired convexity results.  This was studied in detail in~\cite{engquist2016optimal}, where the following theorem was proved.

\begin{theorem}[Convexity of squared Wasserstein metric~{\cite{engquist2016optimal}}]\label{thm:convexity}
The squared Wasserstein metric $W_2^2(f(m),g)$ is convex with respect to the model parameters $m$ corresponding to a shift~$s$ in~\eqref{eq:shift}, the eigenvalues of a dilation matrix~$A$ in~\eqref{eq:dilation}, or the local rescaling parameter~$\beta$ in~\eqref{eq:rescaleLocal}.
\end{theorem}

\subsection{Insensitivity to noise}
When performing FWI with real data, it is natural to experience noise in the measured signal.  Consequently, it is imperative that a misfit function be robust with respect to noise. As demonstrated in~\cite{engquist2016optimal}, the Wasserstein metric is substantially less sensitive to noise than the traditional $L^2$ norm.

This again follows from the interpretation of $W_2^2$ as a transportation cost.  Intuitively, noise added to the data will increase the distance $\abs{T(x)-x}$ that mass moves at some points $x$, but will also decrease this distance at other points.  Thus the overall effect of noise on the total transportation cost
\bq \int_X f(x)\abs{T(x)-x}^2\,dx \eq
will be {negligible}.

This is simplest to calculate in one dimension.  For example, we can consider the setting from~\cite{engquist2016optimal}.  Here the data $f$ and $g$ are given on a grid with a total of $N$ data points along each dimension.  At each grid point, the difference $f-g$ is given by a random variable drawn from a uniform distribution of the form $U[-c,c]$ for some constant $c$.    Regardless of the number of data points, noise of this type is expected to have a large effect on the $L^2$ distance,
\bq\label{eq:noiseL}
\E\|f-g\|_{L^2} = \bO(1).
\eq
Using the exact formula for the one-dimensional optimal transport plan, we can also directly compute the expected value of the squared Wasserstein metric.
\bq \E W_2^2(f,g) = \bO\left(\frac{1}{N}\right). \eq
Thus even if the noise is very strong (with order-one amplitude), its effect on the misfit is negligible if there are a large number of data points.

\begin{figure}
\centering
  \subfloat[]{\includegraphics[width=0.45\textwidth]{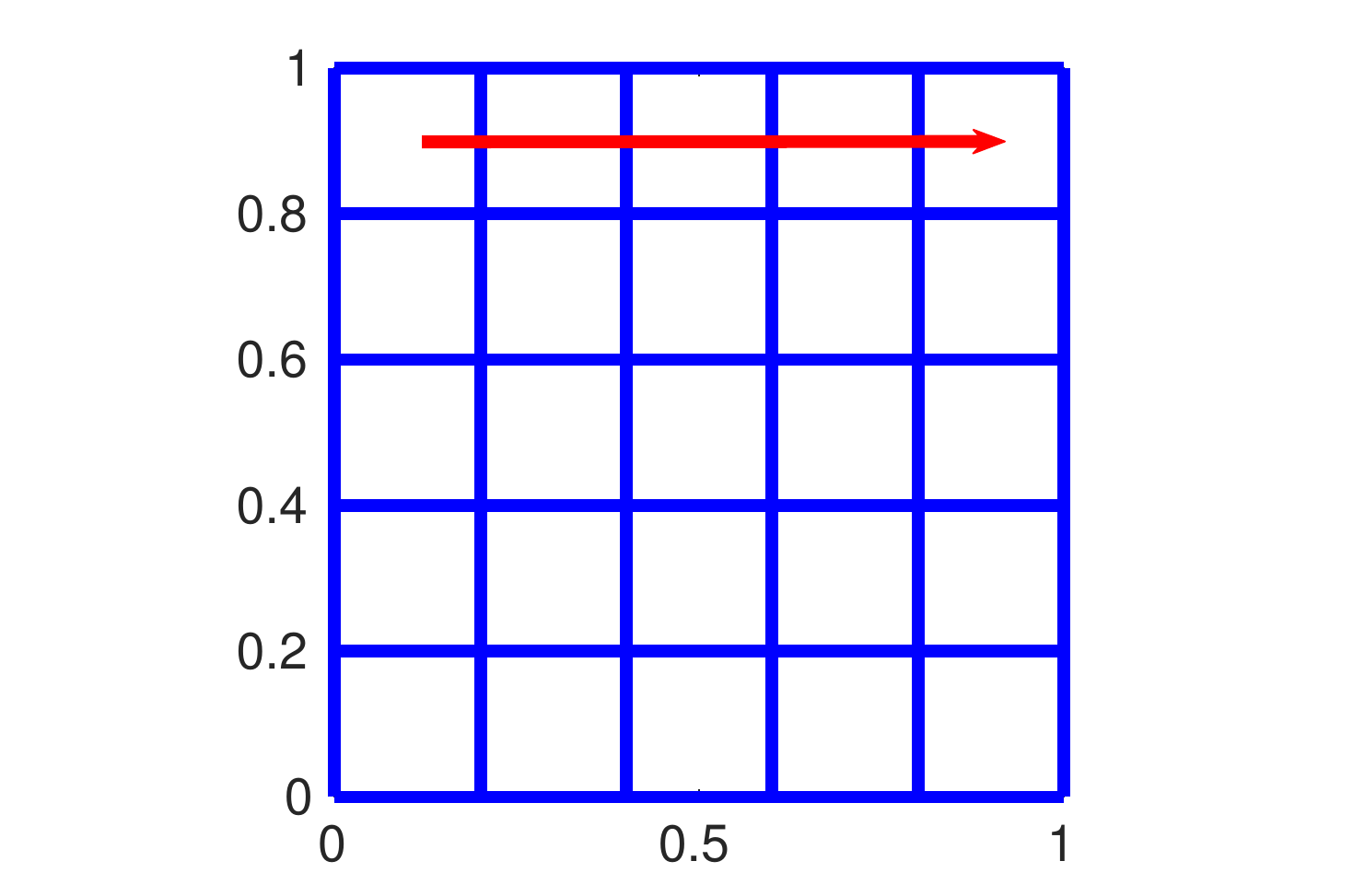}\label{fig:T1}}
  \subfloat[]{\includegraphics[width=0.45\textwidth]{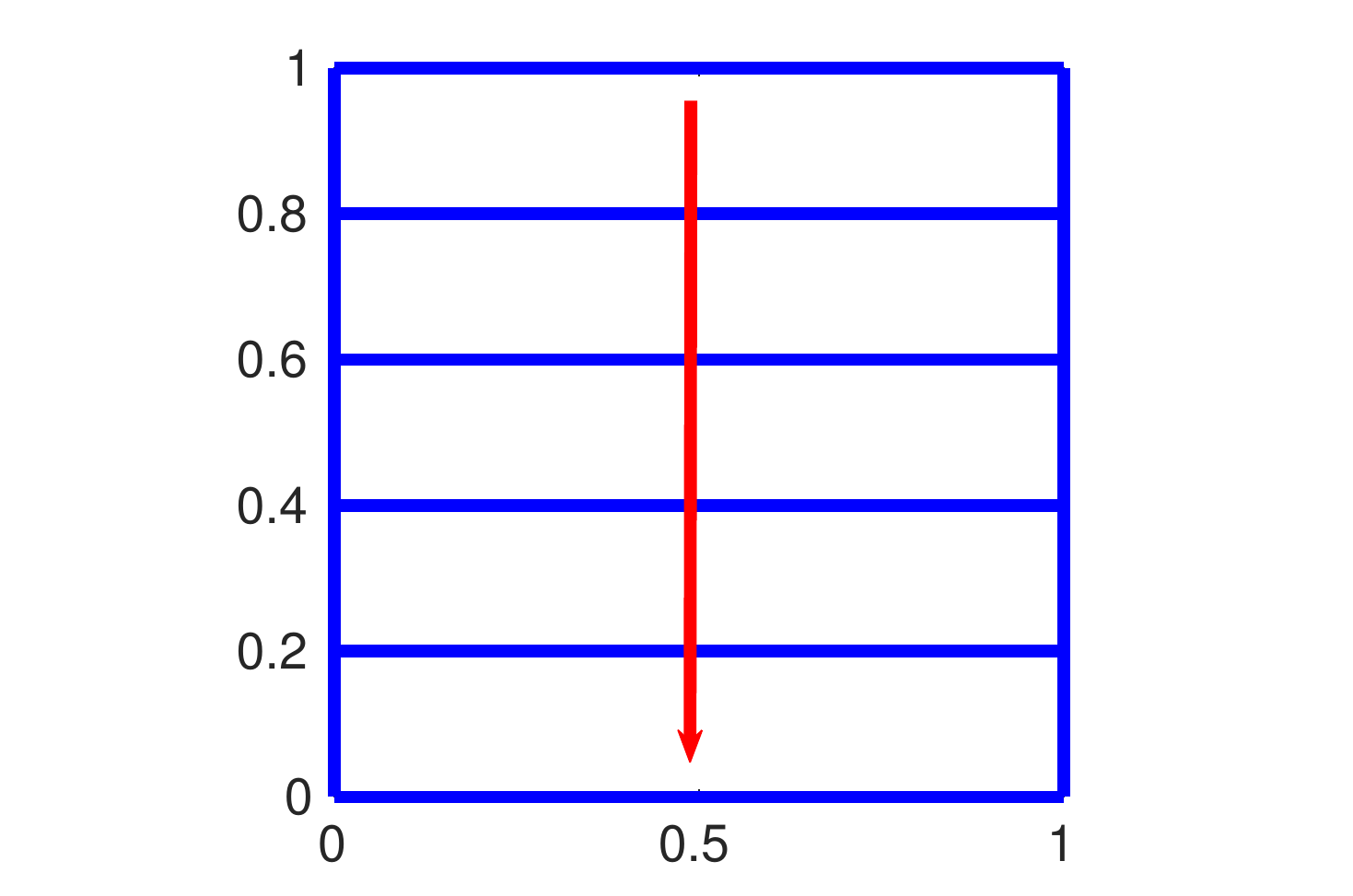}\label{fig:T2}}
  \caption{(A)~The optimal map for each row: $T_x = T_i$ for $x_i < x \leq x_{i+1}$ and  (B)~the optimal map in y direction: $T_y$}
\label{fig:noise}
\end{figure}

While there is no exact formula to exploit in higher-dimensions, we can place a bound on the expected effects of noise by considering a sequence of one-dimensional optimal transport problems.  That is, we can produce a sequence of mappings $T_j(x)$, $j = 1,\ldots, n$ that optimally rearrange the mass along the $jth$ dimension; see Figure~\ref{fig:noise}.  These one-dimensional maps can again be expressed exactly.  The resulting composite map
\bq \tilde{T}(x) = T_n\circ T_{n-1}\circ\cdots\circ T_1(x)\eq
will be mass-preserving, but not optimal.  As in~\cite{engquist2016optimal}, this leads to the estimate
\bq \E\tilde{W}(f,g) = \E\int f(x)\abs{x-T(x)}^2\,dx \leq \E\int f(x)\abs{x-\tilde{T}(x)}^2 = \bO\left(\frac{1}{N}\right). \eq
Thus for typical seismic data, the effect of noise is expected have a negligible effect on the behavior of the squared Wasserstein metric.

\section{Numerical scheme}
In this section, we describe the numerical schemes we use to compute the $W_2$ misfit.  We also explain the adjoint source that is needed for efficient inversion on geophysical data.

\subsection{Data normalization}\label{sec:normalization}
In optimal transport theory, there are two main requirements for signals $f$ and $g$: positivity and mass balance.  Since these are not expected for seismic signals, some data pre-processing is needed before we can implement Wasserstein-based FWI.
In~\cite{EFWass,engquist2016optimal}, the signals were separated into positive and negative parts $f^+ = \max\{f,0\}$, $f^- = \max\{-f,0\}$ and scaled by the total mass $\langle f \rangle = \int_X f(x)\,dx$.  Inversion was accomplished using the modified misfit function
\bq W_2^2\left(\frac{f^+}{\langle f^+ \rangle},\frac{g^+}{\langle g^+ \rangle} \right) + W_2^2\left(\frac{f^-}{\langle f^- \rangle}, \frac{g^-}{\langle g^- \rangle}\right). \eq

While this approach preserves the desirable theoretical properties of convexity to shifts and noise insensitivity, it is not easy to combine with the adjoint-state method and more realistic examples. We require the scaling function to be differentiable so that it is easy to apply chain rule when calculating the Fr\'{e}chet derivative and also better suited for the \MA solver. 

There are other different ways to rescale the data sets so that they become positive. For example, we can square the data as $\tilde{f} = f^2$ or extract the envelope of the data. The convexity with respect to shifts is preserved by these methods but we have lost some information in the gradient. In the squaring case, the gradient of $W_2$ with respect to $f$ is zero when $f$ is zero, which can cause severe difficulties in recovering reflections. The envelope approach, on the other hand, loses important phase information.

In this paper we propose normalization via a linear transformation and rescaling. We begin by selecting a constant $c$ such that both $f+c > 0$ and $g+c>0$. In experiments, $c$ is chosen around 1.1 times $|g_{min}|$. This constant is fixed in inversion. After shifting the signals to ensure positivity, we rescale so all signals share a common total mass. Thus we obtain the modified data $\tilde{f} = P(f)$ and $\tilde{g} = P(g)$ where
\bq\label{eq:rescale} P(f) \equiv \frac{f +c}{\langle f+c \rangle}. \eq


This normalization has several advantages. First, it preserves the phase in that the number and location of local maximum and minimum are maintained and the regularity is good for the adjoint-state method. The normalization function $P(f)$ does not change significantly from iteration to iteration because of the mean zero property of the data, which helps in convergence. There is, however, a serious problem in that this normalization results in a misfit function that is not convex with respect to simple shifts. 
Nevertheless, the full inverse problem is very complex and convexity with respect to shifts is just one aspect. 
Our empirical experience is that this linear normalization works remarkably well on realistic examples, but we believe further research is desirable in solving this problem and increasing the understanding of convergence properties.

To simplify notation, we will hereafter use $f$ and $g$ denoting their normalized version $\tilde{f}$ and $\tilde{g}$ by \eqref{eq:rescale}.

\subsection{Compare trace by trace: $W_2^2(f,g)$ in 1D}
We first describe the scheme used for the one-dimensional Wasserstein metric, which we use to compare the data trace by trace for an overall misfit:
\bq
d(f,g) = \sum\limits_{r=1}^R W_2^2(f(x_r,t),g(x_r,t)),
\eq
where $x_r$ denotes the receiver location.

\subsubsection{Computation of the objective function}
In this setting, if the last time record for a receiver is at $T_0$, we can use the exact formula~\eqref{myOT1D} to express the 1D quadratic Wasserstein metric as
\bq\label{myOT1D2}
W_2^2(f,g) = \int\limits_0^{T_0}|t-G^{-1}(F(t))|^2 f(t)dt
\eq
where $F$ and $G$ are the cumulative distribution functions for $f$ and $g$ respectively: \(
F(t) = \int_{0}^t f, \quad G(t) = \int_{0}^t g\).

This will be approximated in a discrete setting; that is, assuming that $f$ and $g$ are given at a discrete set of points $t = (t_0, t_1 ,\dots,t_n)^T$ in the time domain. We compute $F$ and $G$ using numerical integration.
For each value $y$, since $G$ is monotone increasing we can find $t_n$ and $t_{n+1}$ such that $G(t_n)< y \leq G(t_{n+1})$ in $\bO(\log(N))$ complexity by binary search and $N$ is the number of data samples in each trace. For $y$ in this range we can estimate $G^{-1}(y) = t_{n+1}$. Here we will also do numerical interpolation between $t_n$ and $t_{n+1}$ for better accuracy. 

Using finite difference matrices, we can express the discrete 1D quadratic Wasserstein metric as
\bq\label{eq:wassDisc}
d_1(f ,g) = (t-G^{-1}\circ F(t))^T\diag(f)(t-G^{-1}\circ F(t)),
\eq
where $G^{-1}\circ F$ is the optimal map that transports $f$ onto $g$.

After summing over all the traces, we obtain the final misfit between the synthetic data and observed data: $d(f,g) = \sum_{r=1}^R d_1(f_r,g_r)$.
By exploiting the explicit solution for optimal transport on the real line, we are able to compute the misfit in $\bO(N\log(N))$ complexity.

\subsubsection{Computation of adjoint source}
\label{sec:linear}
We also require the Fr\'{e}chet gradient of the misfit, which acts as the adjoint source in the adjoint-state method.

The first variation of the squared Wasserstein metric for the 1D case is
\begin{align*}
\delta d_1 & = -2\  \diag\left(\frac{d G^{-1}(y)}{dy}\biggr\rvert_{F(t)}\right) \diag(f) \diag(\delta f) L (t-G^{-1}\circ F(t))    \numberthis \label{eqn:DW2_1D_1}   \\ 
&\quad + \diag(t-G^{-1}\circ F(t)) \diag(\delta f) (t-G^{-1}\circ F(t)), 
\end{align*}
where $L$ is the lower triangular matrix whose non-zero components are 1.

By the inverse function theorem, we have:
\bq
\frac{d G^{-1}(y)}{dy}\biggr\rvert_{F(t)} = \frac{1}{\frac{dG(t)}{dt}\biggr\rvert_{G^{-1}\circ F (t)}} = \frac{1}{g(G^{-1}\circ F (t))}
\eq
Then the adjoint source term for the discrete 1D quadratic Wasserstein metric can be expressed as
\bq\label{eqn:DW2_1D_2}
\nabla d_1(t) = \left[-2 \ \diag \left(\frac{f(t)}{g(G^{-1}\circ F (t))}  \right)L +  \diag(t-G^{-1}\circ F(t)) \right] (t-G^{-1}\circ F(t)). \eq

\subsection{Compare globally: $W_2^2(f,g)$ in higher dimensions}
Secondly, we wish to examine the effects of comparing the data $f$ and $g$ globally via a single, higher-dimensional optimal transportation computation.

\subsubsection{Computation of the objective function}

In this case there is no simple exact formula for the Wasserstein metric.  Instead, we will compute it via the solution of the \MA equation:

\bq\label{eq:MAA}
\begin{cases}
\det(D^2u(x)) = f(x)/g(\nabla u(x)) + \langle u \rangle,& x\in X\\
\nabla u(x) \cdot n = x\cdot n, & x \in \partial X\\
u \text{ is convex.}
\end{cases}
\eq

The squared quadratic Wasserstein metric is then given by
\bq\label{eq:WassMA}
W_2^2(f,g) = \int_X f(x)\abs{x-\nabla u(x)}^2\,dx.
\eq

We solve the \MA equation numerically using an almost-monotone finite difference method relying on the following reformulation of the \MA operator, which automatically enforces the convexity constraint~\cite{FroeseTransport}.
\begin{multline}\label{eq:MA_convex}
{\det}^+(D^2u) = \\ \min\limits_{\{v_1,v_2\}\in V}\left\{\max\{u_{v_1,v_1},0\} \max\{u_{v_2,v_2},0\}+\min\{u_{v_1,v_1},0\} + \min\{u_{v_2,v_2},0\}\right\}
\end{multline}
where $V$ is the set of all orthonormal bases for $\R^2$.  

Equation~\eqref{eq:MA_convex} can be discretized by computing the minimum over finitely many directions $\{\nu_1,\nu_2\}$, which may require the use of a wide stencil.  For simplicity and brevity, we describe a low-order version of the scheme and refer to~\cite{FroeseTransport,FOFiltered} for complete details.  In practice, this simplified scheme is sufficient for obtaining accurate inversion results.

The scheme relies on the finite difference operators
\begin{align*}
[\Dt_{x_1x_1}u]_{ij} &= \frac{1}{dx^2} 
\left(
{u_{i+1,j}+u_{i-1,j}-2u_{i,j}}
\right)
\\
[\Dt_{x_2x_2}u]_{ij} &= \frac{1}{dx^2}
\left(
u_{i,j+1}+u_{i,j-1}-2u_{i,j}
\right)
\\
[\Dt_{x_1x_2}u]_{ij} &= \frac{1}{4dx^2}
\left(
u_{i+1,j+1}+u_{i-1,j-1}-u_{i+1,j-1}-u_{i-1,j+1}
\right)
\\
[\Dt_{x_1}u]_{ij} &= \frac{1}{2dx}
\left(
u_{i+1,j}-u_{i-1,j}
\right)\\
[\Dt_{x_2}u]_{ij} &= \frac{1}{2dx}
\left(
u_{i,j+1}-u_{i,j-1}
\right)\\
[\Dt_{vv}u]_{ij} &= \frac{1}{2dx^2}\left(u_{i+1,j+1}+u_{i-1,j-1}-2u_{i,j}\right)\\
[\Dt_{\vp\vp}u]_{ij} &= \frac{1}{2dx^2}\left(u_{i+1,j-1}+u_{i+1,j-1}-2u_{i,j}\right)\\
[\Dt_{v}u]_{ij} &= \frac{1}{2\sqrt{2}dx}\left(u_{i+1,j+1}-u_{i-1,j-1}\right)\\
[\Dt_{\vp}u]_{ij} &= \frac{1}{2\sqrt{2}dx}\left(u_{i+1,j-1}-u_{i-1,j+1}\right).
\end{align*}

In the low-order version of the scheme, the minimum in~\eqref{eq:MA_convex} is approximated using only two possible values.  The first uses directions aligning with the grid axes.
\begin{multline}\label{MA1}
MA_1[u] = \max\left\{\Dt_{x_1x_1}u,\delta\right\}\max\left\{\Dt_{x_2x_2}u,\delta\right\} \\+ \min\left\{\Dt_{x_1x_1}u,\delta\right\} + \min\left\{\Dt_{x_2x_2}u,\delta\right\} - f / g\left(\Dt_{x_1}u, \Dt_{x_2}u\right) - u_0.
\end{multline}
Here $dx$ is the resolution of the grid, $\delta>K\Delta x/2$ is a small parameter that bounds second derivatives away from zero, $u_0$ is the solution value at a fixed point in the domain, and $K$ is the Lipschitz constant in the $y$-variable of $f(x)/g(y)$.

For the second value, we rotate the axes to align with the corner points in the stencil, which leads to
\begin{multline}\label{MA2}
MA_2[u] = \max\left\{\Dt_{vv}u,\delta\right\}\max\left\{\Dt_{\vp\vp}u,\delta\right\} + \min\left\{\Dt_{vv}u,\delta\right\} + \min\left\{\Dt_{\vp\vp}u,\delta\right\}\\ - f / g\left(\frac{1}{\sqrt{2}}(\Dt_{v}u+\Dt_{\vp}u), \frac{1}{\sqrt{2}}(\Dt_{v}u-\Dt_{\vp}u)\right) - u_0.
\end{multline}

Then the monotone approximation of the \MA equation is
\bq\label{eq:MA_compact} M_M[u] \equiv -\min\{MA_1[u],MA_2[u]\} = 0. \eq
We also define a second-order approximation, obtained from a standard centred difference discretisation,
\bq\label{eq:MA_nonmon} M_N[u] \equiv -\left((\Dt_{x_1x_1}u)(\Dt_{x_2x_2}u)-(\Dt_{x_1x_2}u^2)\right) + f/g\left(\Dt_{x_1}u,\Dt_{x_2}u\right) + u_0 = 0.\eq

These are combined into an almost-monotone approximation of the form
\bq\label{eq:MA_filtered} M_F[u] \equiv M_M[u] + \epsilon S\left(\frac{M_N[u]-M_M[u]}{\epsilon}\right) \eq
where $\epsilon$ is a small parameter and the filter $S$ is given by
\bq\label{eq:filter}
S(x) = \begin{cases}
x & \abs{x} \leq 1 \\
0 & \abs{x} \ge 2\\
-x+ 2  & 1\le x \le 2 \\
-x-2  & -2\le x\le -1.
\end{cases} 
\eq

The Neumann boundary condition is implemented using standard one-sided differences.  As described in~\cite{engquist2016optimal,FroeseTransport}, the (formal) Jacobian $\nabla M_F[u]$ of the scheme can be obtained exactly.  In particular, it is known to be sparse and diagonally dominant.

This finite difference approximation effectively replaces the \MA equation with a large system of nonlinear algebraic equations, which can be solved using Newton's method.
Computing the Newton updates requires inverting sparse $M$-matrices, which can be done efficiently.  The number of Newton iterations required depends weakly on the smoothness of the data and resulting solution $u$.  In numerical experiments carried out in~\cite{FroeseTransport}, the total computational complexity required to solve the Monge-Amp\`ere equation varied from $\bO(N)$ to $\bO(N^{1.3})$ where $N$ was the total number of grid points.

Once the discrete solution $u_h$ is computed, the squared Wasserstein metric is approximated via
\bq\label{eq:WassDiscrete}  W_2^2(f,g) \approx \sum\limits_{j=1}^n (x_j-D_{x_j}u_h)^T\diag(f)(x_j-D_{x_j}u_h). \eq

\subsubsection{Computation of adjoint source}
\label{sec:linear}
In \cite{engquist2016optimal}, we consider the linearisation of the discretised version of the Wasserstein metric.  Using the finite difference matrices introduced, we can express the discrete Wasserstein metric as
\bq\label{eq:wassDisc}
d(f) = \sum\limits_{j=1}^n (x_j-D_{x_j}u_f)^T\diag(f)(x_j-D_{x_j}u_f),
\eq
where $n$ is the data dimension, the potential $u_f$ satisfies the discrete \MA equation
\bq\text{M}[u_f] = 0.\eq
The first variation of the squared Wasserstein metric as
\bq \delta d = -2\sum\limits_{j=1}^n (D_{x_j} \delta u)^T \diag(f) (x_j - D_{x_j}u_f) + \sum\limits_{j=1}^n (x_j-D_{x_j}u_f)^T \diag(\delta f) (x_j - D_{x_j}u_f). \eq

Linearising the \MA equation, we have to first order
\bq \nabla M_F[u_f] \delta u = \delta f. \eq
Here $\nabla M_F$ is the (formal) Jacobian of the discrete \MA equation, which is already being inverted in the process of solving the \MA equation via Newton's method.  Then the gradient of the discrete squared Wasserstein metric can be expressed as
\bq \nabla d = \sum\limits_{j=1}^n\left[-2\nabla M_F^{-1}[u_f]^TD_{x_j}^T\diag(f) + \diag(x_j-D_{x_j}u_f)\right](x_j - D_{x_j}u_f). \eq
Notice that once the \MA equation itself has been solved, this gradient is easy to compute as it only requires the inversion of a single matrix that is already being inverted as a part of the solution of the \MA equation.

\section{Computational results}
In this section, we apply the quadratic Wasserstein metric ($W_2$) to several synthetic data models. We provide results for two approaches to using $W_2$: trace-by-trace comparison and using the entire data sets as objects.  These are compared with results produced by using the least-squares norm $L^2$ to measure the misfit. 
Due to limitations of the current \MA solver, we will present global $W_2$ based FWI on smaller scale models.
In the inversion process, we avoid the use of techniques such as adding regularization and smoothing the gradient in order to see the effectiveness of this new misfit.

\subsection{1D case study}\label{sec:numerical_1D}
\begin{figure}
\centering
\subfloat[]{\includegraphics[width=0.45\textwidth]{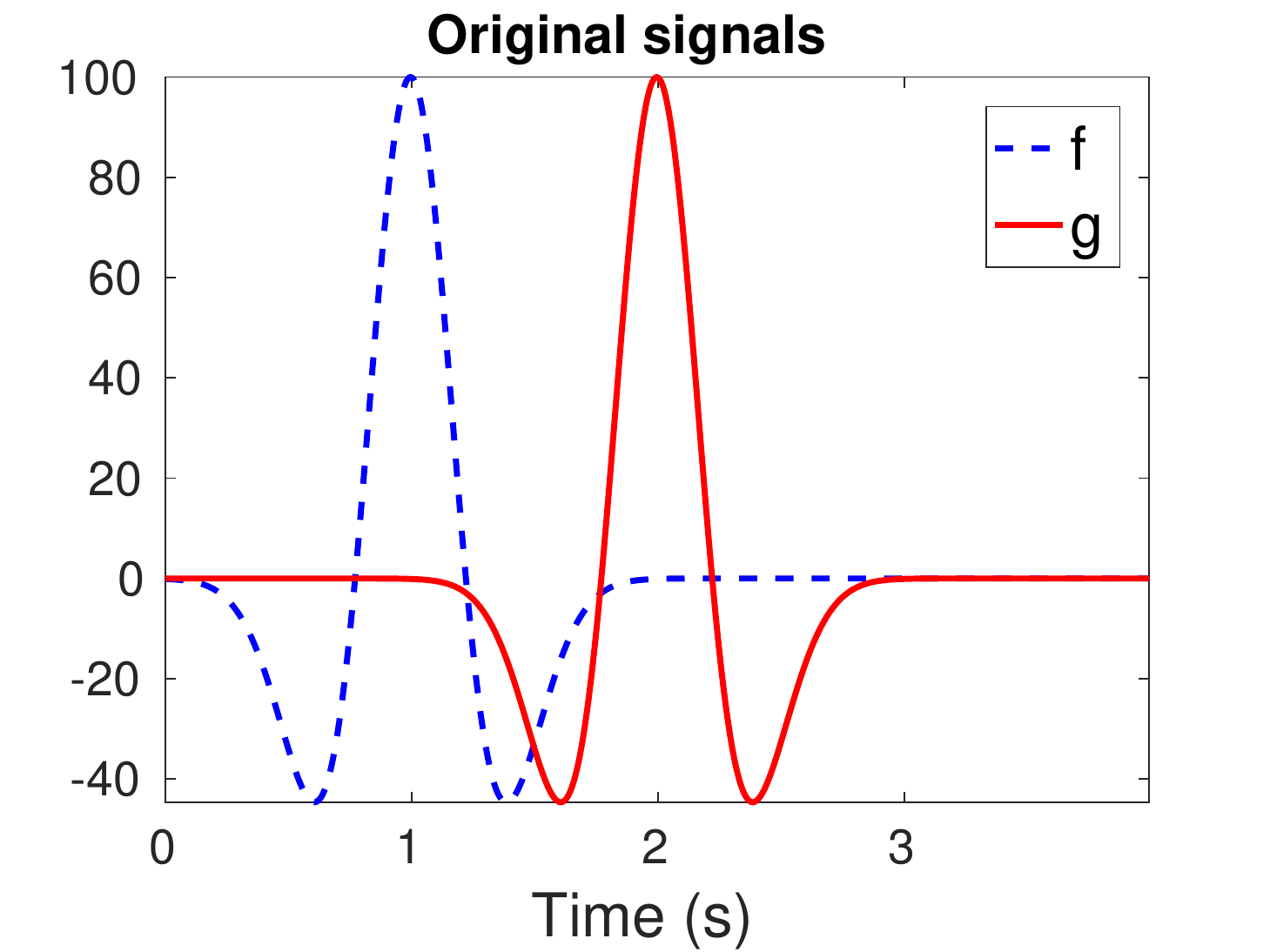}\label{fig:raw1D}}
\subfloat[]{\includegraphics[width=0.45\textwidth]{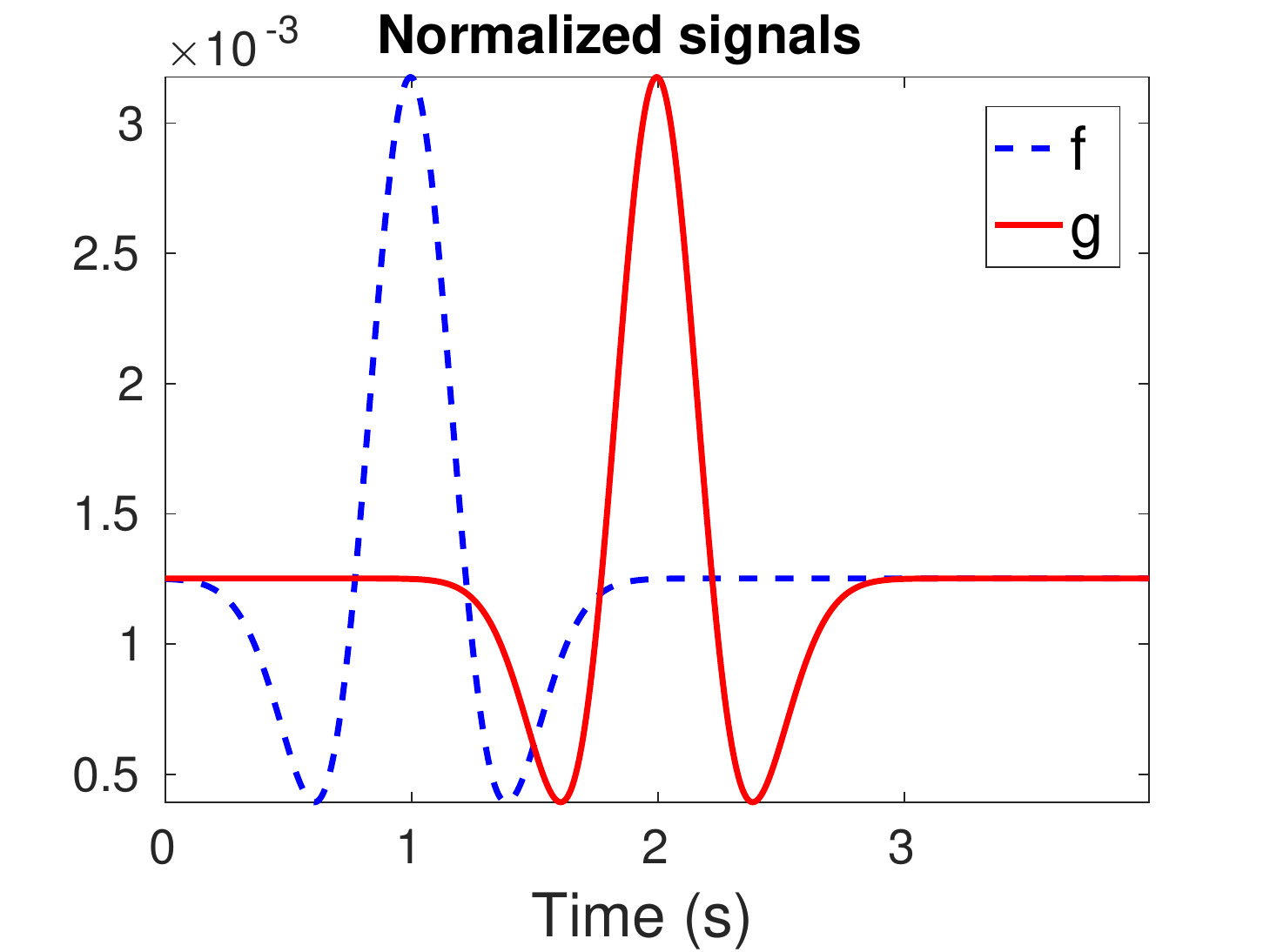}\label{fig:norm1D}}
\caption{(A)~Original synthetic signal $f$ and observed signal $g$ and (B)~normalized synthetic signal $f$ and observed signal $g$ that satisfy the requirements of optimal transport. }\label{fig:1D_data}
\end{figure}

We begin with a simple test case from~\cite{EFWass} and focus on two Ricker wavelet signals, one a time shift of the other. We regard these two signals as observed data $g(t)$ and synthetic data $f(t;s) = g(t-s)$  as shown in Figure~\ref{fig:1D_data}. This is a case in which quadratic Wasserstein metric ($W_2$) is applied to 1D signals. 

\begin{figure}
\centering
\subfloat[]{\includegraphics[width=0.45\textwidth]{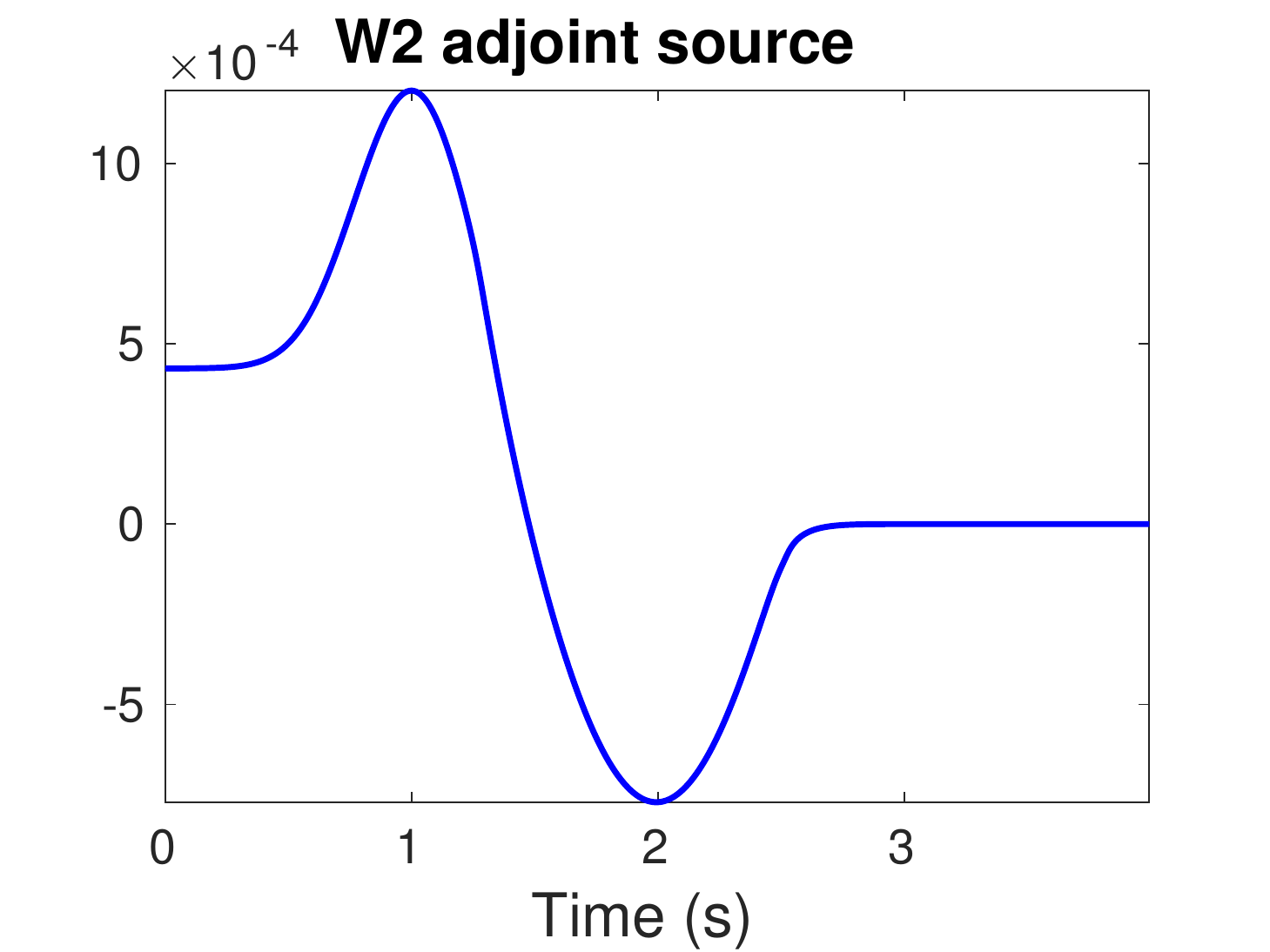}\label{fig:raw1D}}
\subfloat[]{\includegraphics[width=0.45\textwidth]{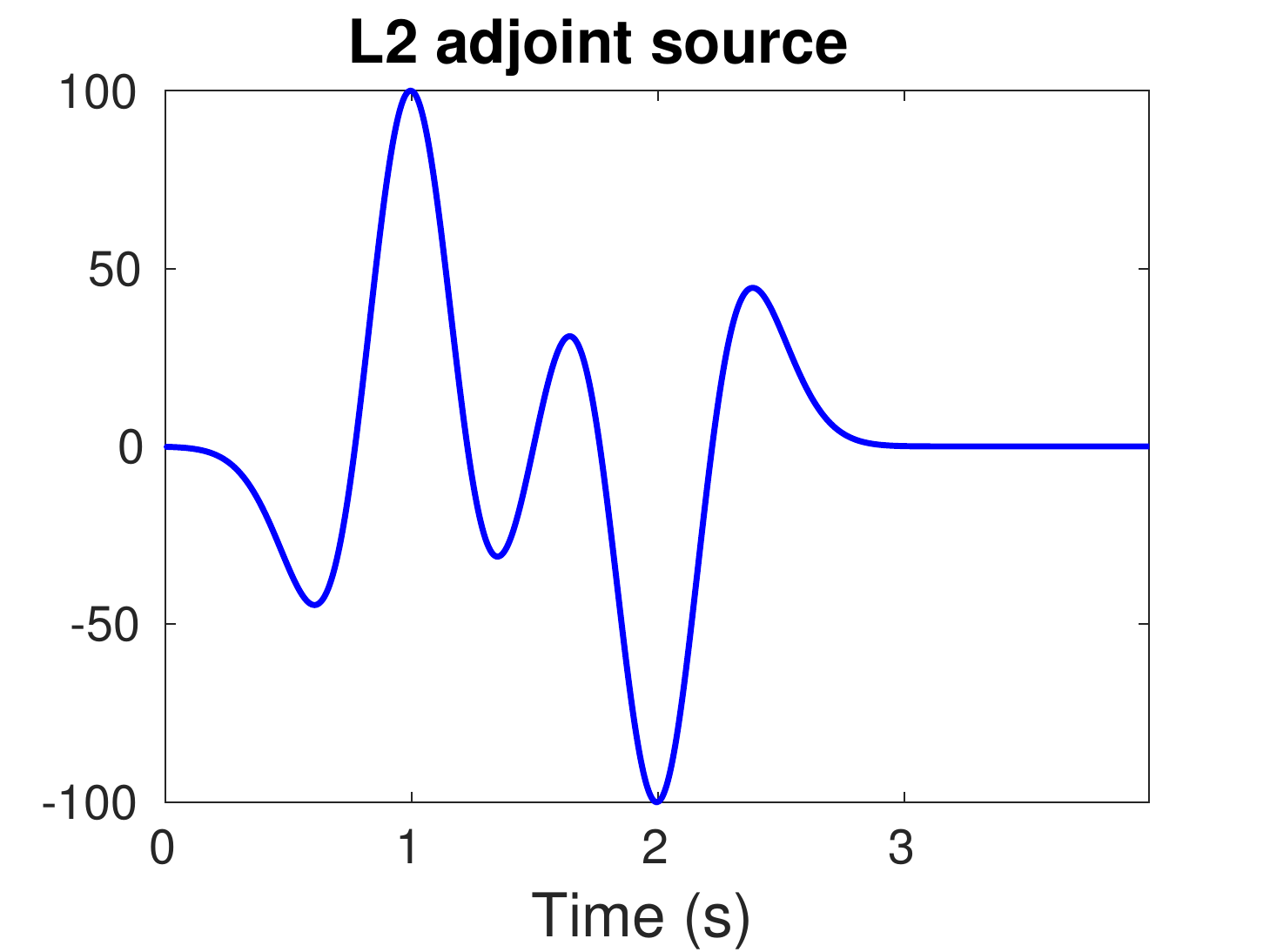}\label{fig:norm1D}}
\caption{(A)~Adjoint source of $W_2^2(f,g)$ with respect to $f$ and (B)~adjoint source of $L^2(f,g)$ with respect to $f$ }
\label{fig:1D_AS}
\end{figure}

The adjoint source for $L^2$ and $W_2$ misfits between these two signals are very different as shown by Figure~\ref{fig:1D_AS}. The adjoint source for $W_2$ is very similar to the adjoint source of the KR norm applied on this 1D case; see Figure~4 of \cite{W1_3D} for more detail. This illustrates the character of optimal transport based misfit functions, which shift mass from the synthetic data to observed data in a way that corrects the phase difference between $f$ and $g$. The $L^2$ norm, on the other hand, only seeks to correct the amplitude difference, which is the origin of the cycle skipping. 

We observe that the adjoint source of $W_2$ is smoother than the adjoint source of the KR norm (Figure 4 in \cite{W1_3D}) and has no DC component. The smoothness of the adjoint source is ideal for quasi-Newton methods, e.g. the l-BFGS algorithm, which are designed to minimize smooth functions. It is also numerically more stable to back propagate in time in order to compute gradient updates.

\subsection{Camembert model}\label{sec:numerical_Camembert}

\begin{figure}
\centering
  \subfloat[]{\includegraphics[width=0.45\textwidth]{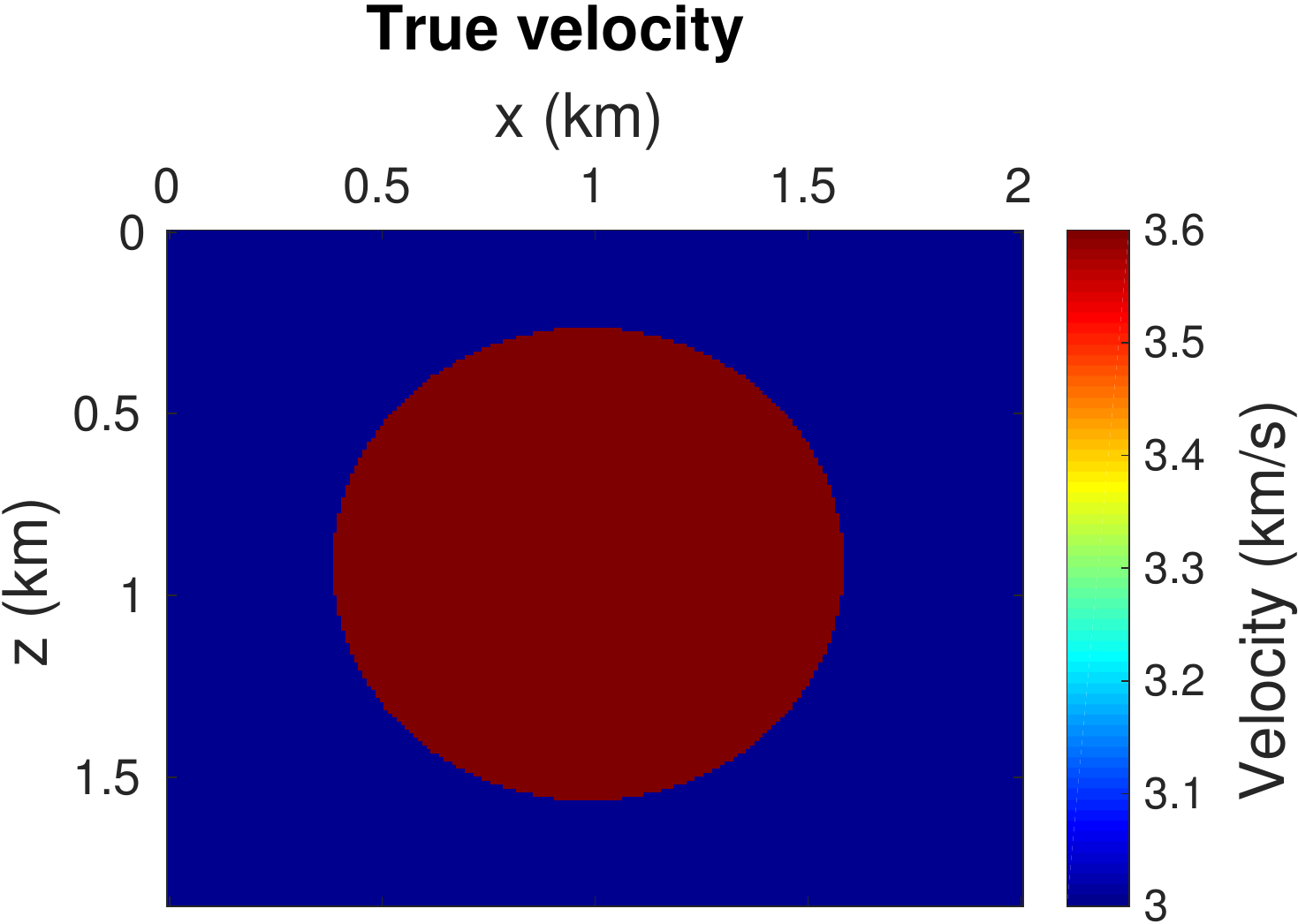}\label{fig:cheese_true}}
  \subfloat[]{\includegraphics[width=0.45\textwidth]{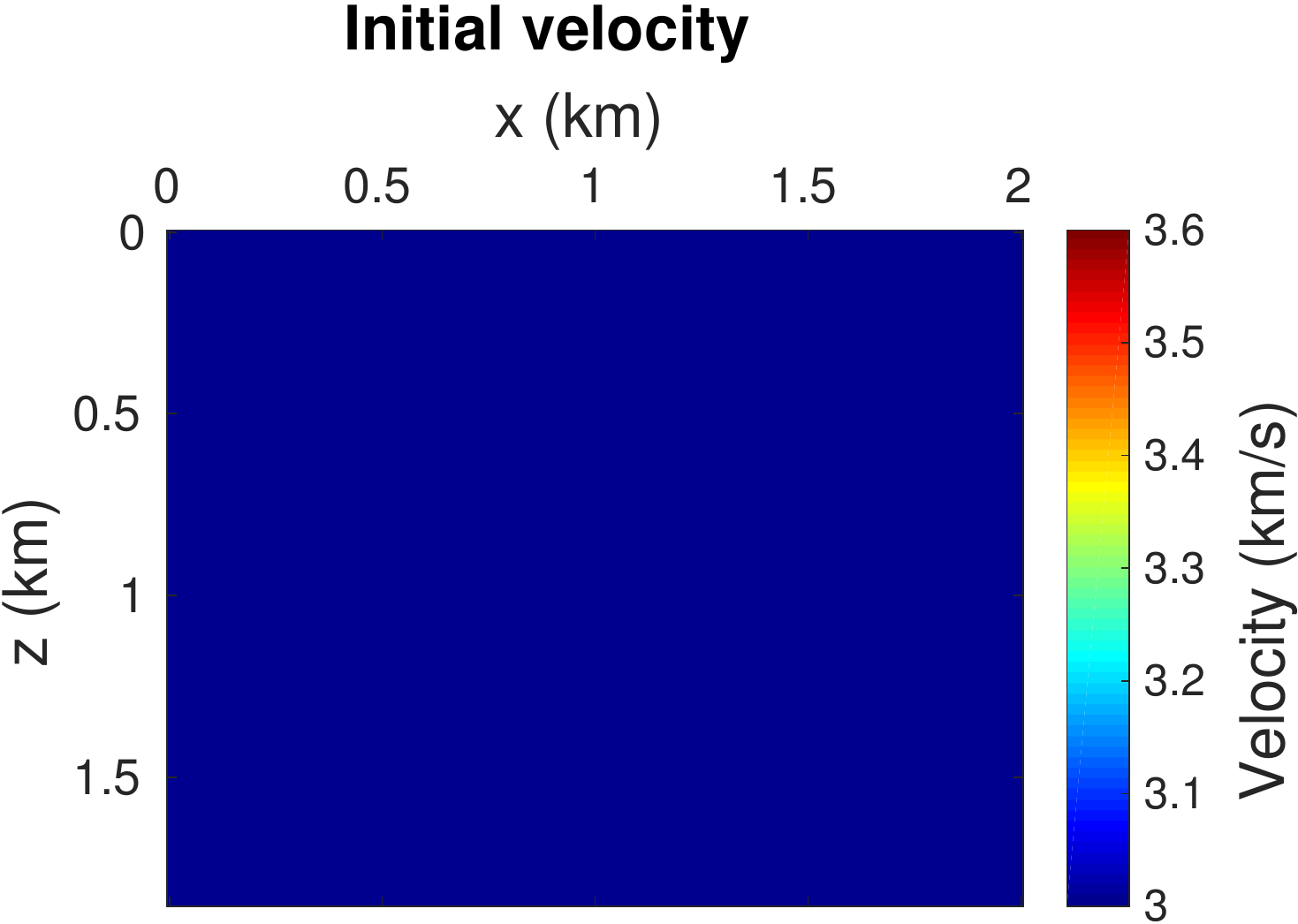}\label{fig:cheese_v0}}
  \caption{(A)~True velocity and (B)~inital velocity for Camembert model}
  \label{fig:cheese_true,cheese_v0}
\end{figure}

\begin{figure}
\centering
  \subfloat[]{\includegraphics[width=0.45\textwidth]{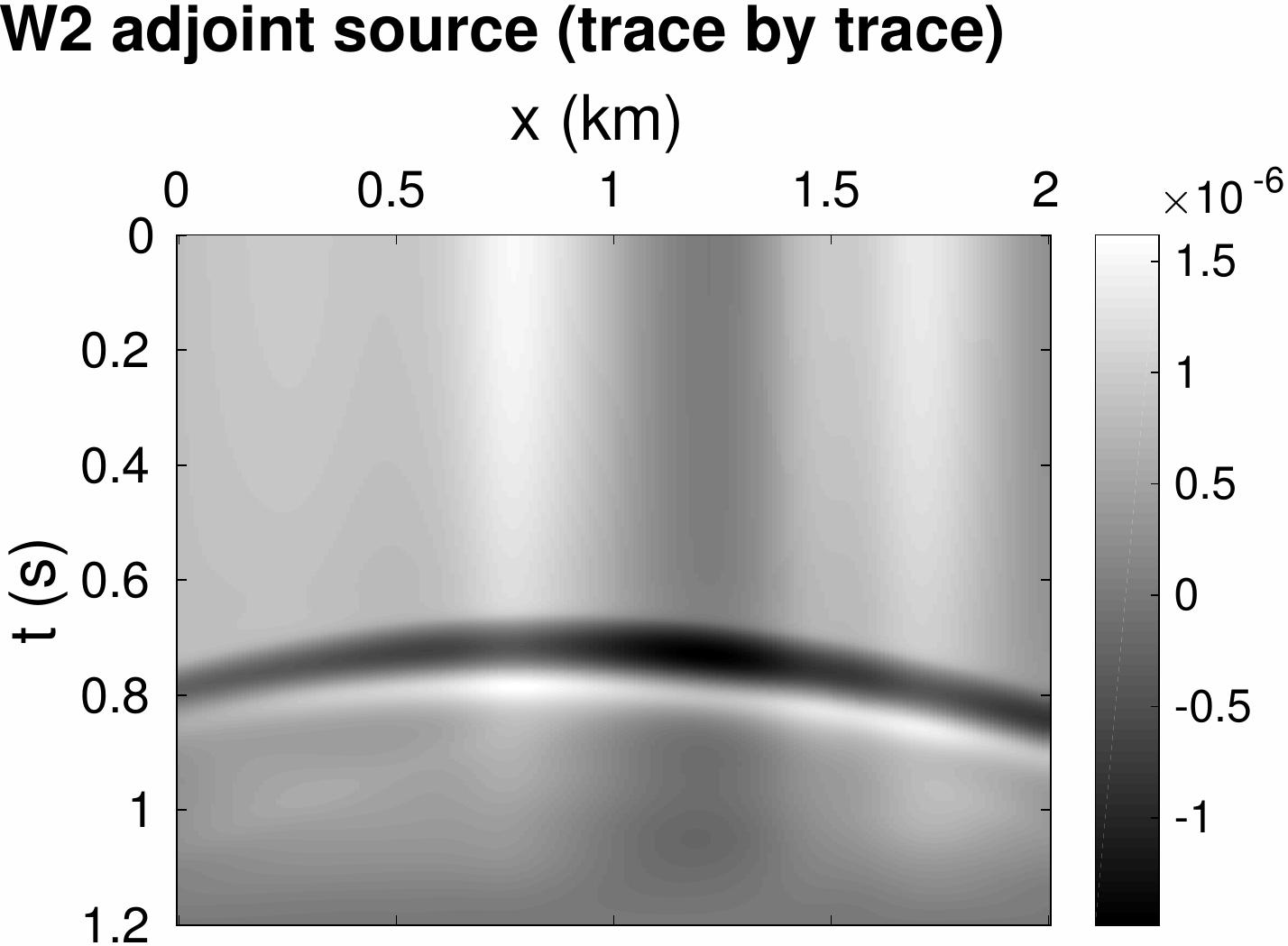}\label{fig:cheese_dw2_1D}}
 \subfloat[]{\includegraphics[width=0.45\textwidth]{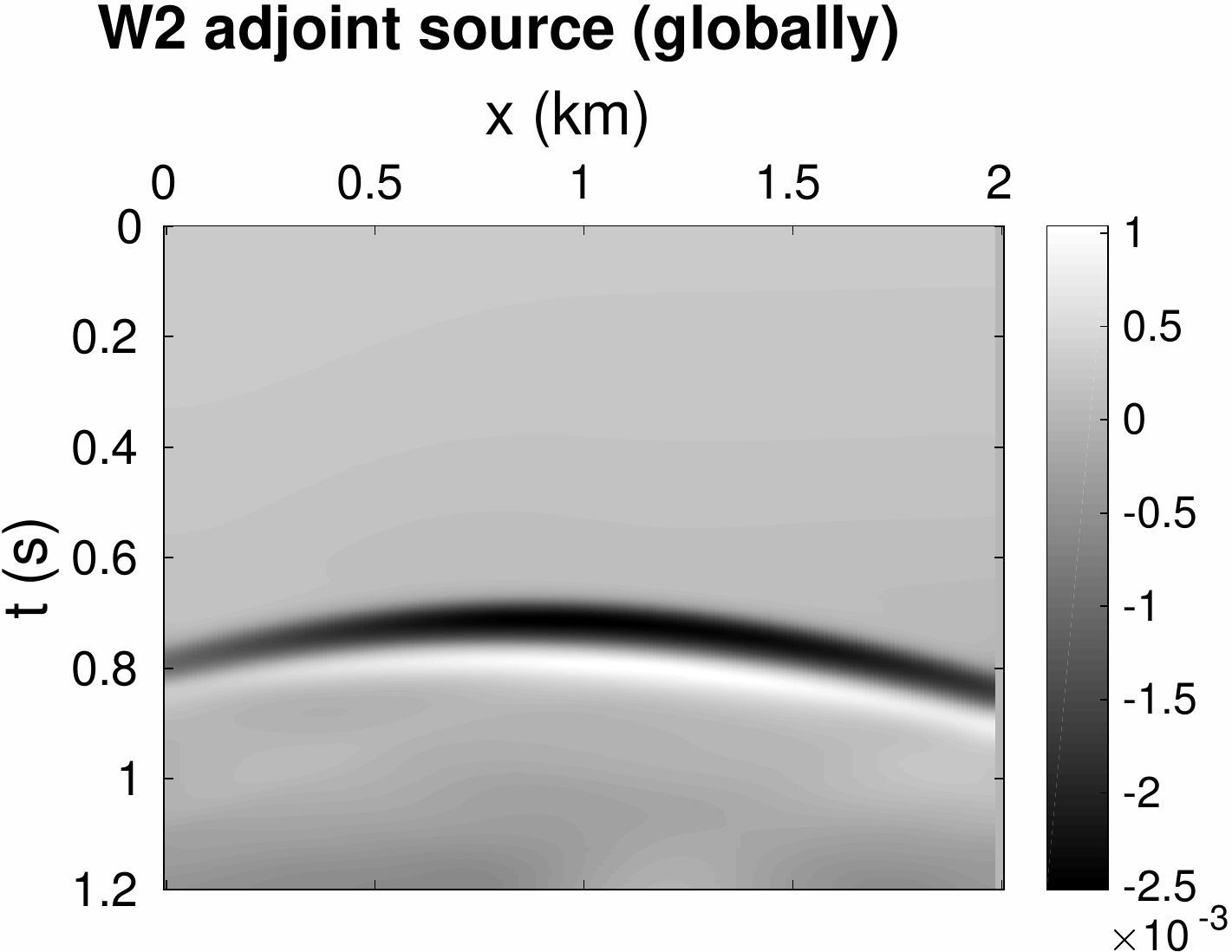}\label{fig:cheese_dw2_2D}}
  \caption{(A)~Adjoint source of $W_2$ processed trace by trace and (B)~adjoint source for global $W_2$ for the Camembert model}
  \label{fig:cheese_dw2}
\end{figure}

\begin{figure}
\centering
\subfloat[]{\includegraphics[width=0.45\textwidth]{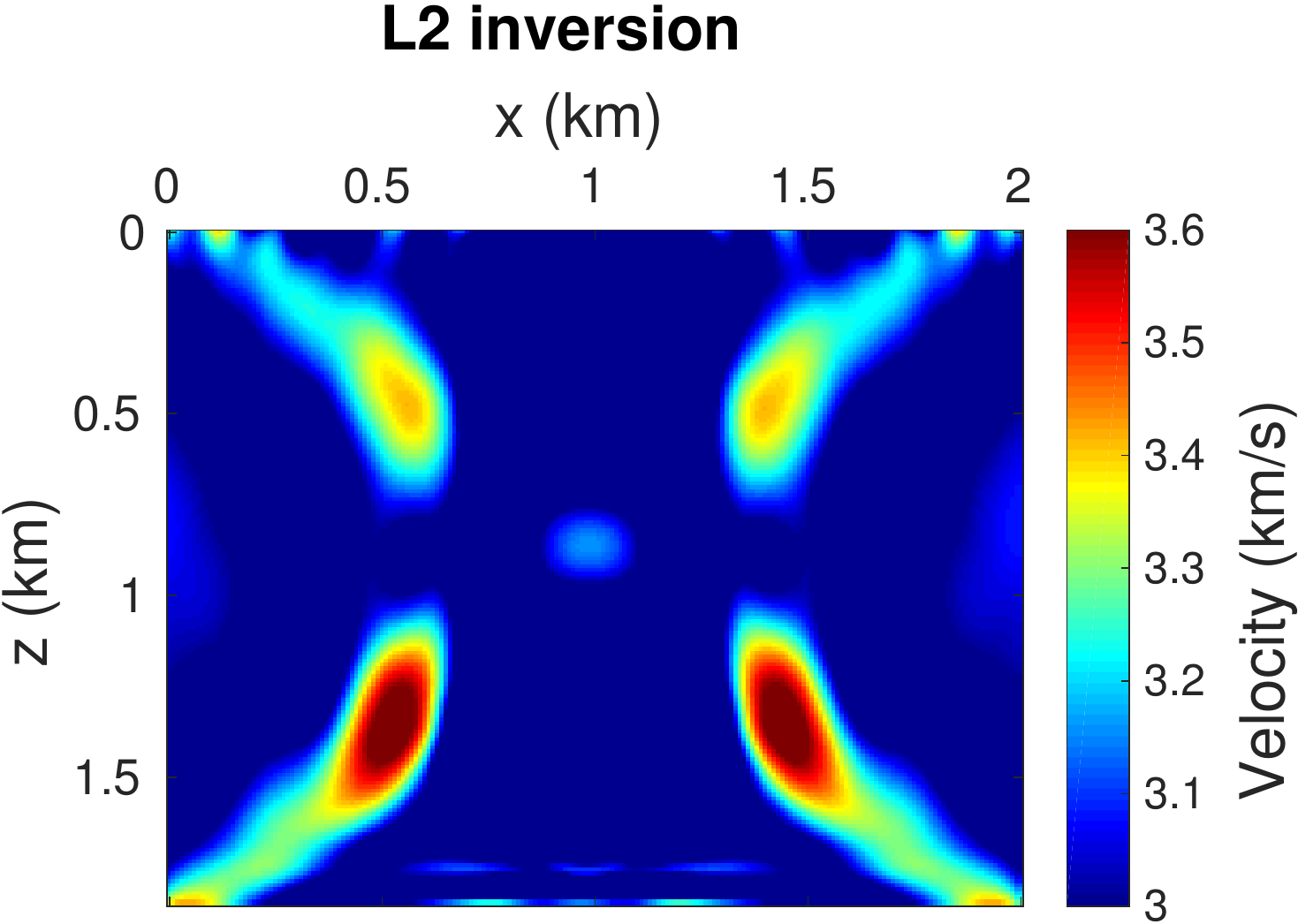}\label{fig:cheese_L2}}  
\subfloat[]{\includegraphics[width=0.45\textwidth]{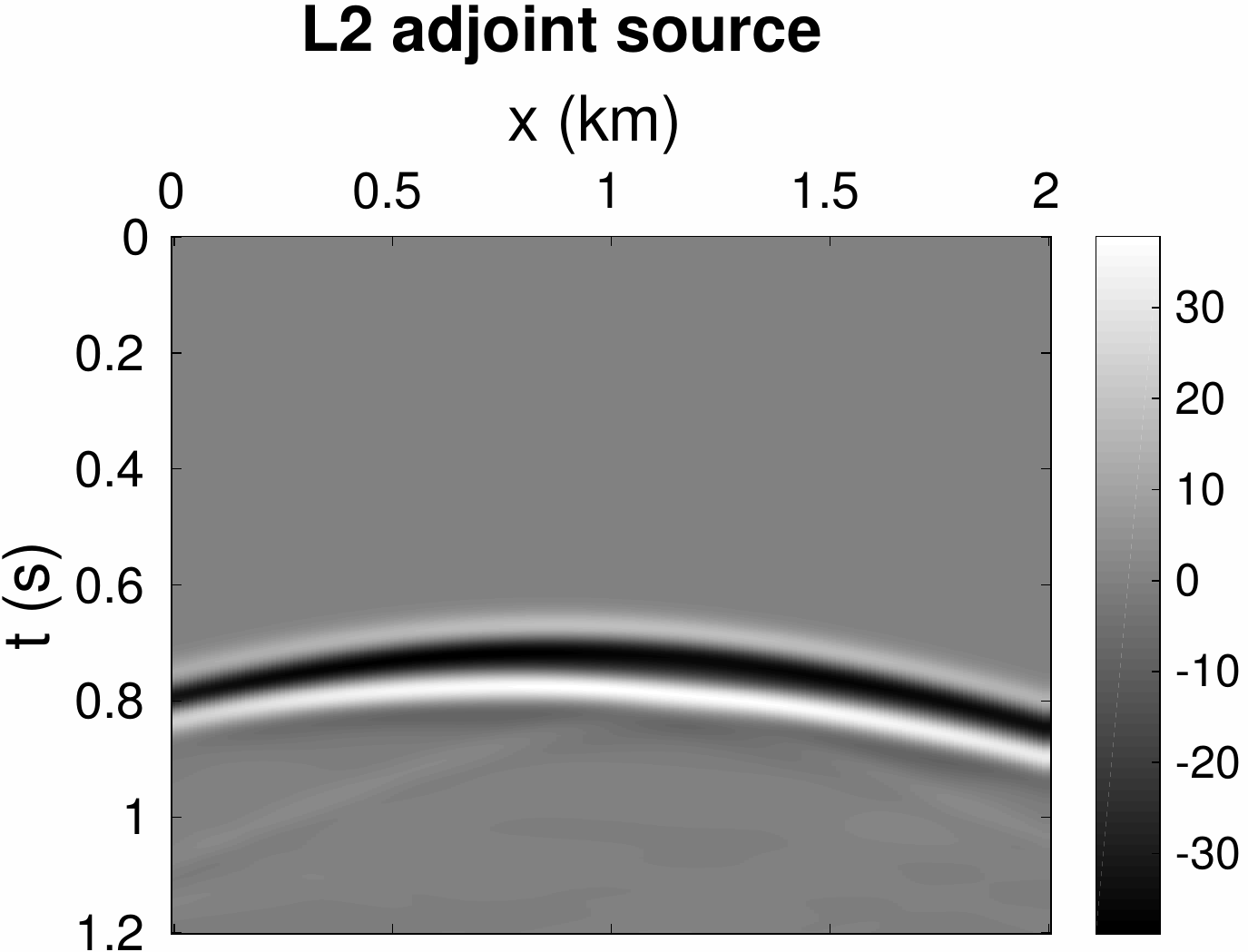}\label{fig:cheese_DL2}}
\caption{(A)~Inversion result using $L^2$ as misfit function and (B)~adjoint source for $L^2$ for the Camembert model}
  \label{fig:cheese_L2}
\end{figure}

\begin{figure}
\centering
  \subfloat[]{\includegraphics[width=0.45\textwidth]{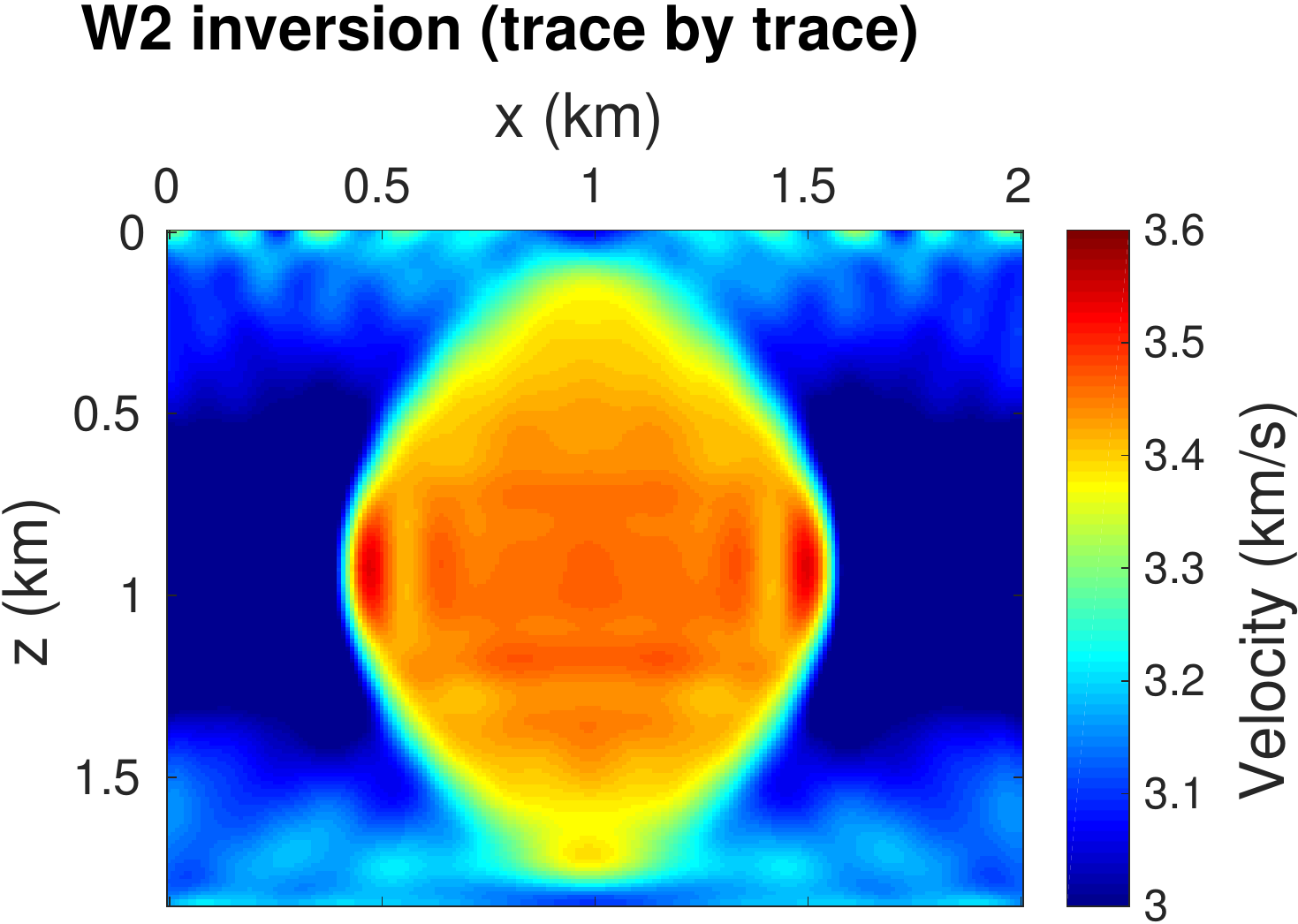}\label{fig:cheese_w2_1D}}
  \subfloat[]{\includegraphics[width=0.45\textwidth]{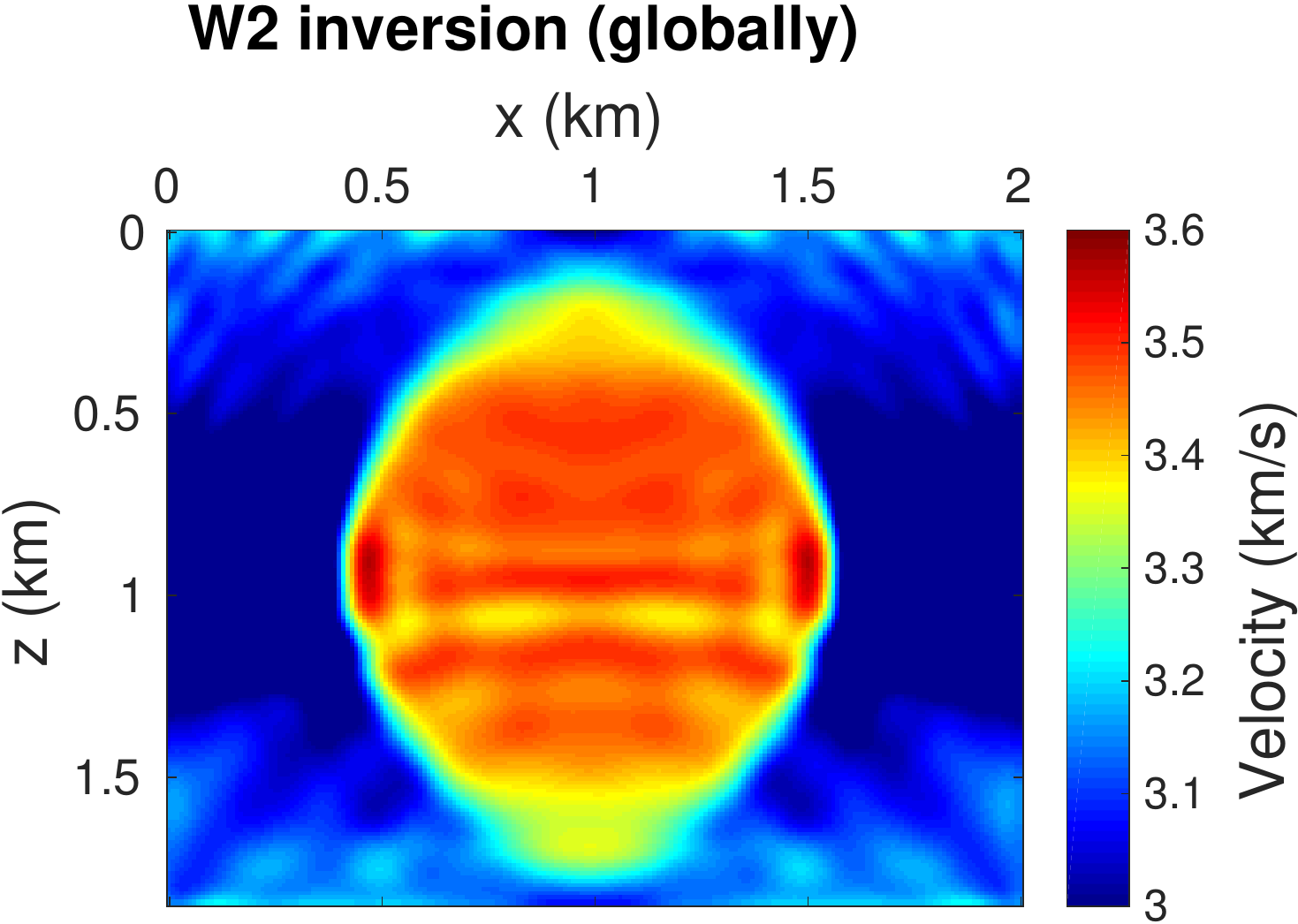}\label{fig:cheese_w2_2D}}
  \caption{(A)~Inversion result for $W_2$ processed trace by trace and (B)~inversion result for global $W_2$  for the Camembert model}
  \label{fig:cheese_w2}
\end{figure}

Full-waveform inversion (FWI) with the least-squares norm ($L^2$) minimization~\cite{tarantola1982generalized} is effective when the initial model is close to the true model. However, if the initial model is far from the true model, the $L^2$ misfit may suffer from local minima since it utilizes a point-by-point comparison that records the oscillatory and nonlinear features of the data. The difficulty of local minima in seismic inversion 
was clearly demonstrated with the so-called Camembert example~\cite{gauthier1986two}.

We repeat the experiments with three different misfit functions for full-waveform inversion: $W_2$ applied trace by trace, $W_2$ applied globally, and the traditional $L^2$ least-squares norm. The comparison among these three different misfit functions illustrate the advantages of quadratic Wasserstein metric ($W_2$).

We set the Camembert-shaped inclusion as a circle with radius 0.6km located in the center of the rectangular velocity model. The velocity is 3.6km/s inside and 3km/s outside the circle as Figure~\ref{fig:cheese_true}. The inversion starts from an initial model with homogeneous velocity 3km/s everywhere as in Figure~\ref{fig:cheese_v0}. We place 11 equally spaced sources on the  top at 50m depth and 201 receivers on the bottom with 10 m fixed acquisition. The discretization  of the forward wave equation is 10m in the $x$ and $z$ directions and 10ms in time. The source is a Ricker wavelet with a peak frequency of 10Hz, and a high-pass filter is applied to remove the frequency components from 0 to 2Hz. 

We first perform the inversion by using 1D optimal transport to calculate the misfit trace by trace. Figure~\ref{fig:cheese_dw2_1D} and Figure~\ref{fig:cheese_w2_1D} show the adjoint source and final inversion results respectively. Since the data is actually two-dimensional (in both time and spatial domain), an alternative approach is to find the optimal transport map between these two data sets instead of slice them into traces. Figure~\ref{fig:cheese_dw2_2D} and Figure~\ref{fig:cheese_w2_2D} are the adjoint source and final inversion result respectively of comparing the two data sets via a global optimal map. Both approaches converge in 10 iterations using the l-BFGS optimization method.

Figure~\ref{fig:cheese_L2} is the inversion result obtained with the traditional $L^2$ least-squares method. It converges to a local minimum after 100 iterations using the l-BFGS optimization method. Although Figure~\ref{fig:cheese_DL2} looks similar in shape to Figure~\ref{fig:cheese_dw2} at first glance, the adjoint source of $W_2$ based misfits functions only have negative-positive components (``black-white'' curves in Figure~\ref{fig:cheese_dw2}), while the adjoint source for $L^2$ has positive-negative-positive components (``white-black-white'' curves in Figure~\ref{fig:cheese_DL2}). Thus it provides $L^2$ based inversion with an incorrect direction for updating the velocity model, which leads to local minima and cycle skipping. In terms of numerical error in the velocity, the global $W_2$ inversion result (Figure~\ref{fig:cheese_w2_2D}) is more accurate than the trace-by-trace $W_2$ results (Figure~\ref{fig:cheese_w2_1D}).

\subsection{Scaled Marmousi model with global $W_2$ misfit computation}\label{sec:scaled_marm}

\begin{figure}
\centering
  \subfloat[]{\includegraphics[width=0.45\textwidth]{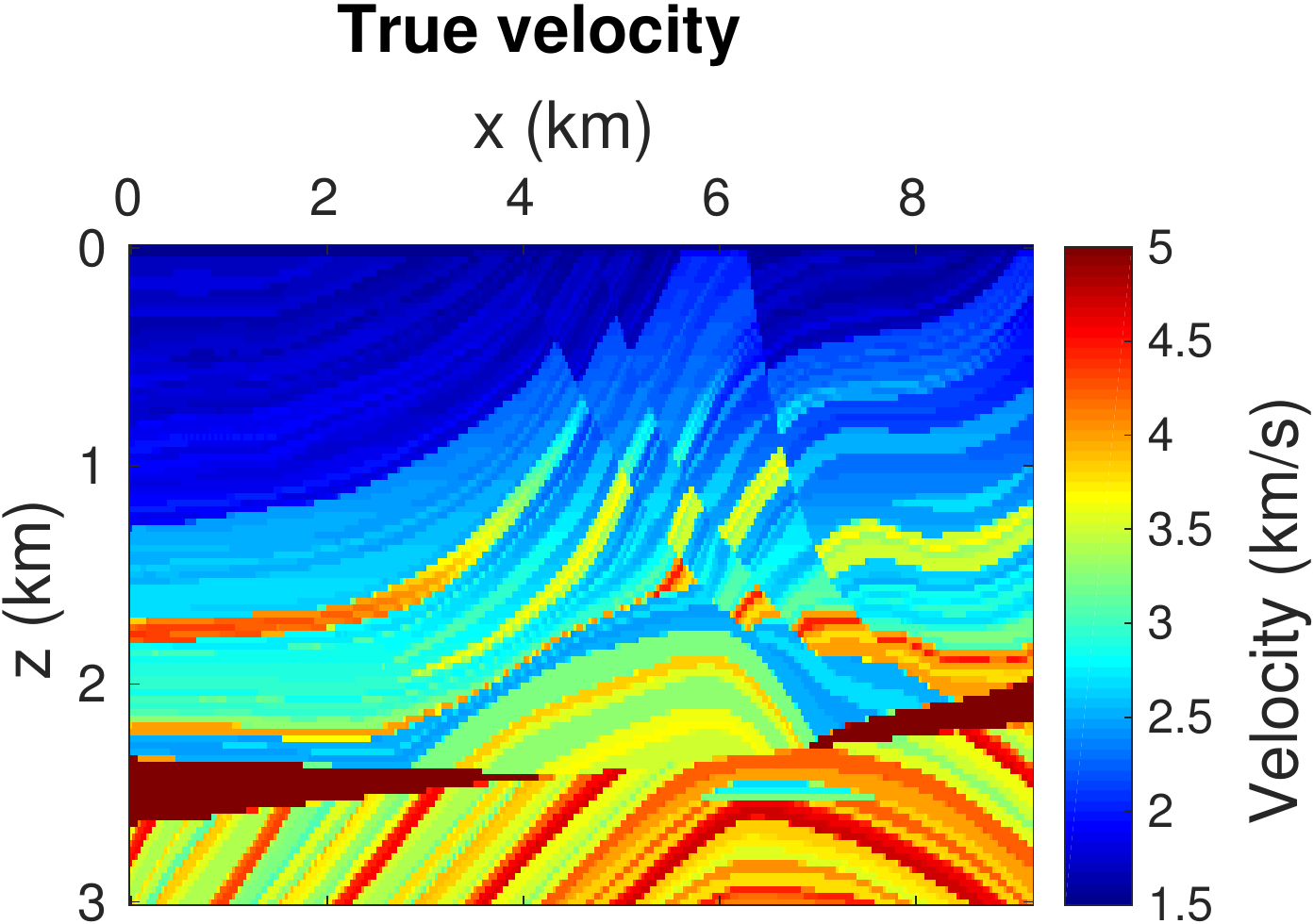}\label{fig:marm2_true}}
  \subfloat[]{\includegraphics[width=0.45\textwidth]{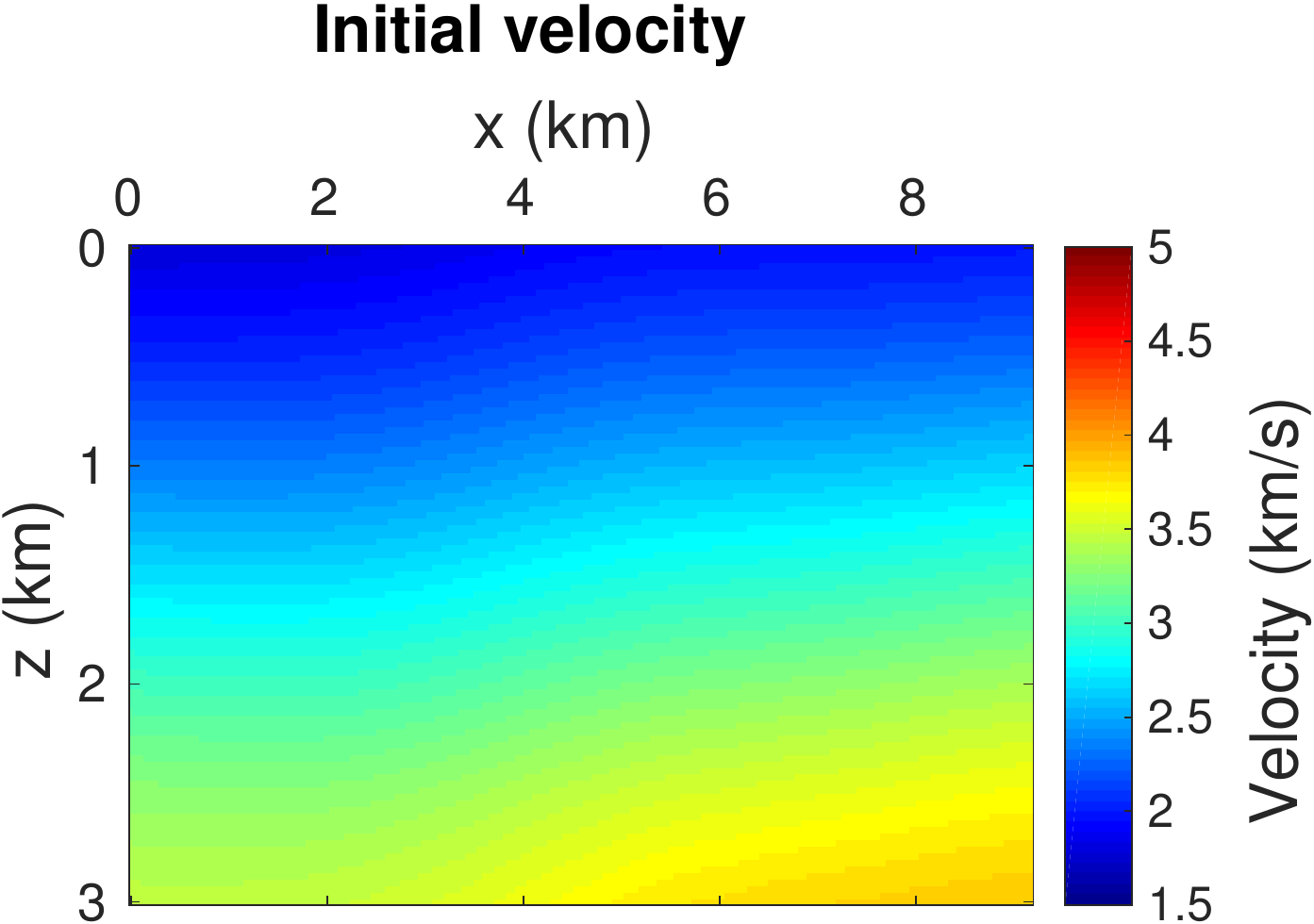}\label{fig:marm2_v0}}
  \caption{(A)~True velocity and (B)~inital velocity for true Marmousi model}
  \label{fig:marm2_true,marm2_v0}
\end{figure}

\begin{figure}
\centering
  \subfloat[]{\includegraphics[width=0.45\textwidth]{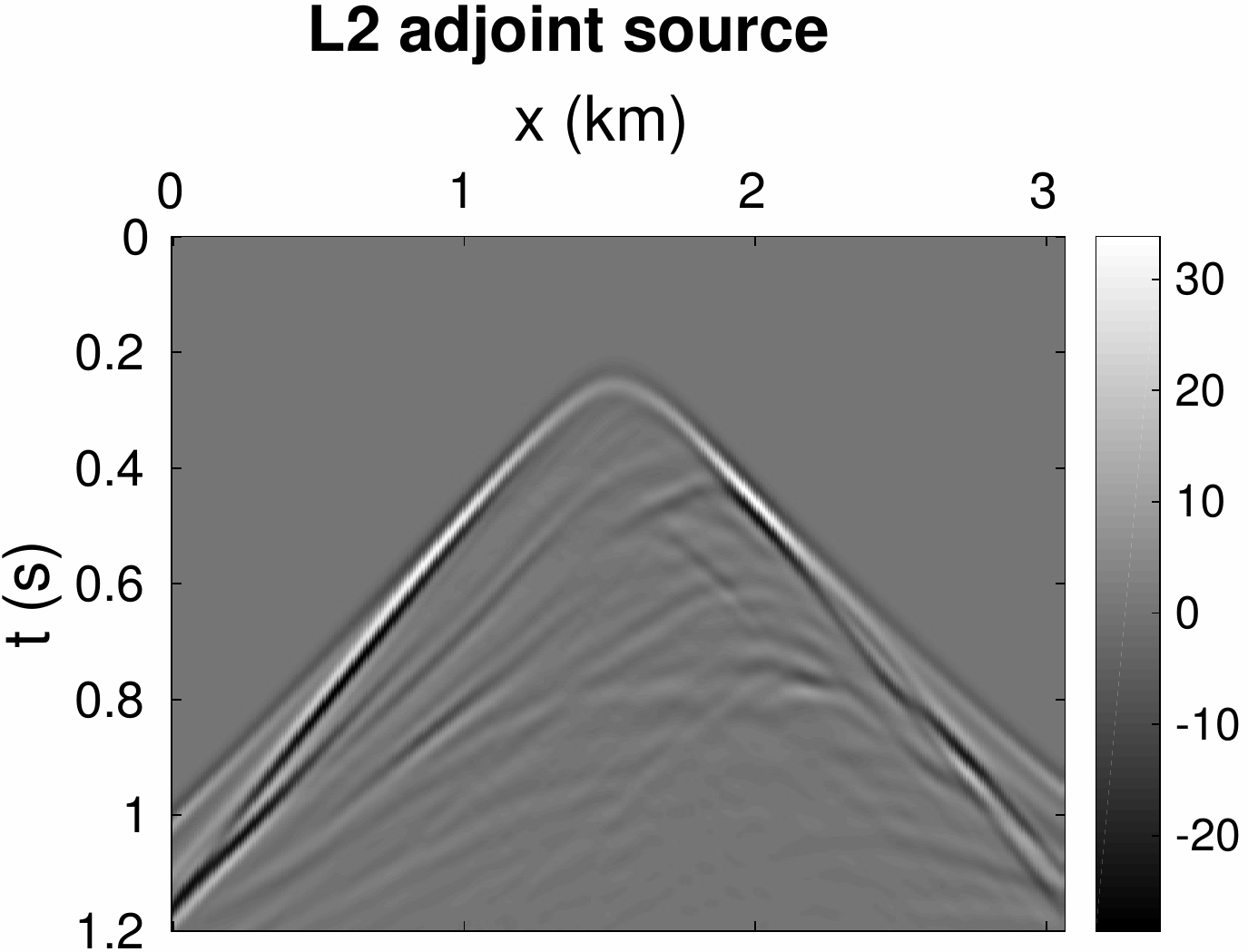}\label{fig:marm_dL2}}
  \subfloat[]{\includegraphics[width=0.45\textwidth]{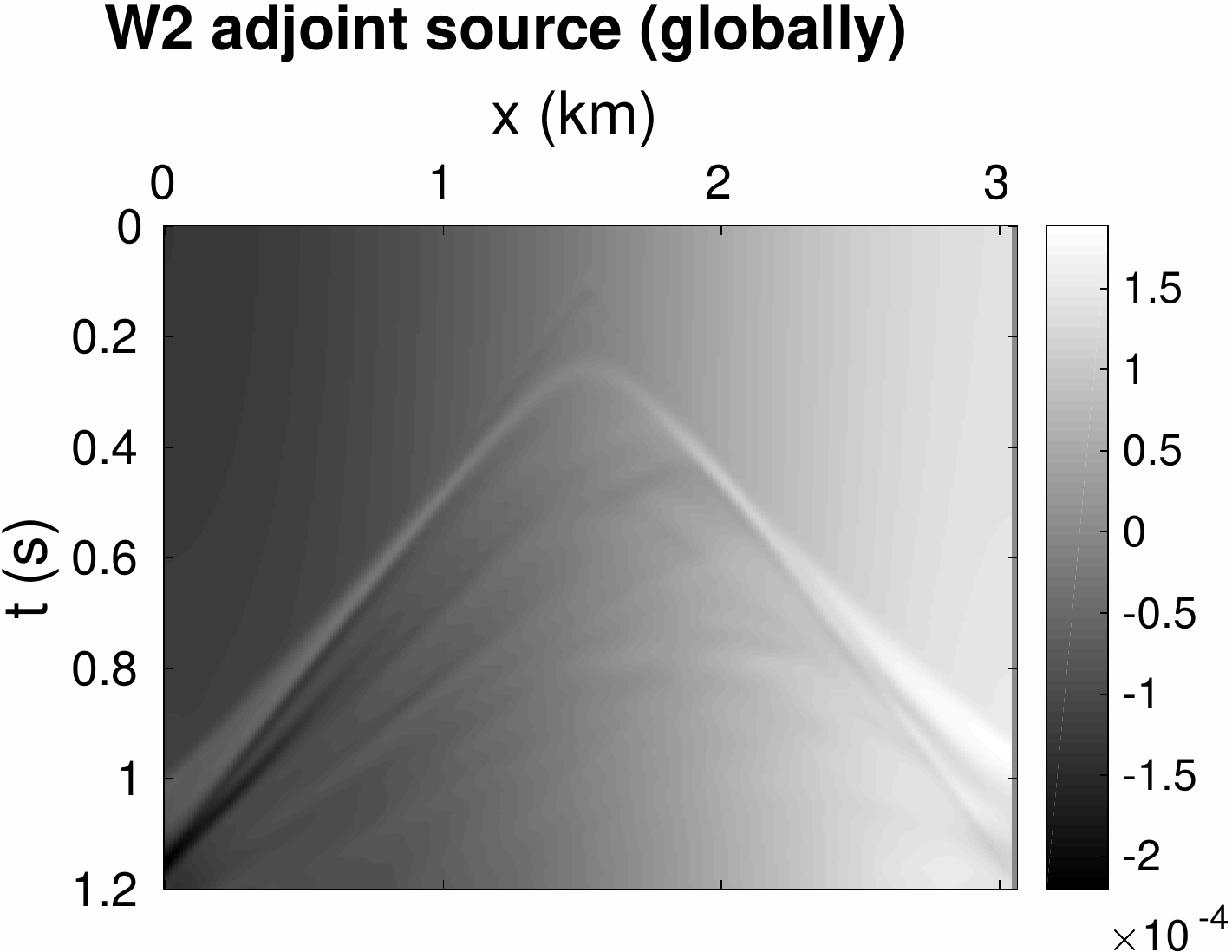}\label{fig:marm_dw2_2D}}
  \caption{Adjoint source of ~(A) $L^2$ and (B)~ global $W_2$ for the scaled Marmousi model}
  \label{fig:marm_dw2}
\end{figure}

\begin{figure}
\centering
  \subfloat[]{\includegraphics[width=0.45\textwidth]{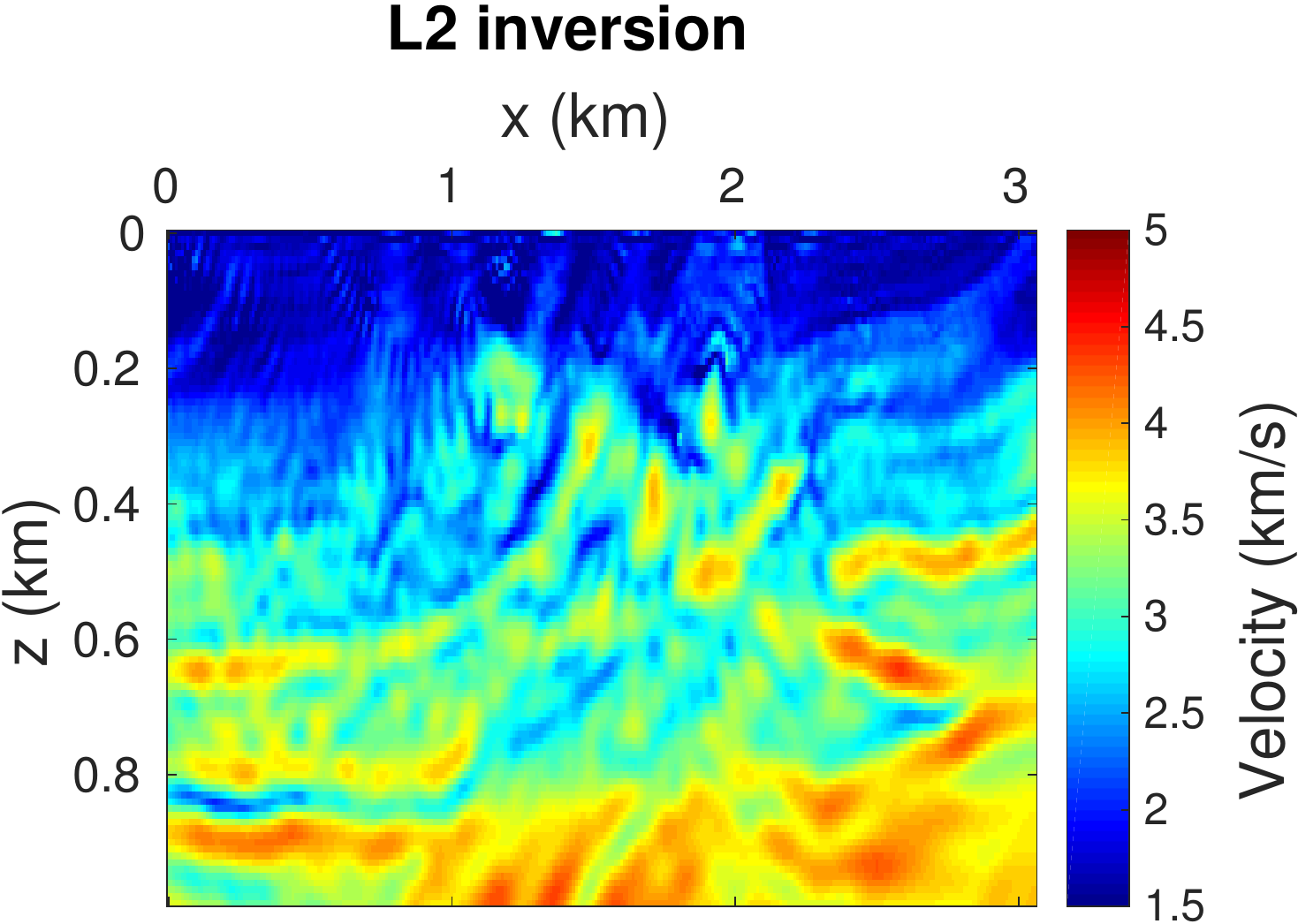}\label{fig:marm_L2}}
  \subfloat[]{\includegraphics[width=0.45\textwidth]{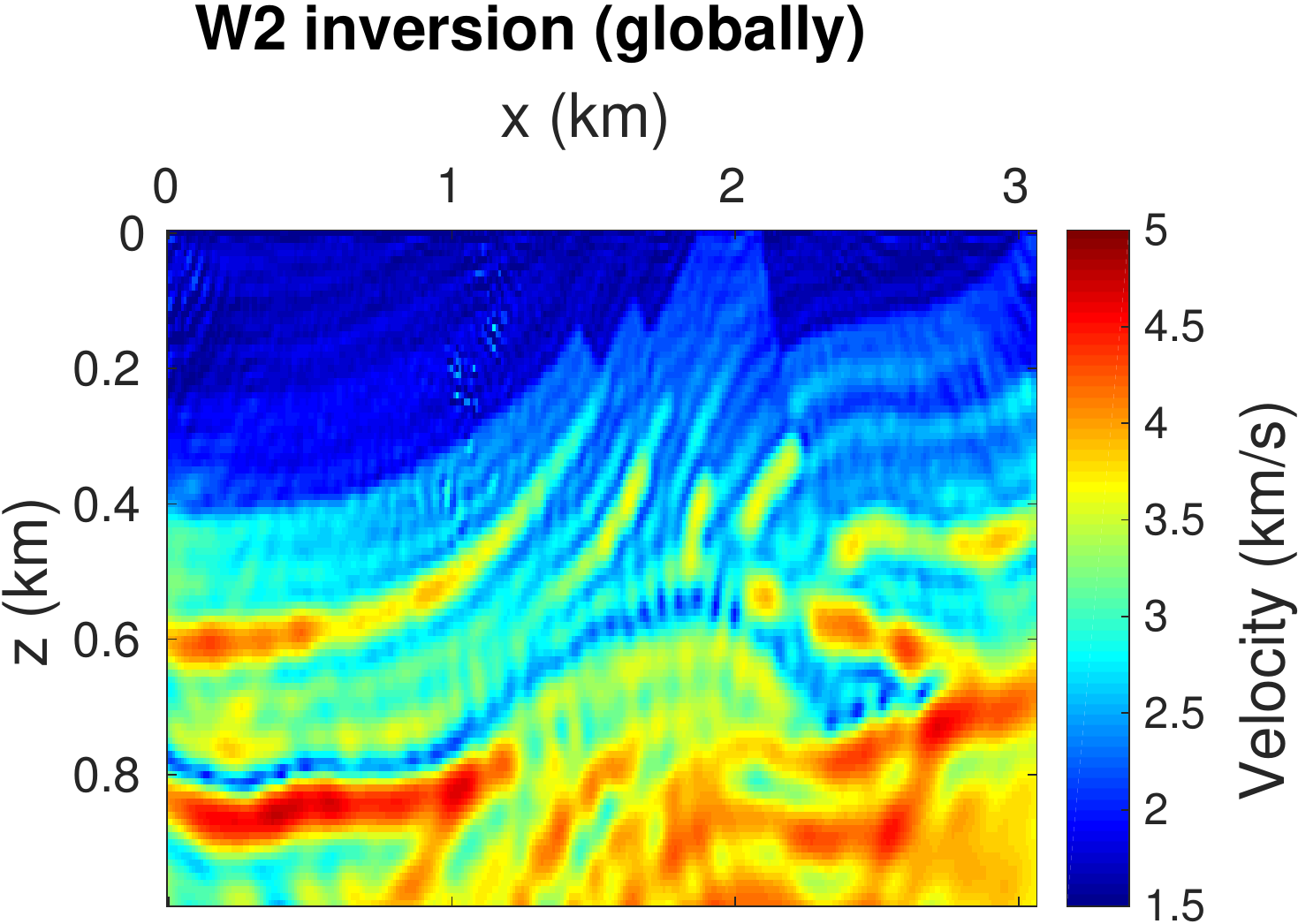}\label{fig:marm_w2_2D}}
  \caption{Inversion results of (A)~$L^2$ and (B)~global $W_2$  for the scaled Marmousi model}
  \label{fig:marm_inv_scaled}
\end{figure}

Our second 2D synthetic experiment is the Marmousi model. First we use a scaled Marmousi model to compare the inversion between global $W_2$ and the conventional $L^2$ misfit function. Figure~\ref{fig:marm2_true} is the P-wave velocity of the true Marmousi model, but in this experiment we use a scaled model which is 1km in depth and 3km in width. The inversion starts from an initial model that is the true velocity smoothed by Gaussian filter with a deviation of 40, which is highly smoothed and far from the true model (a scaled version of Figure~\ref{fig:marm2_v0}). We place 11 evenly spaced sources on top at 50m depth and 307 receivers on top at the same depth with 10m fixed acquisition. The discretization  of the forward wave equation is 10m in the $x$ and $z$ directions and 10ms in time. The source is a Ricker wavelet with a peak frequency of 15Hz, and a bandpass filter is applied to remove the frequency components from 0 to 2Hz. 

We compute the $W_2$ misfit via a global optimal map between the entire 2D data sets by solving the \MA equation. Figure~\ref{fig:marm_dw2_2D} and Figure~\ref{fig:marm_w2_2D} are the adjoint source and final inversion results respectively. Inversions are terminated after 200 l-BFGS iterations. Figure~\ref{fig:marm_L2} shows the inversion result using the traditional $L^2$ least-squares method after 200 l-BFGS iterations. The inversion result of global $W_2$ avoids the problem of local minima suffered by the conventional $L^2$ metric, whose result demonstrates spurious high-frequency artifacts due to a point-by-point comparison of amplitude. 

\subsection{True Marmousi model with trace-by-trace $W_2$ misfit computation}\label{sec:true_marm}

\begin{figure}
\centering
  \subfloat[]{\includegraphics[width=0.45\textwidth]{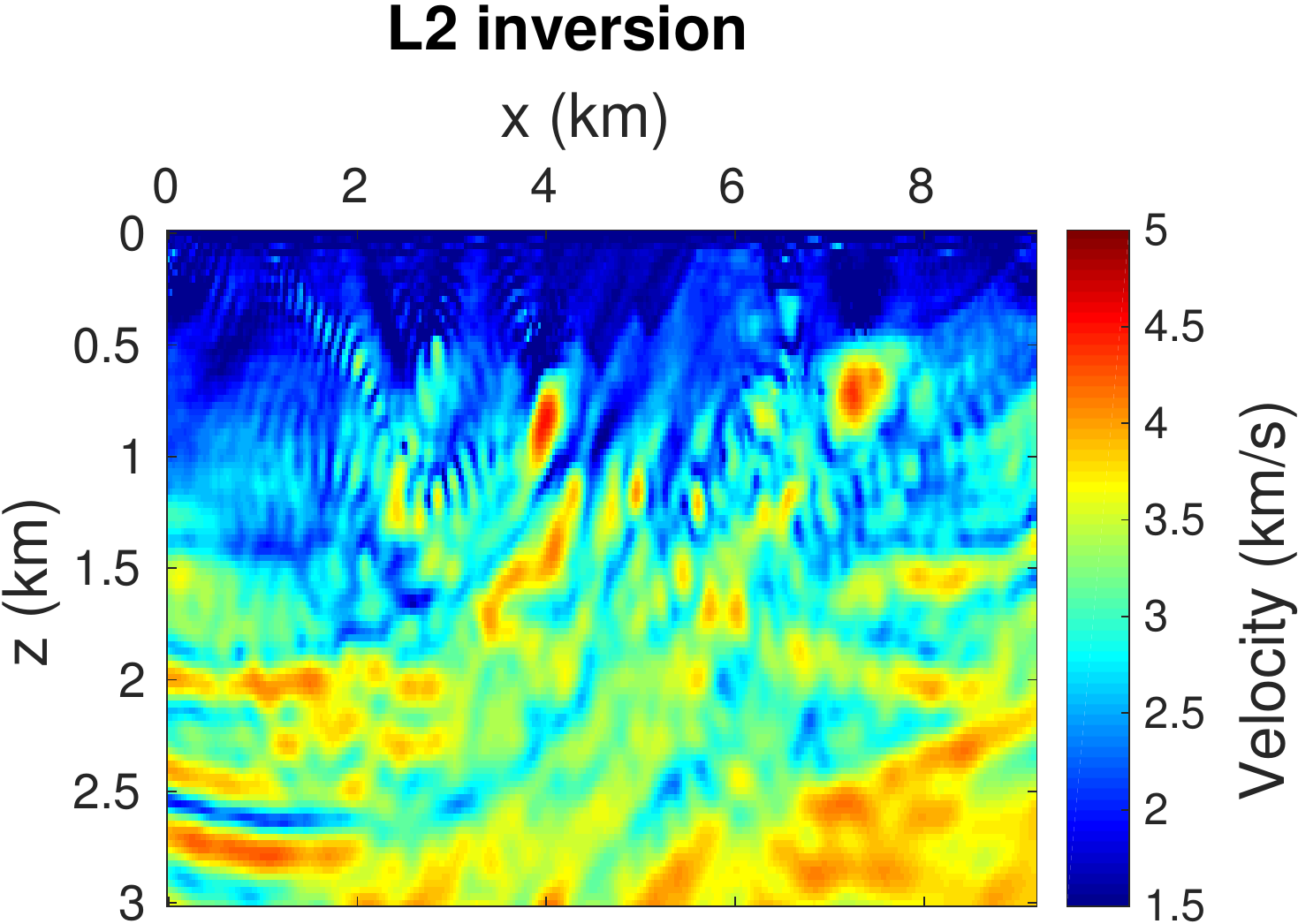}\label{fig:marm2_L2}}
  \subfloat[]{\includegraphics[width=0.45\textwidth]{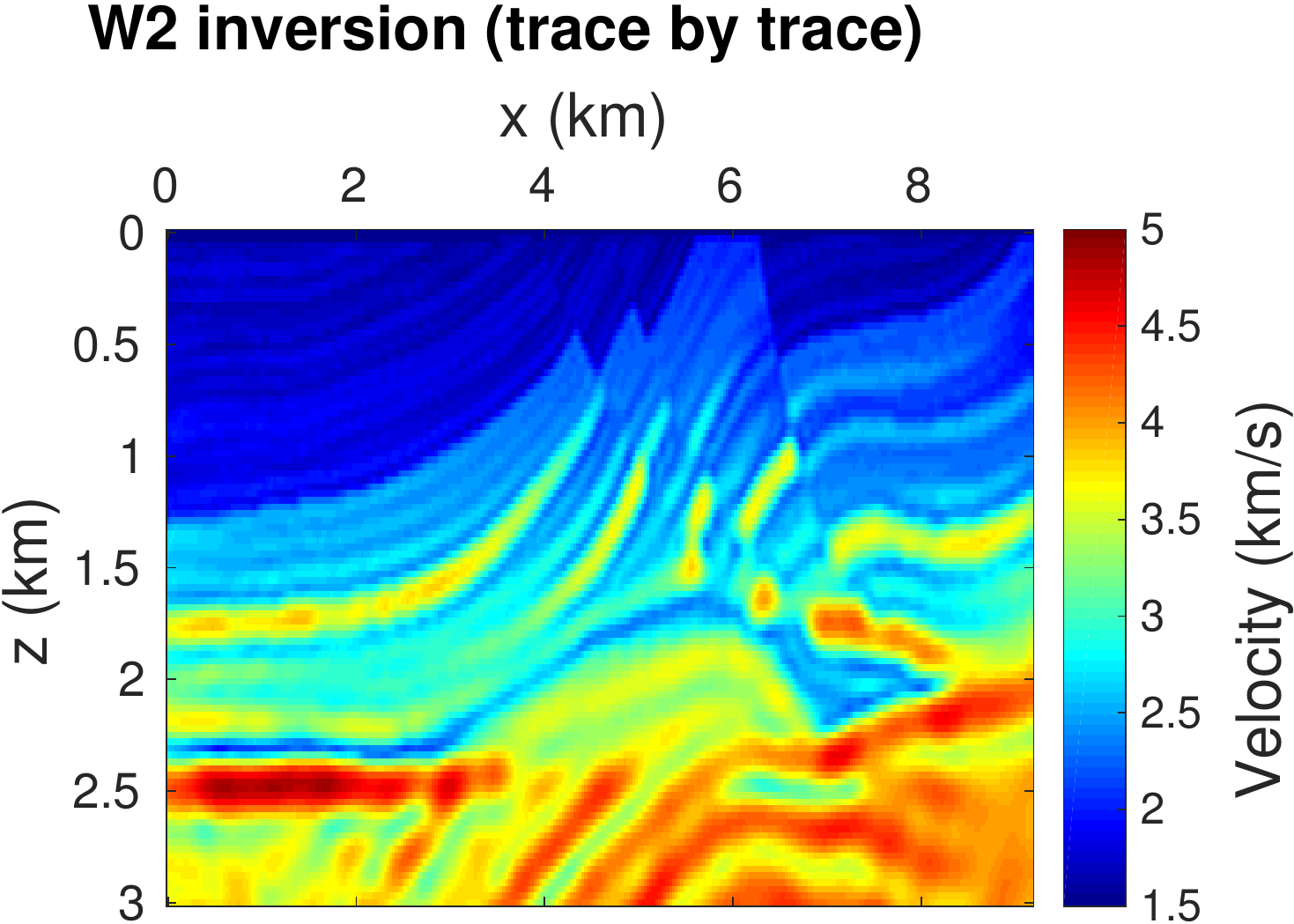}\label{fig:marm2_w2_1D}}
  \caption{Inversion results of (A)~$L^2$ and (B)~trace-by-trace $W_2$ for the true Marmousi model}
  \label{fig:marm2_inv}
\end{figure}

\begin{figure}
\centering
  \subfloat[]{\includegraphics[width=0.45\textwidth]{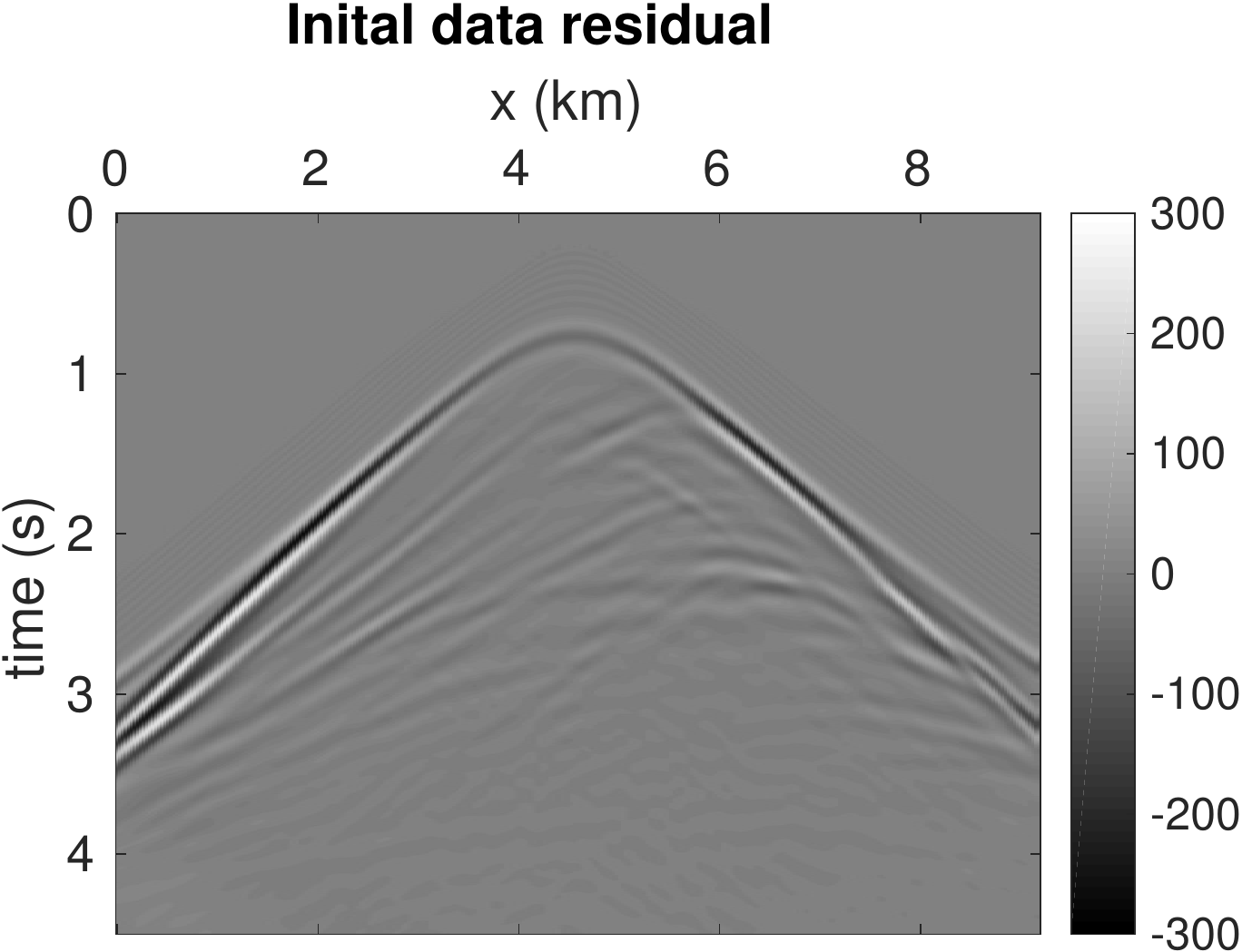}\label{fig:MARM2_res0}}
  \subfloat[]{\includegraphics[width=0.45\textwidth]{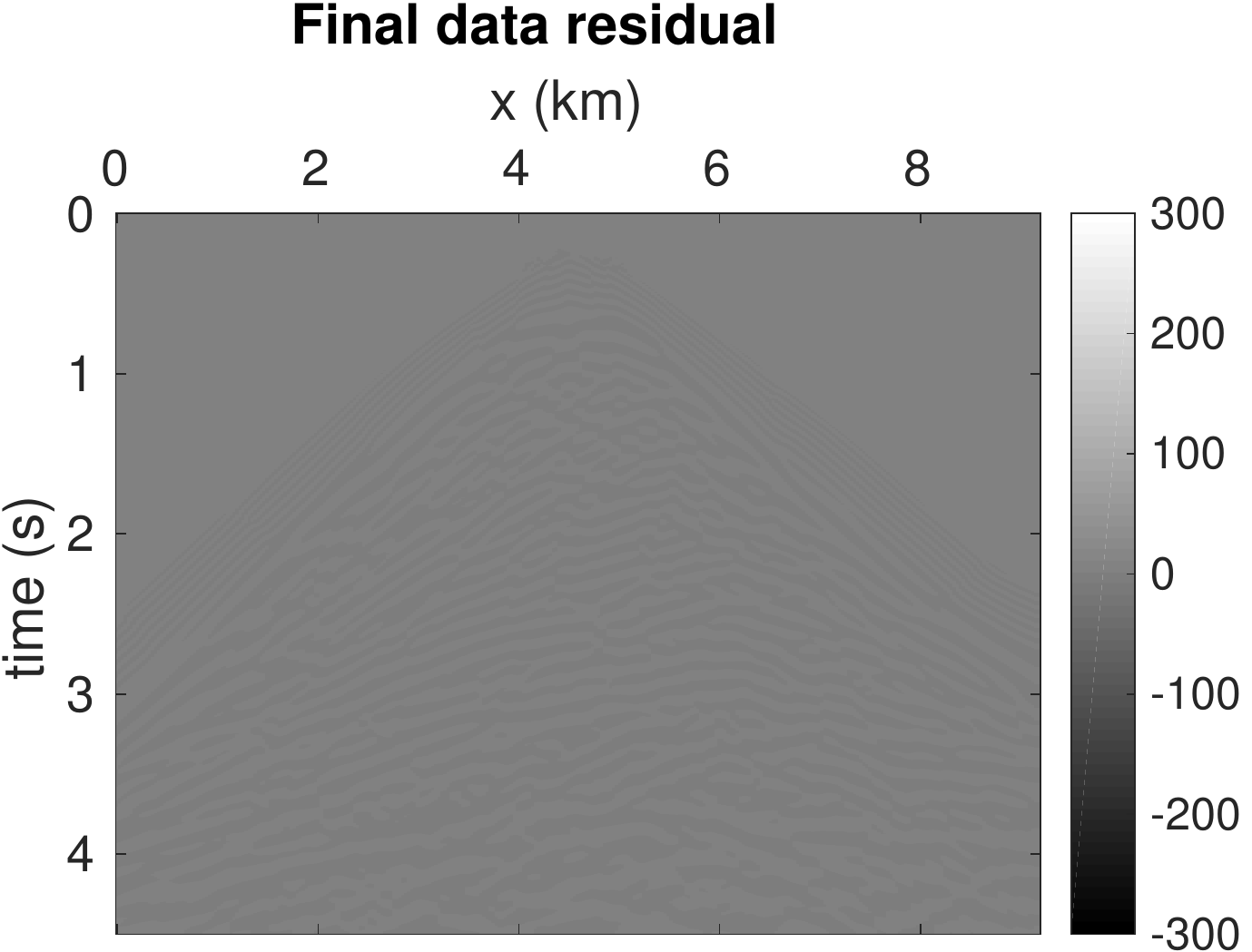}\label{fig:MARM2_W2res}}
  \caption{(A)~The difference of data to be fit and the prediction with initial model and (B)~the final data residual of trace-by-trace $W_2$ for the true Marmousi model}
  \label{fig:MARM2_res}
\end{figure}

\begin{figure}
\centering
  \subfloat[]{\includegraphics[width=0.45\textwidth]{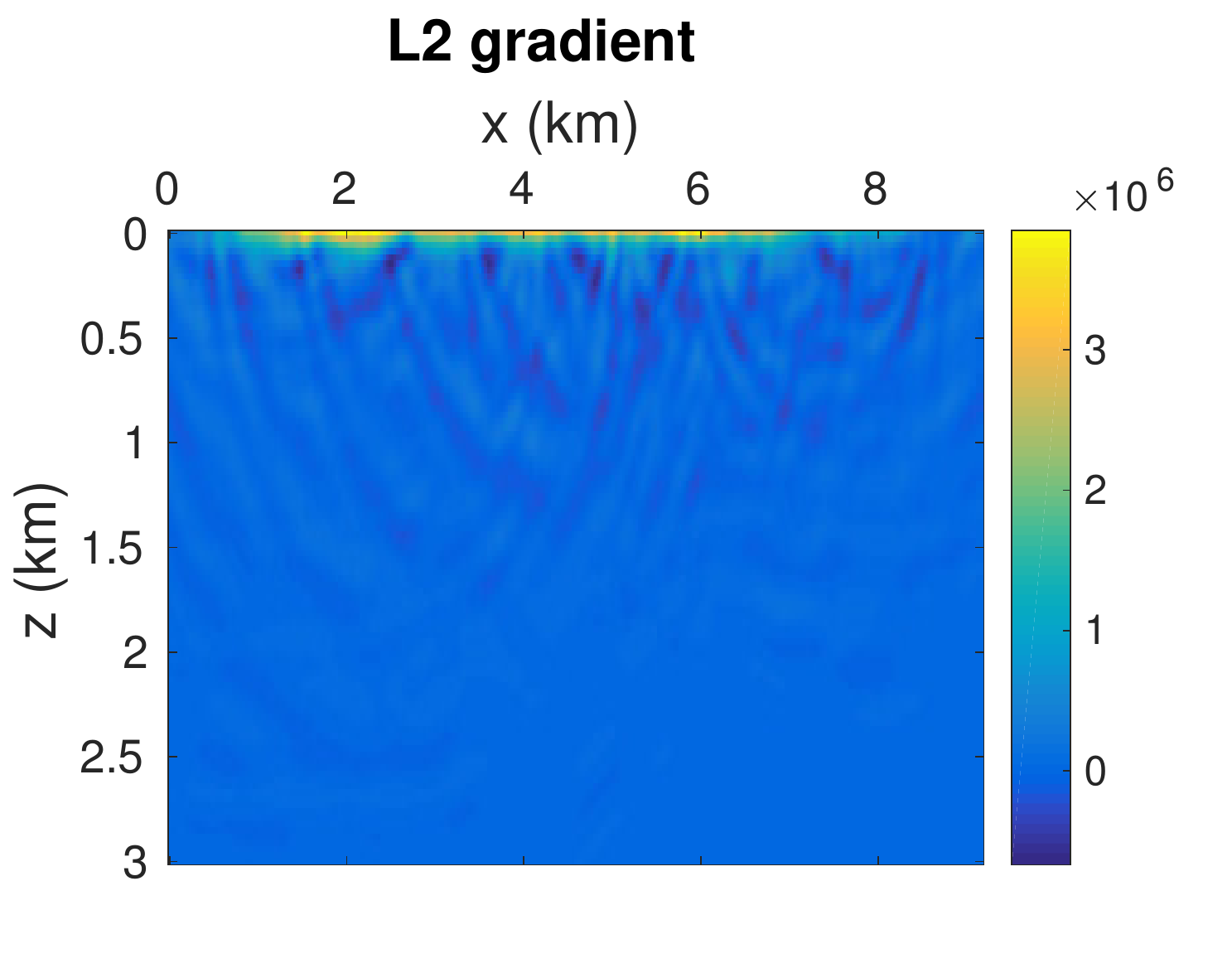}\label{fig:marm2_L2_grad}}
  \subfloat[]{\includegraphics[width=0.45\textwidth]{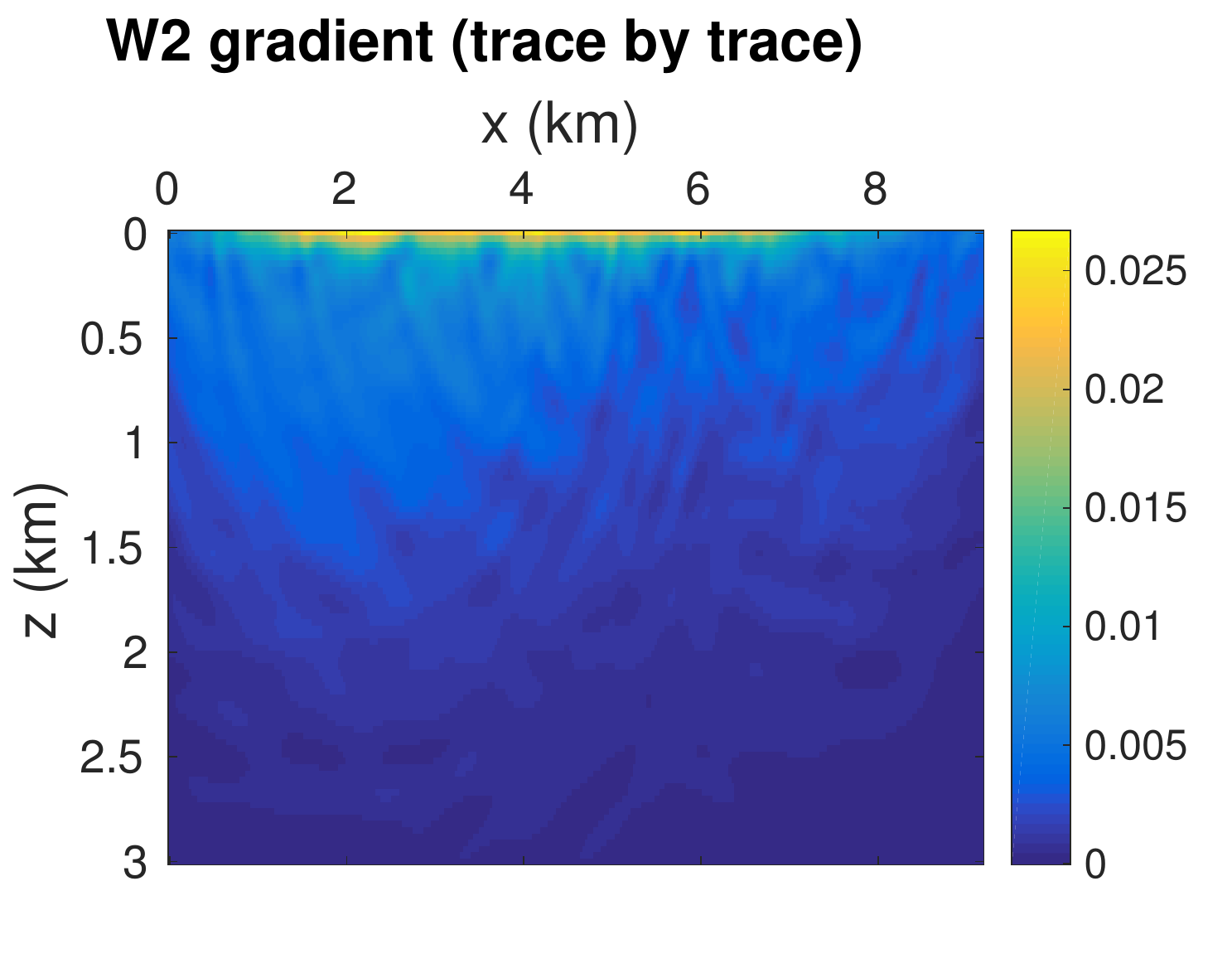}\label{fig:marm2_W2_grad}}
  \caption{The gradient in the first iteration of (A)~$L^2$ and (B)~trace-by-trace $W_2$ inversion for the true Marmousi model}
  \label{fig:marm2_grad}
\end{figure}

\begin{figure}
\centering
 {\includegraphics[width=1.0\textwidth]{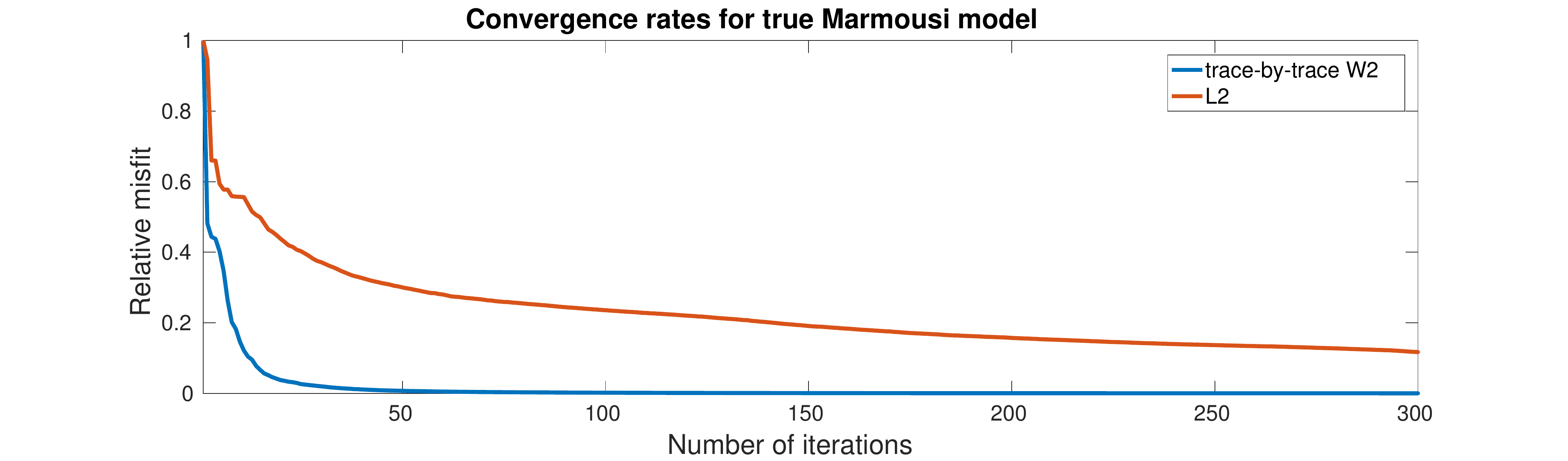}}
   \caption{The convergence curves for trace-by-trace $W_2$ and $L^2$ based inversion of the true Marmousi model}
	\label{fig:marm2_conv}
\end{figure}

The next experiment is to invert true Marmousi model by conventional $L^2$ and trace-by-trace $W_2$ misfit. Figure~\ref{fig:marm2_true} is the P-wave velocity of the true Marmousi model, which is 3km in depth and 9km in width. The inversion starts from an initial model that is the true velocity smoothed by Gaussian filter with a deviation of 40 (Figure~\ref{fig:marm2_v0}). We place 11 evenly spaced sources on top at 150m depth in the water layer and 307 receivers on top at the same depth with 30m fixed acquisition. The discretization  of the forward wave equation is 30m in the $x$ and $z$ directions and 30ms in time. The source is a Ricker wavelet with a peak frequency of 5Hz, and a high-pass filter is applied to remove the frequency components from 0 to 2Hz. 

We compute the $W_2$ misfit trace by trace. For each receiver, we first normalize the data sets and then solve the optimal transport problem in 1D. With the explicit formula, the computation time is close to $L^2$. The final adjoint source $\frac{dW_2^2(f,g)}{df}$ is a combination of the Fr\'{e}chet derivative $\frac{dW_2^2(f(x_r),g(x_r))}{df(x_r)}$ of all the receivers. We propagate it backward to generate the gradient. Figure~\ref{fig:marm2_L2_grad} and Figure~\ref{fig:marm2_W2_grad} are the gradients in the first iteration of two misfits respectively. 

Starting from a highly smoothed initial model, in the first iteration $W_2$ already focuses on the ``peak'' of the Marmousi model as seen from Figure~\ref{fig:marm2_W2_grad}. The darker area in the gradient matches many features in the velocity model~(Figure~\ref{fig:marm2_true}). However, the gradient of $L^2$ is quite uniform contrary to the model features. Inversions are terminated after 300 l-BFGS iterations. Figure~\ref{fig:marm2_L2} shows the inversion result using the traditional $L^2$ least-squares method and figure~\ref{fig:marm2_w2_1D} shows the final result using trace-by-trace $W_2$ misfit function. Again, the result of $L^2$ metric has spurious high-frequency artifacts while $W_2$ correctly inverts most details in the true Marmousi model. The data residuals before and after trace-by-trace $W_2$ based FWI are presented in Figure~\ref{fig:MARM2_res}. The convergence curves in Figure~\ref{fig:marm2_conv} show that $W_2$ reduces the relative misfit to 0.1 in 20 iterations while $L^2$ converges slowly to a local minimum.

\subsection{Inversion with the noisy data}

\begin{figure}
\centering
  \subfloat[]{\includegraphics[height=0.3\textwidth]{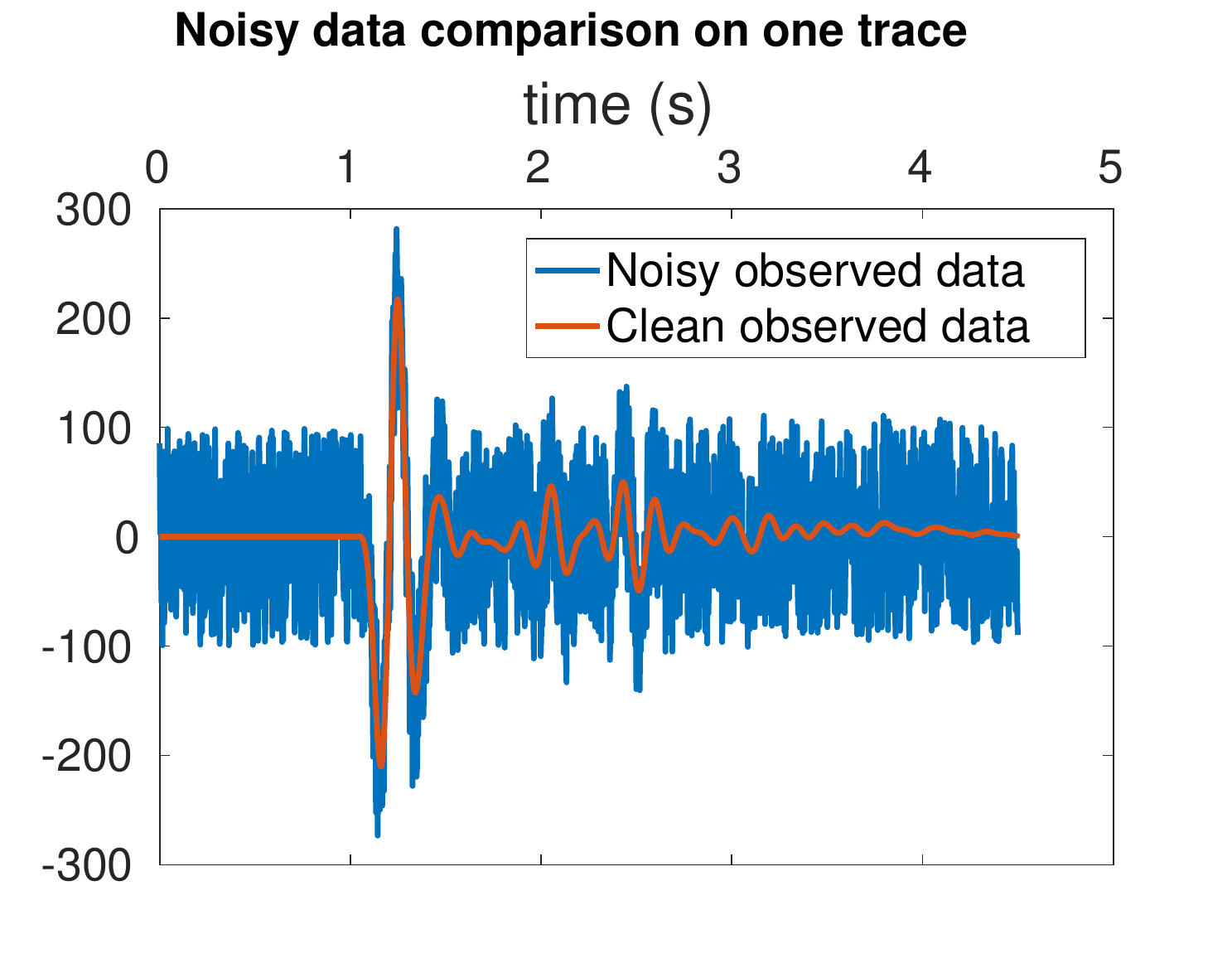}\label{fig:MARM2_noisy_trace}}
  \subfloat[]{\includegraphics[width=0.45\textwidth]{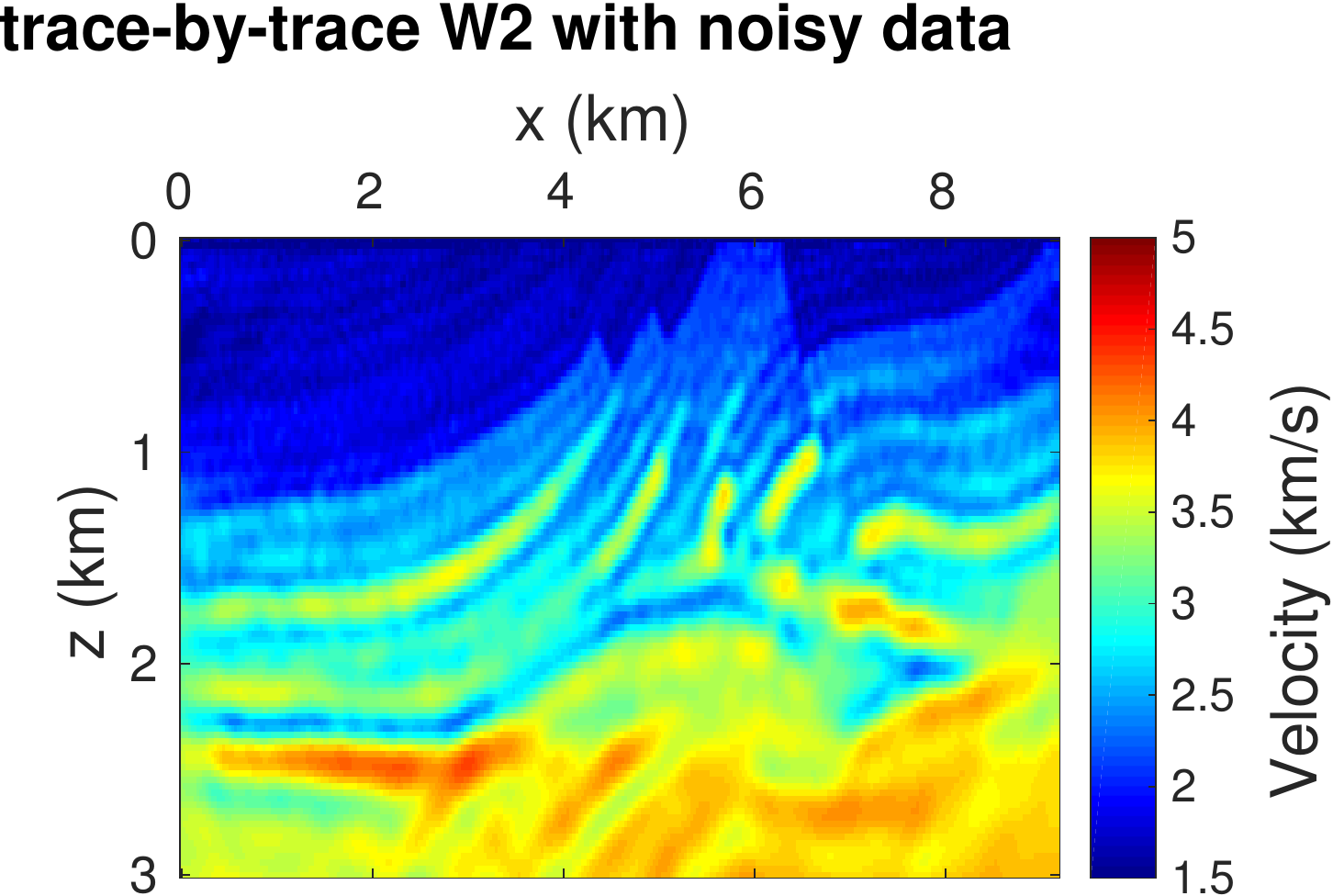}\label{fig:MARM2_noisy_vel}}
  \caption{(A)~Noisy and clean data and (B)~inversion result with the noisy data}
  \label{fig:marm2_noisy}
\end{figure}

One of the ideal properties of the \QW is the insensitivity to noise~\cite{engquist2016optimal}. We repeat the previous experiment with a noisy reference by adding a uniform random iid noise to the data from the true velocity (Figure~\ref{fig:MARM2_noisy_trace}). The signal-to-noise ratio (SNR) is $-3.4716$ dB. In optimal transport the effect of noise is in theory negligible due to the strong cancellation between the nearby positive and negative noise values.

All the settings remain the same as the previous experiment except the observed data. After 96 iterations, the optimization converges to a velocity presented in Figure~\ref{fig:MARM2_noisy_vel}. Although the result has less resolution than Figure~\ref{fig:marm2_w2_1D}, it still recovers most features of Marmousi model correctly. When the noise power is much larger than the signal power, the \QW still converges reasonably well, which again demonstrates its insensitivity to noise.

\subsection{$L^2$ based FWI after $W_2$ building the initial model.}
\begin{figure}
\centering
  \subfloat[]{\includegraphics[width=0.45\textwidth]{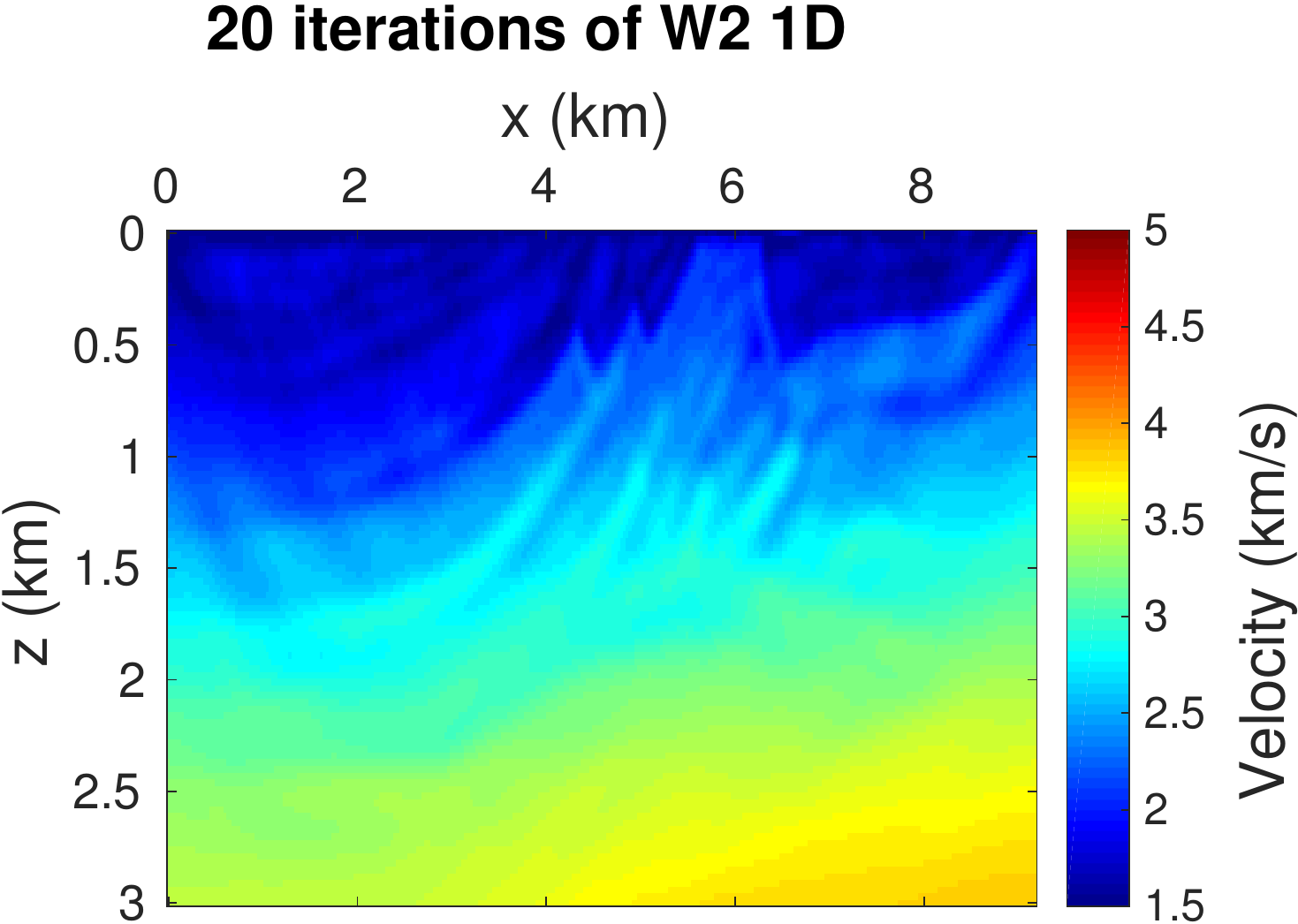}\label{fig:marm2_W2_21}}
  \subfloat[]{\includegraphics[width=0.45\textwidth]{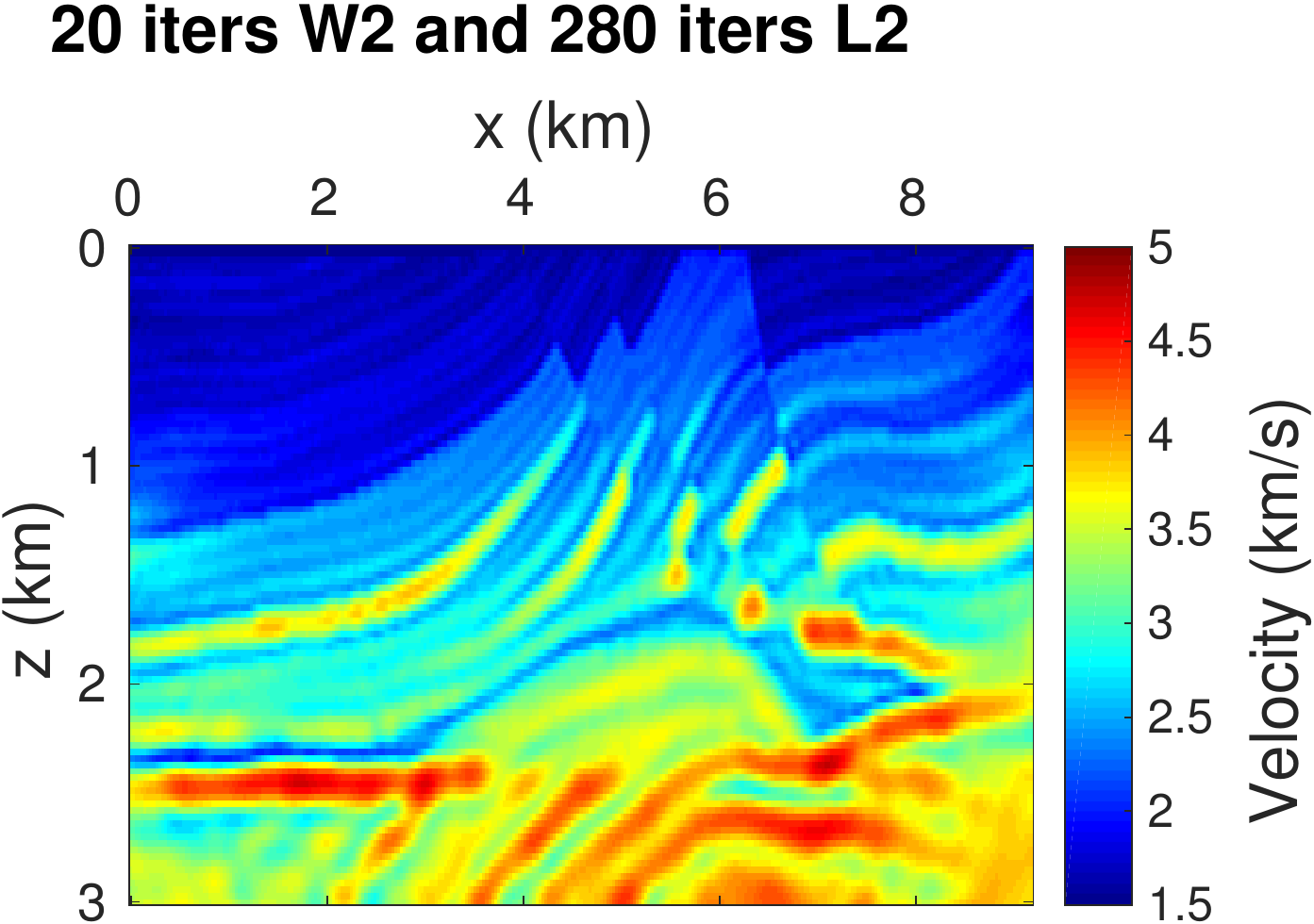}\label{fig:marm2_1stW2_2ndL2}}
  \caption{(A)~Inversion result of 20 iterations of trace-by-trace $W_2$, as the initial model for $L^2$ based inversion and (B)~ 280 iterations $L^2$ based inversion result starting from (A)}
  \label{fig:marm2_W2L2}
\end{figure}

Next we did another $L^2$ based FWI. The initial model (Figure~\ref{fig:marm2_W2_21}) is the inversion result after 20 iterations of trace-by-trace $W_2$ starting from Figure~\ref{fig:marm2_v0}. After running another 280 iterations of $L^2$ based inversion, we get the inversion result Figure~\ref{fig:marm2_1stW2_2ndL2}. 

We compare Figure~\ref{fig:marm2_1stW2_2ndL2} with Figure~\ref{fig:marm2_w2_1D} which are 280 iterations of $L^2$ based inversion and trace-by-trace $W_2$ based inversion respectively starting from Figure~\ref{fig:marm2_W2_21}. These two results both recover most features of the Marmousi model (Figure~\ref{fig:marm2_true}) correctly. However, Figure~\ref{fig:marm2_w2_1D} has only 0.2664 relative error while Figure~\ref{fig:marm2_1stW2_2ndL2} has 0.4211 relative error computed by $\frac{||v_\text{inv}-v_\text{true}||_2^2}{||v_0-v_\text{true}||_2^2}$. Thus we observe the $L^2$ approach converging more slowly than trace-by-trace $W_2$ even with a good initial model that prevents cycle skipping.

\subsection{2004 BP Model with global $W_2$ misfit computation}\label{sec:scaled_BP}

\begin{figure}
\centering
  \subfloat[]{\includegraphics[width=0.45\textwidth]{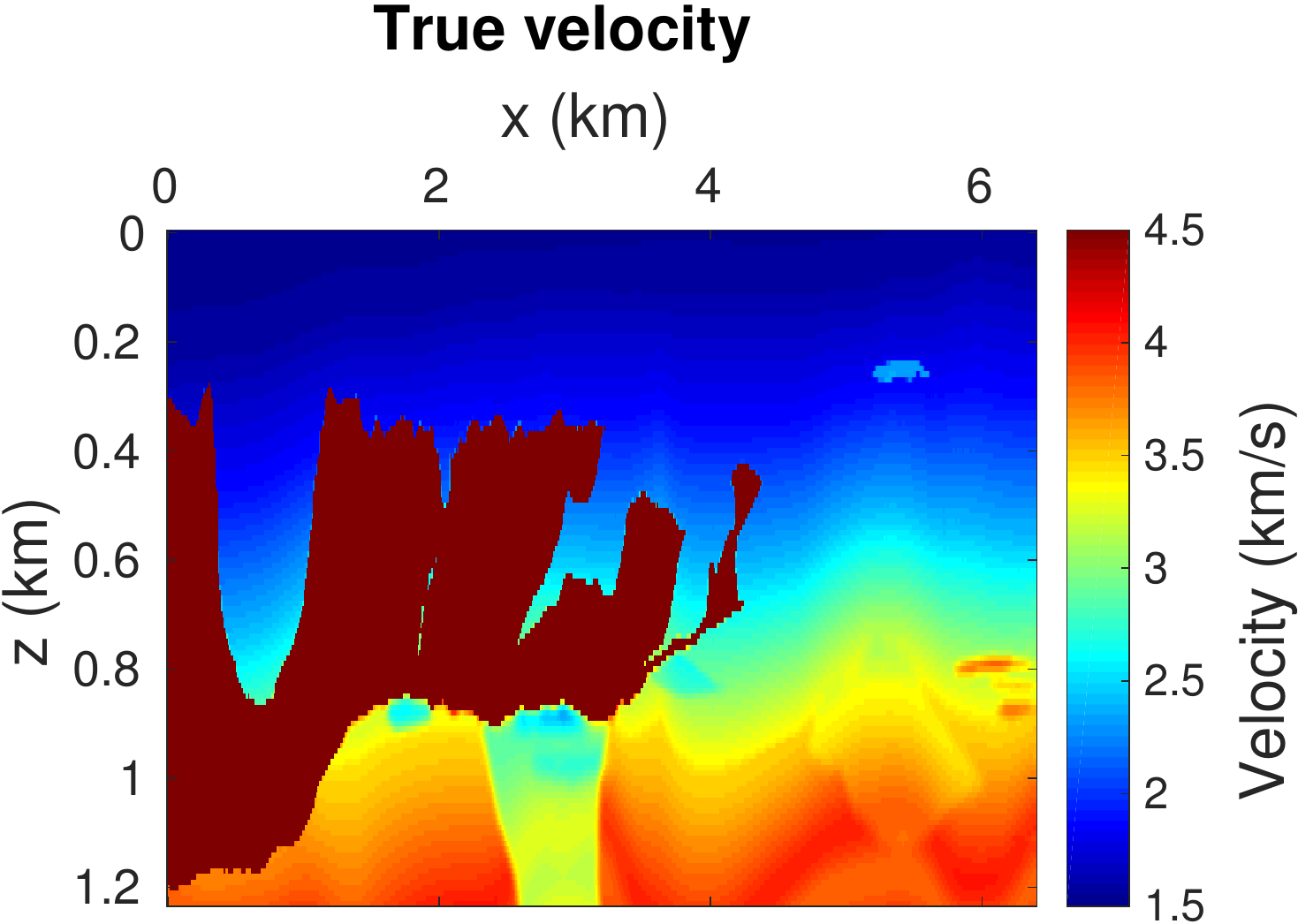}\label{fig:BP_true}}
  \subfloat[]{\includegraphics[width=0.45\textwidth]{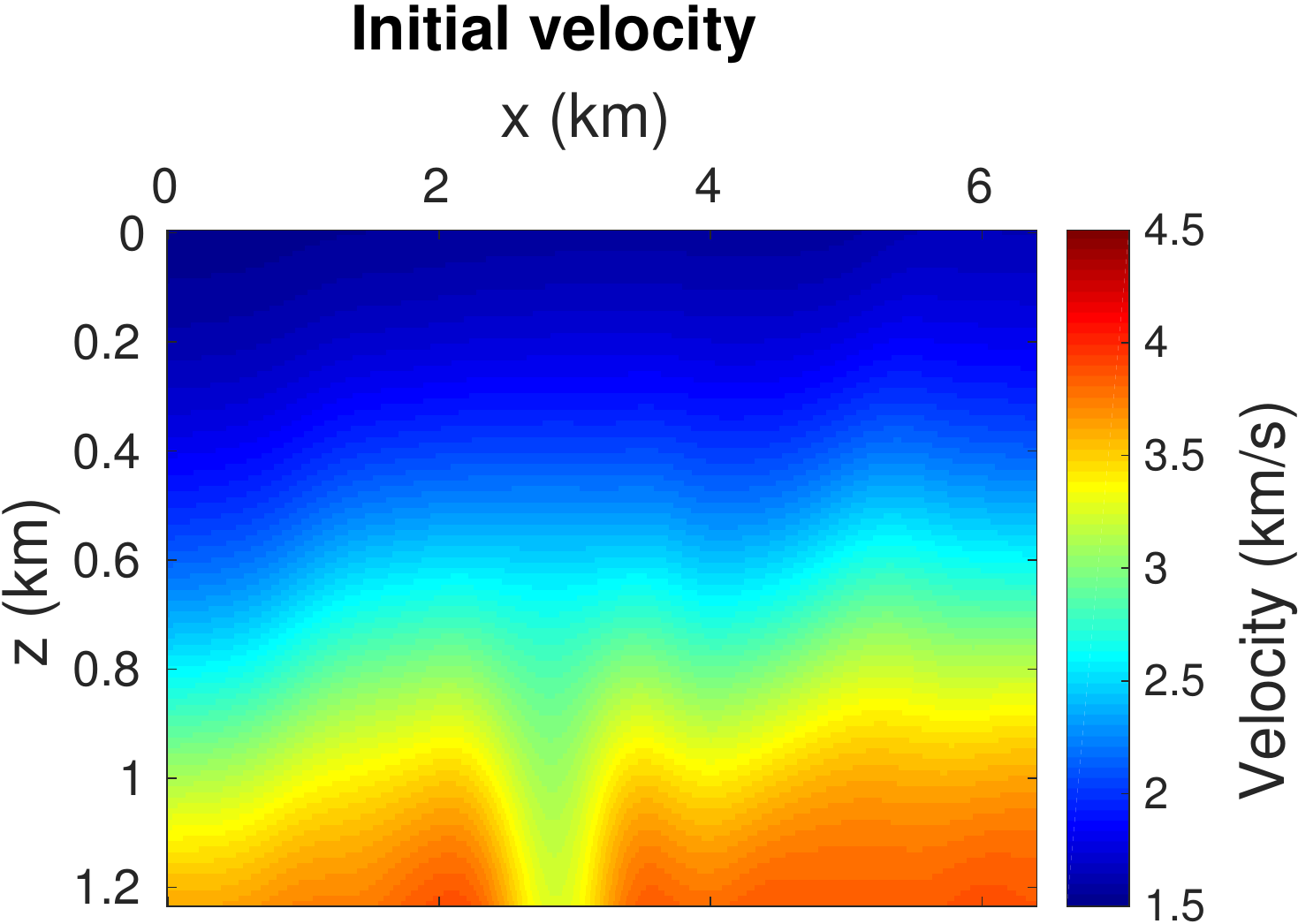}\label{fig:BP_v0}}
  \caption{(A)~True velocity and (B)~inital velocity for the BP model}
  \label{fig:BP_true,BP_v0}\end{figure}

\begin{figure}
\centering
\subfloat[]{\includegraphics[width=0.45\textwidth]{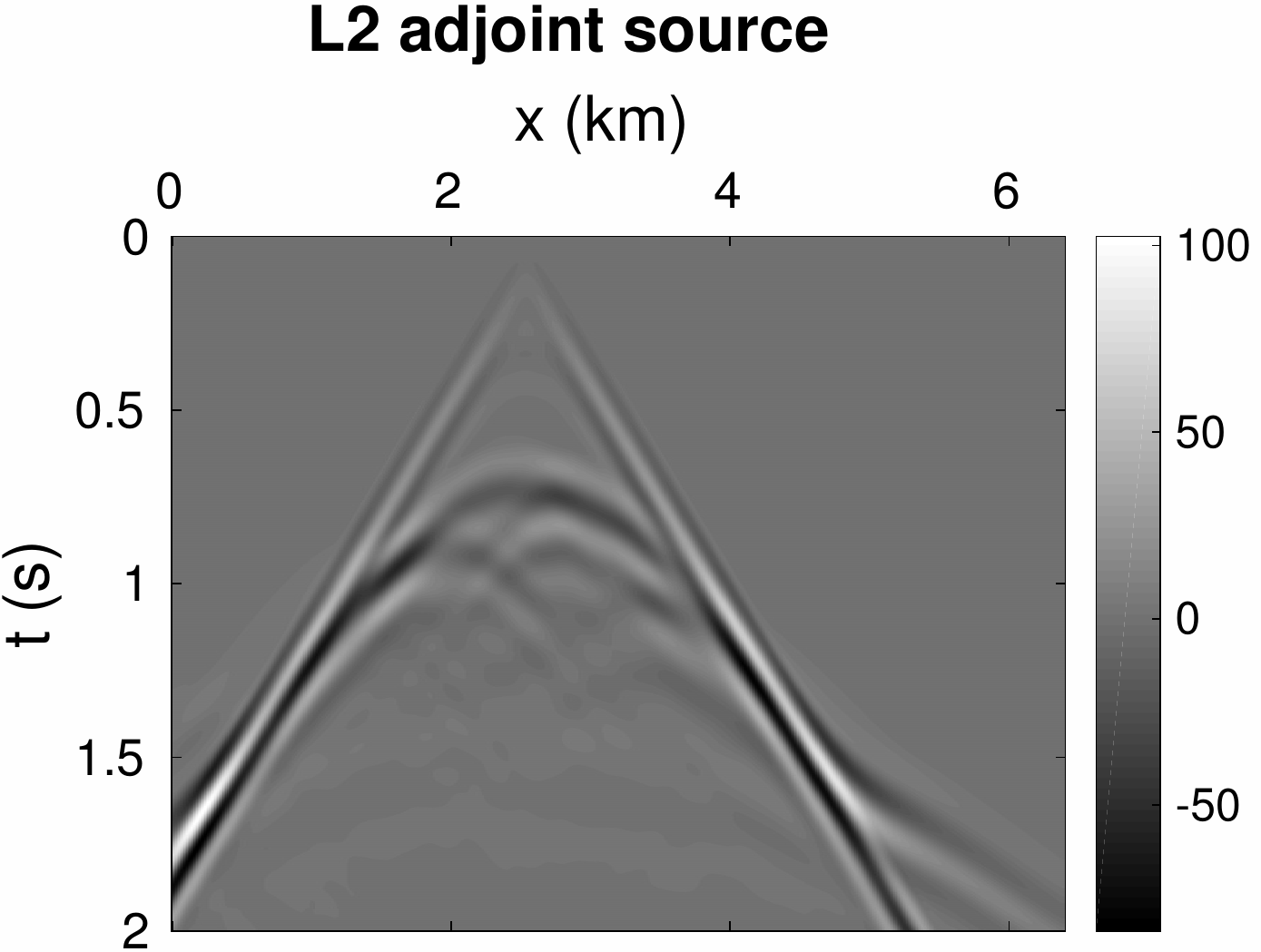}\label{fig:BP_DL2}}
  \subfloat[]{\includegraphics[width=0.45\textwidth]{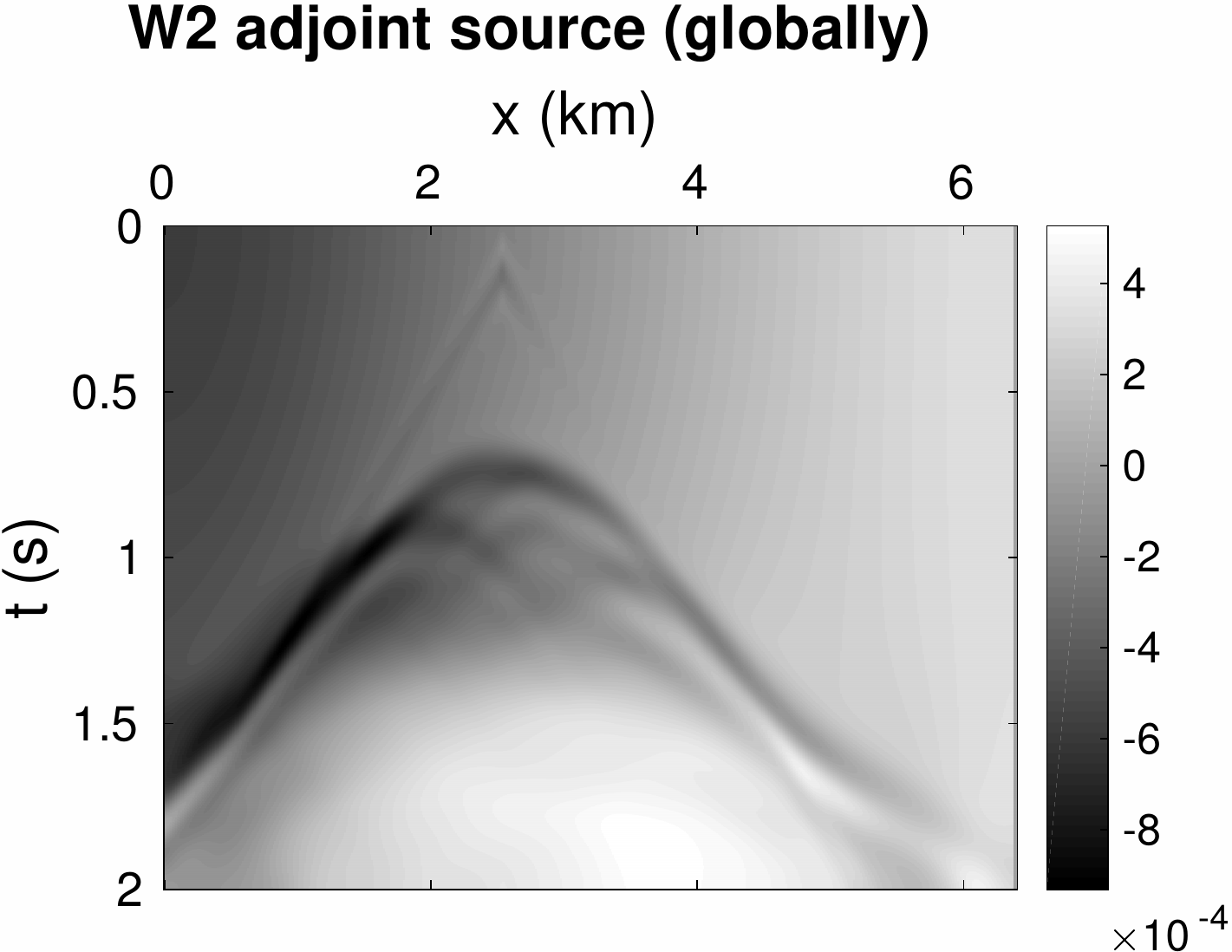}\label{fig:BP_dw2_2D}}
  \caption{Adjoint sources of (A) ~the $L^2$ for the BP model and (B)~the global $W_2$}
  \label{fig:BP_dw2}
\end{figure}

\begin{figure}
\centering
\subfloat[]{\includegraphics[width=0.45\textwidth]{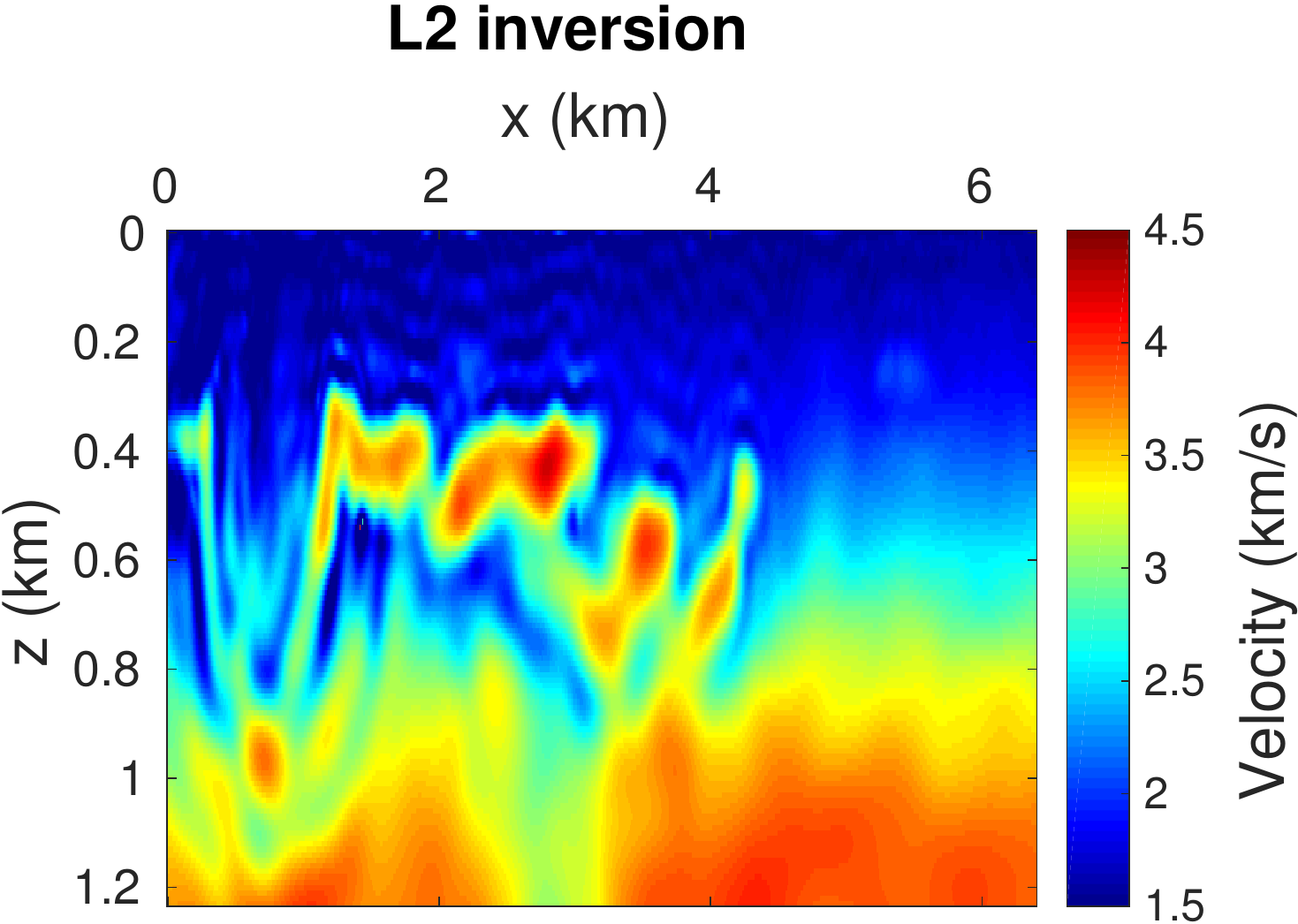}\label{fig:BP_L2}}  
  \subfloat[]{\includegraphics[width=0.45\textwidth]{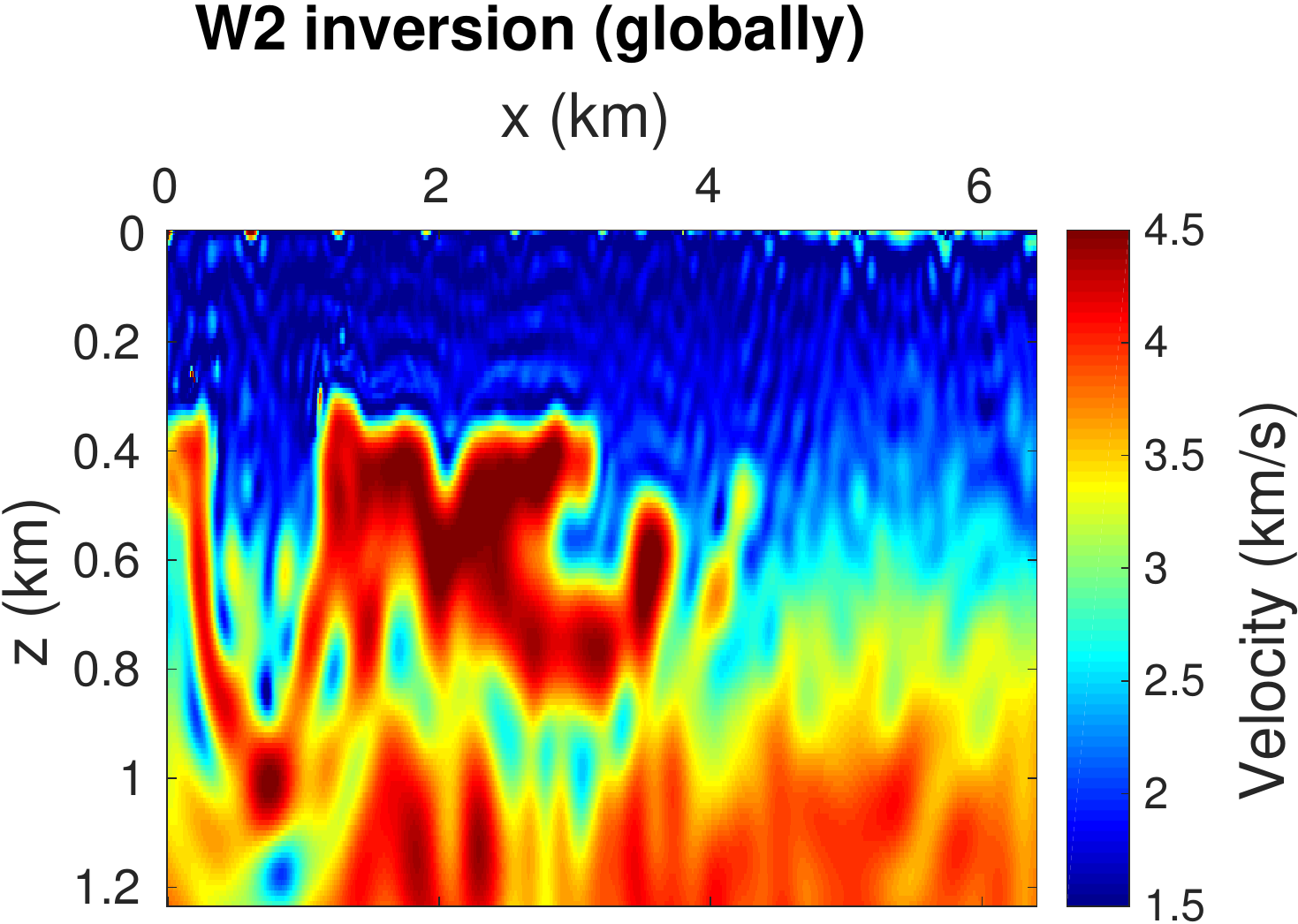}\label{fig:BP_w2_2D}}
  \caption{Inversion results (A) for $L^2$ and (B)~for global $W_2$ for the BP model}
  \label{fig:BP_w2}
\end{figure}

For this experiment, we compare global $W_2$ and conventional $L^2$ as misfit functions for a modified BP 2004 model (Figure~\ref{fig:BP_true}). Part of the model is representative of the complex geology in the deep water Gulf of Mexico. The main challenges in this area are related to obtaining a precise delineation of the salt and recover information on the sub-salt velocity variations~\cite{billette20052004}.  The inversion starts from an initial model with smoothed background without the salt (Figure~\ref{fig:BP_v0}). We put 11 equally spaced sources on top at 50m depth and 641 receivers on top at 50m depth with 10m fixed acquisition. The discretization  of the forward wave equation is 10m in the $x$ and $z$ directions and 10ms in time. The source is a Ricker wavelet with a peak frequency of 5Hz, and a band-pass filter is applied to keep the frequency components from 3 to 9Hz. The total acquisition time is restricted to 2s in order to focus on recovering the upper portion of the salt structure.

As before, we solve the \MA equation numerically to compute the global $W_2$ misfit. Figure~\ref{fig:BP_DL2} and Figure~\ref{fig:BP_dw2_2D} are the adjoint source for two misfit functions. Inversions are stopped after 100 l-BFGS iterations.  Figure~\ref{fig:BP_w2_2D} is the inversion result for $W_2$, which recovered the top salt reasonably well. The $L^2$ metric (Figure~\ref{fig:BP_L2}), on the other hand, converged to a model that has low-velocity anomaly immediately beneath the top salt, which is typical of the cycle skipping commonly encountered in FWI.

\subsection{2004 BP Model with trace-by-trace $W_2$ misfit computation}\label{sec:true_BP}

\begin{figure}
\centering
  \subfloat[]{\includegraphics[width=0.45\textwidth]{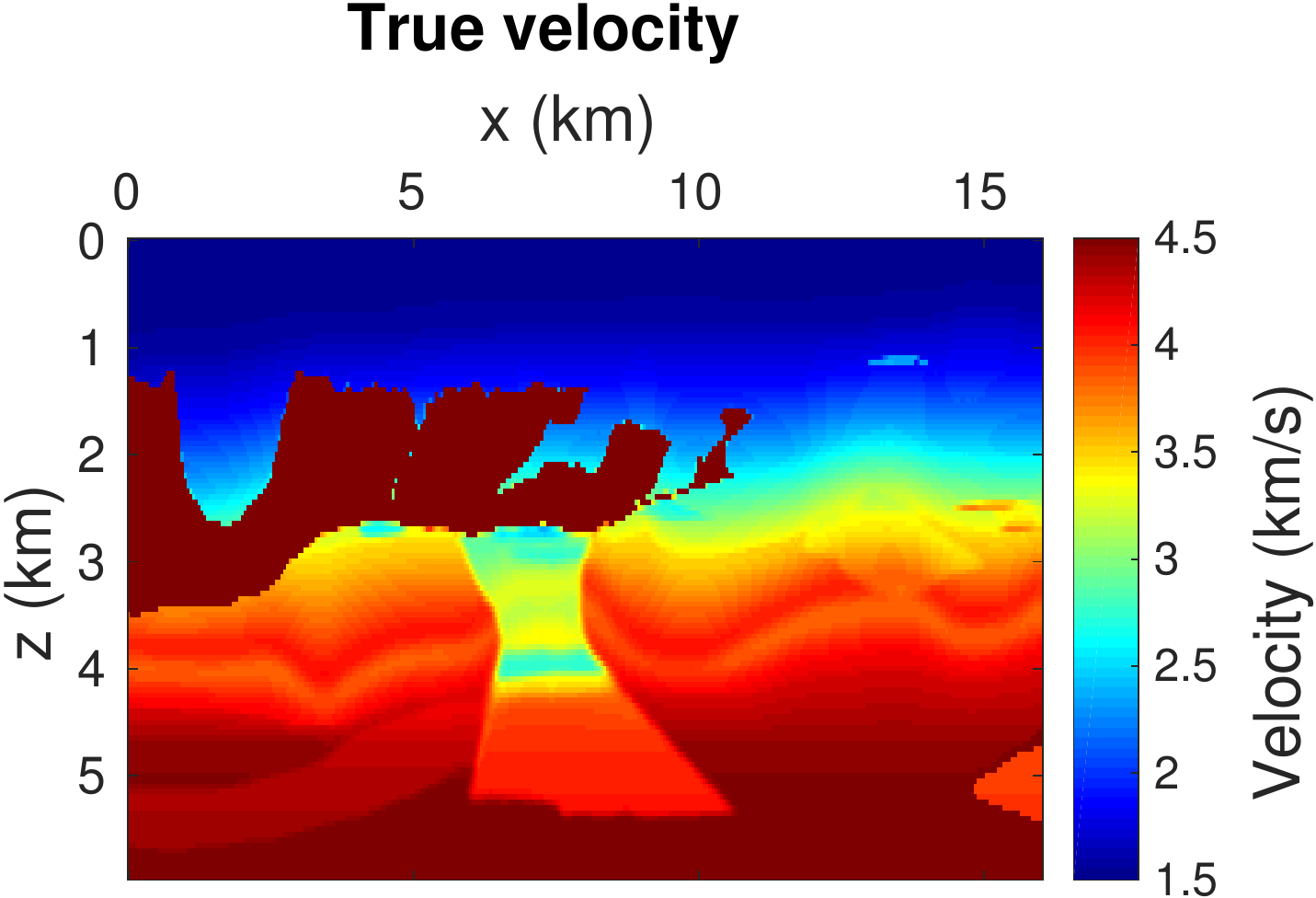}\label{fig:BP2_true}}
  \subfloat[]{\includegraphics[width=0.45\textwidth]{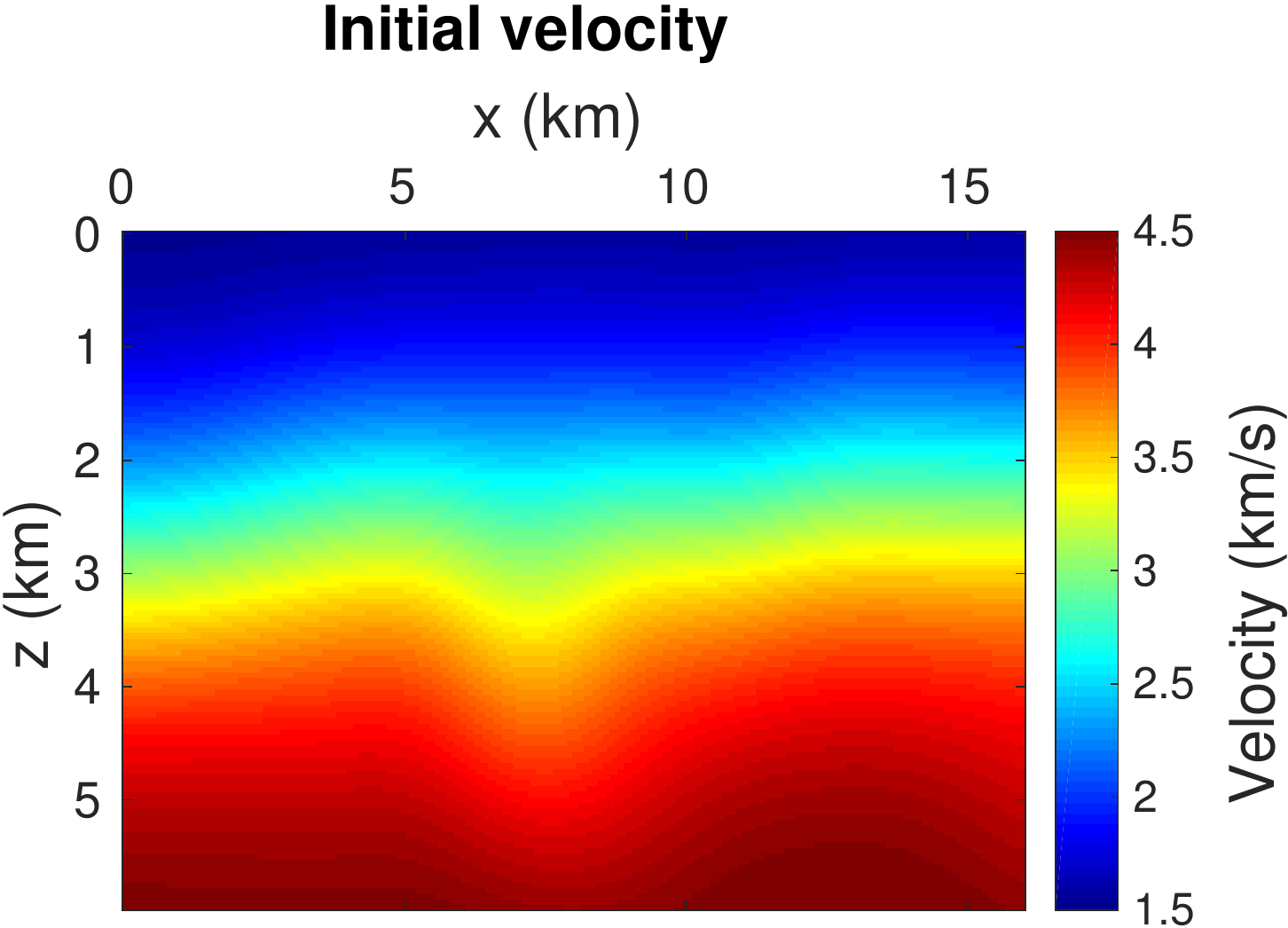}\label{fig:BP2_v0}}
  \caption{(A)~True velocity and (B)~inital velocity for the second BP model}
  \label{fig:BP2_true,BP2_v0}
\end{figure}

\begin{figure}
\centering
  \subfloat[]{\includegraphics[width=0.45\textwidth]{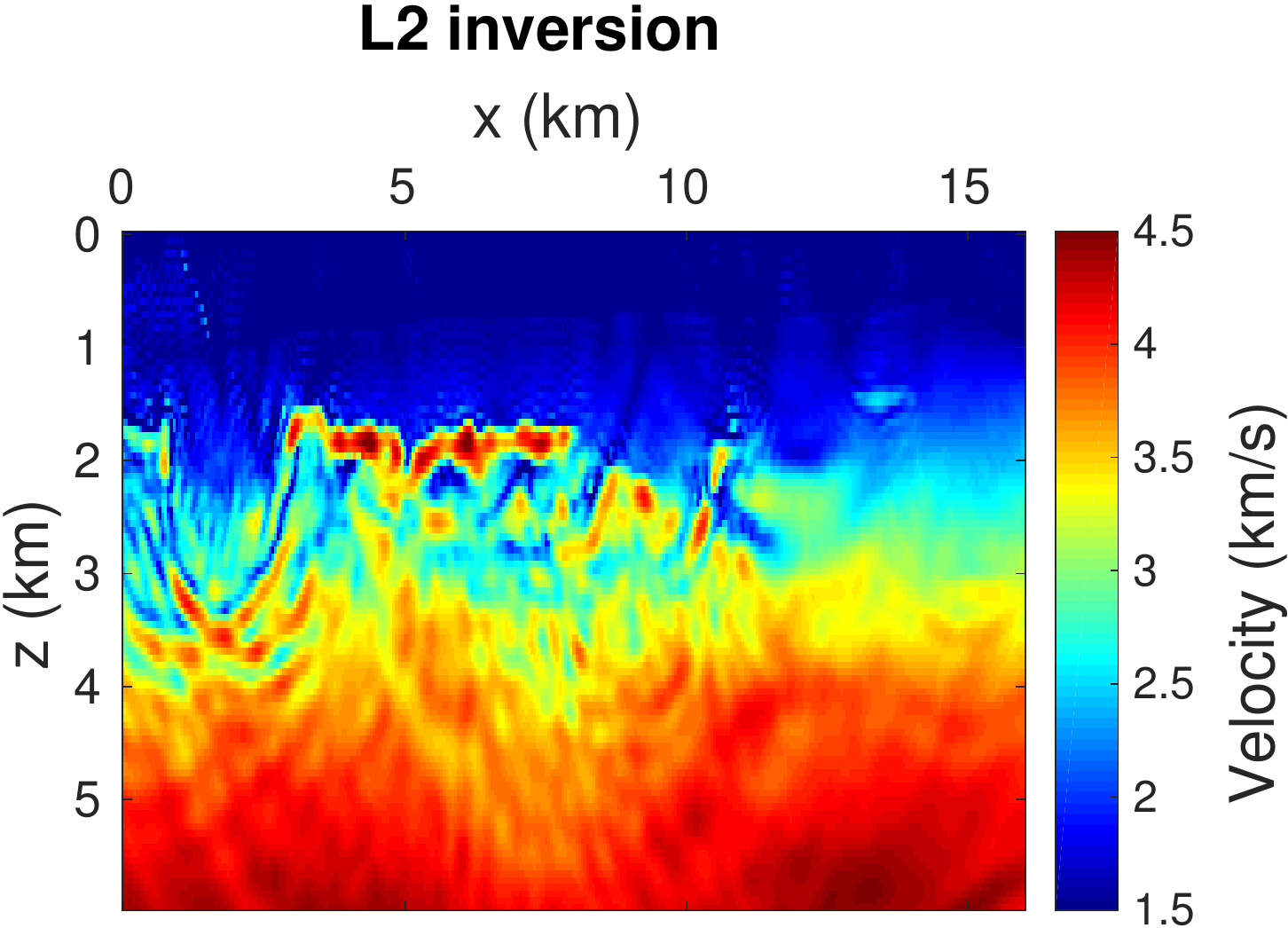}\label{fig:BP2_L2}}
  \subfloat[]{\includegraphics[width=0.45\textwidth]{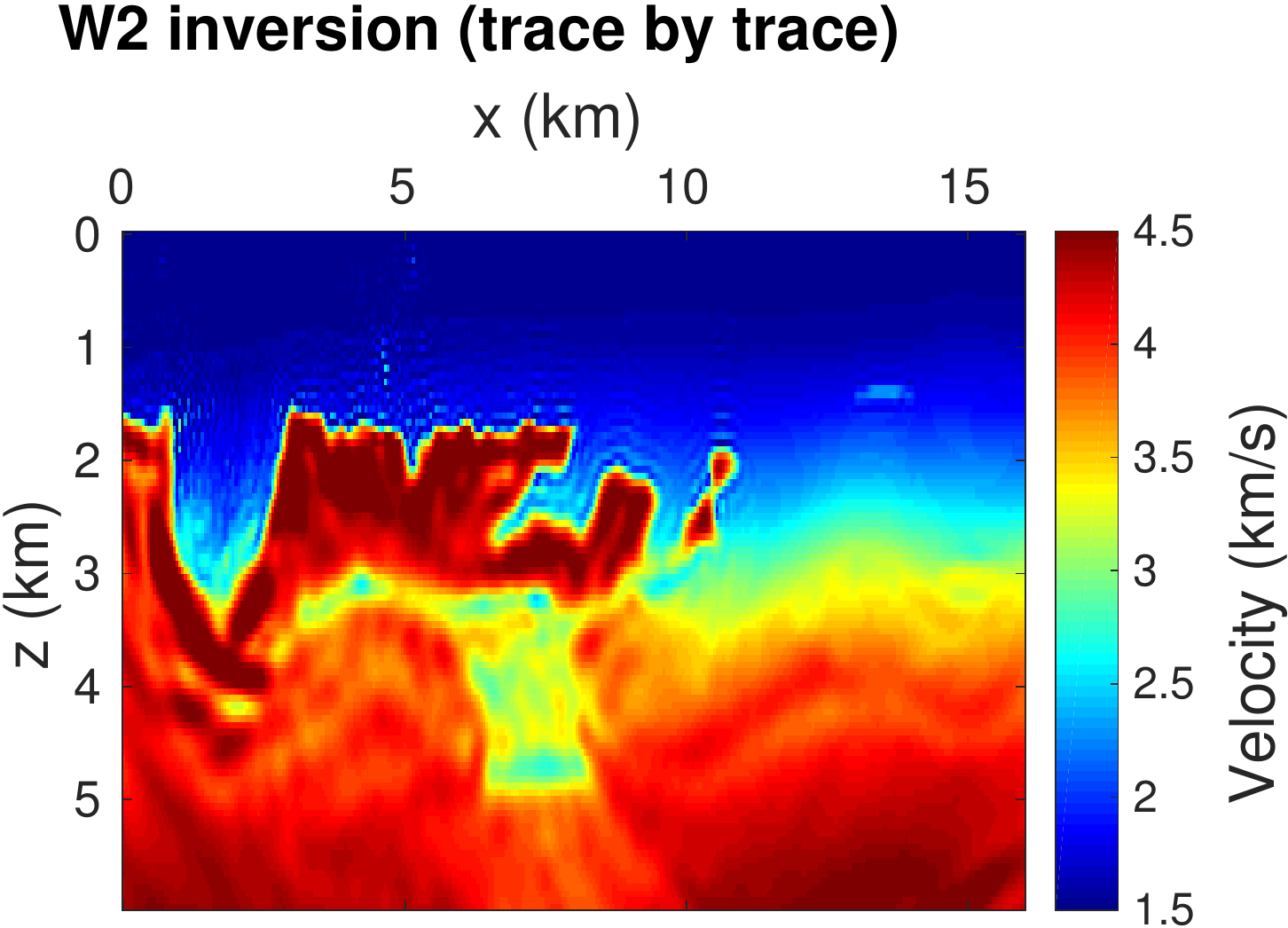}\label{fig:BP2_w2_1D}}
  \caption{Inversion results of (A)~$L^2$ and (B)~trace-by-trace $W_2$ for the second BP model}
  \label{fig:BP2_inv}
\end{figure}

\begin{figure}
\centering
  \subfloat[]{\includegraphics[width=0.45\textwidth]{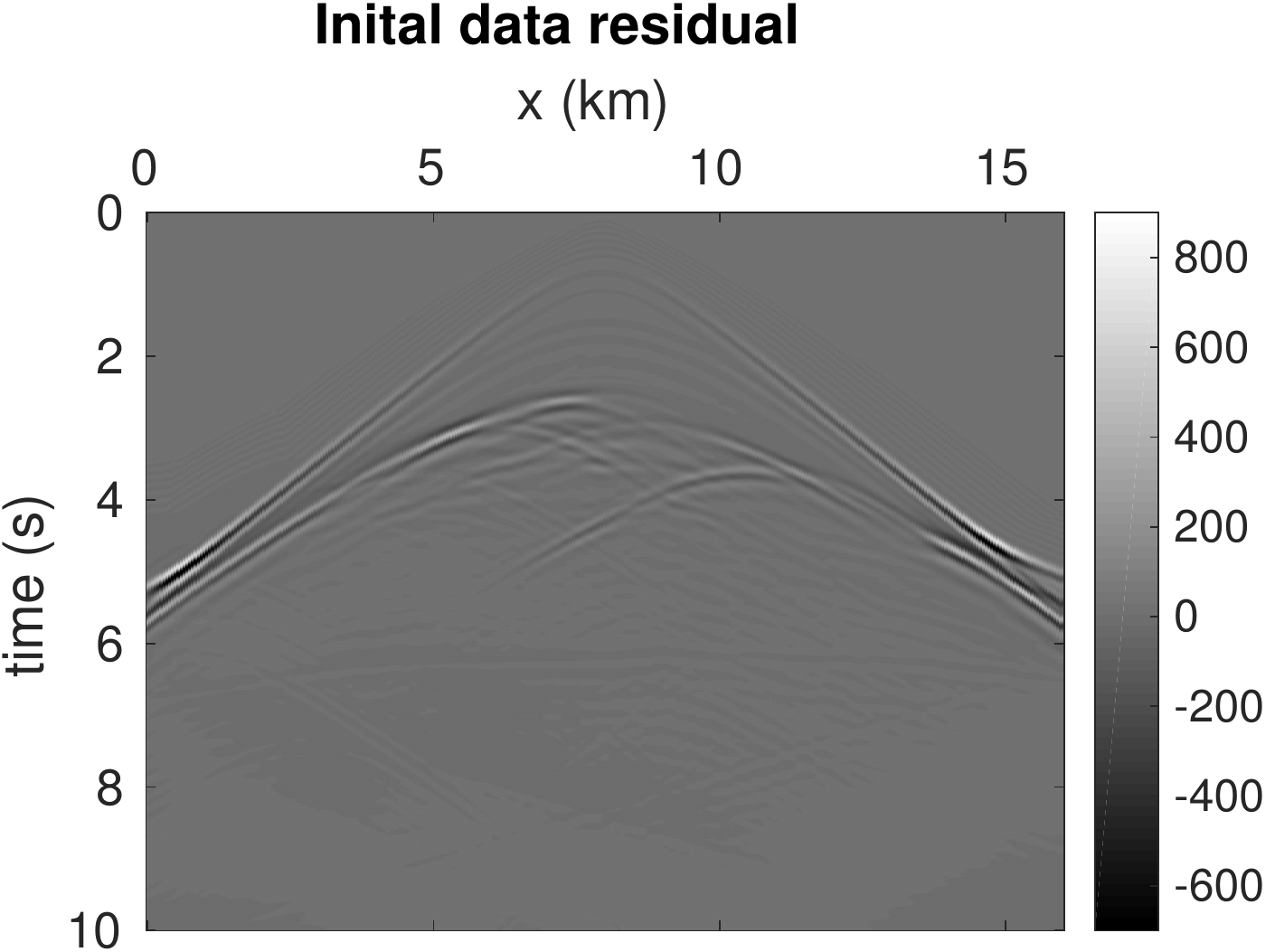}\label{fig:BP2_res0}}
  \subfloat[]{\includegraphics[width=0.45\textwidth]{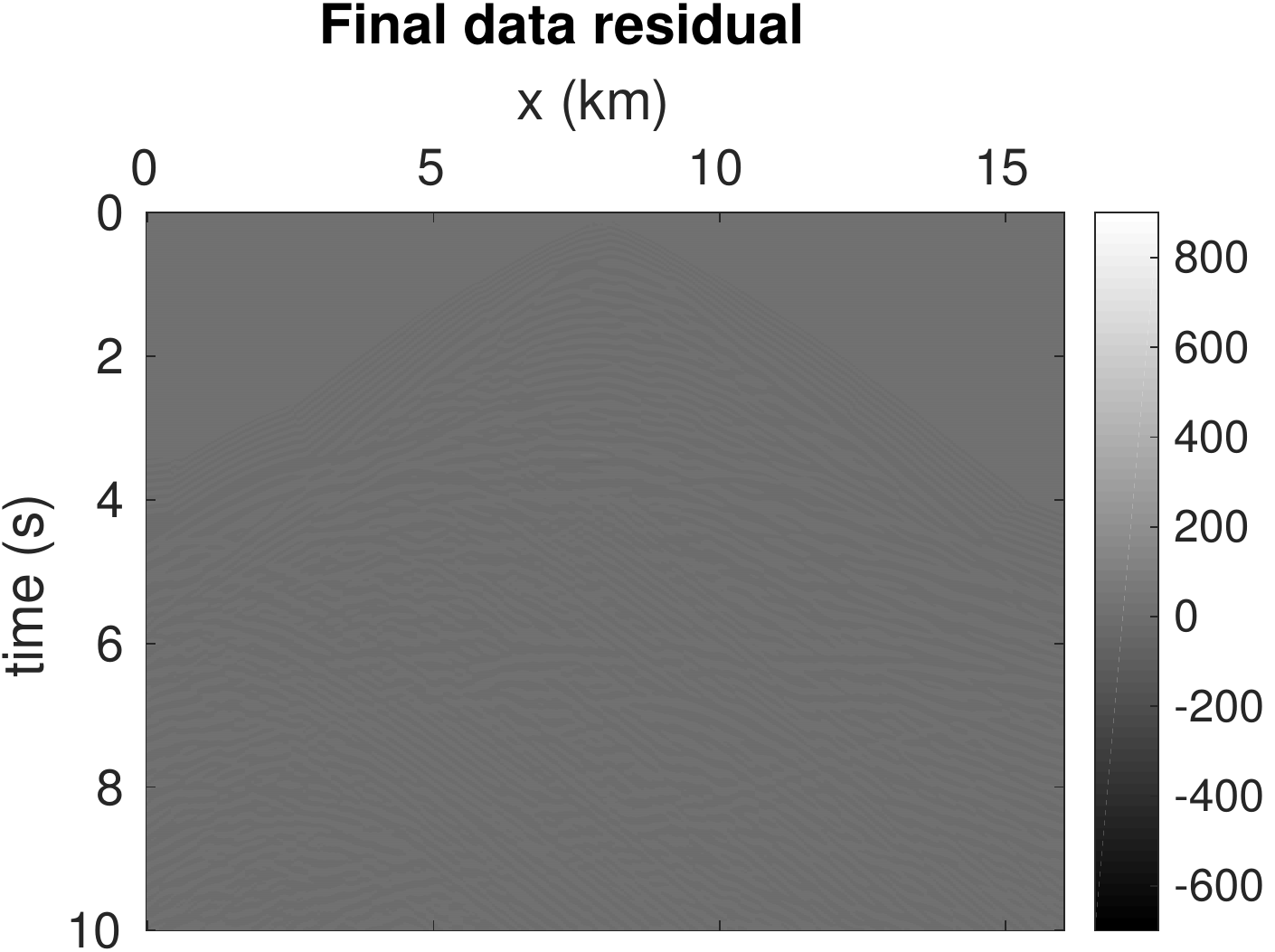}\label{fig:BP2_W2res}}
  \caption{(A)~The difference of data to be fit and the prediction with initial model and (B)~the final data residual of trace-by-trace $W_2$ for the second BP model}
  \label{fig:BP2_res}
\end{figure}

\begin{figure}
\centering
  \subfloat[]{\includegraphics[width=0.45\textwidth]{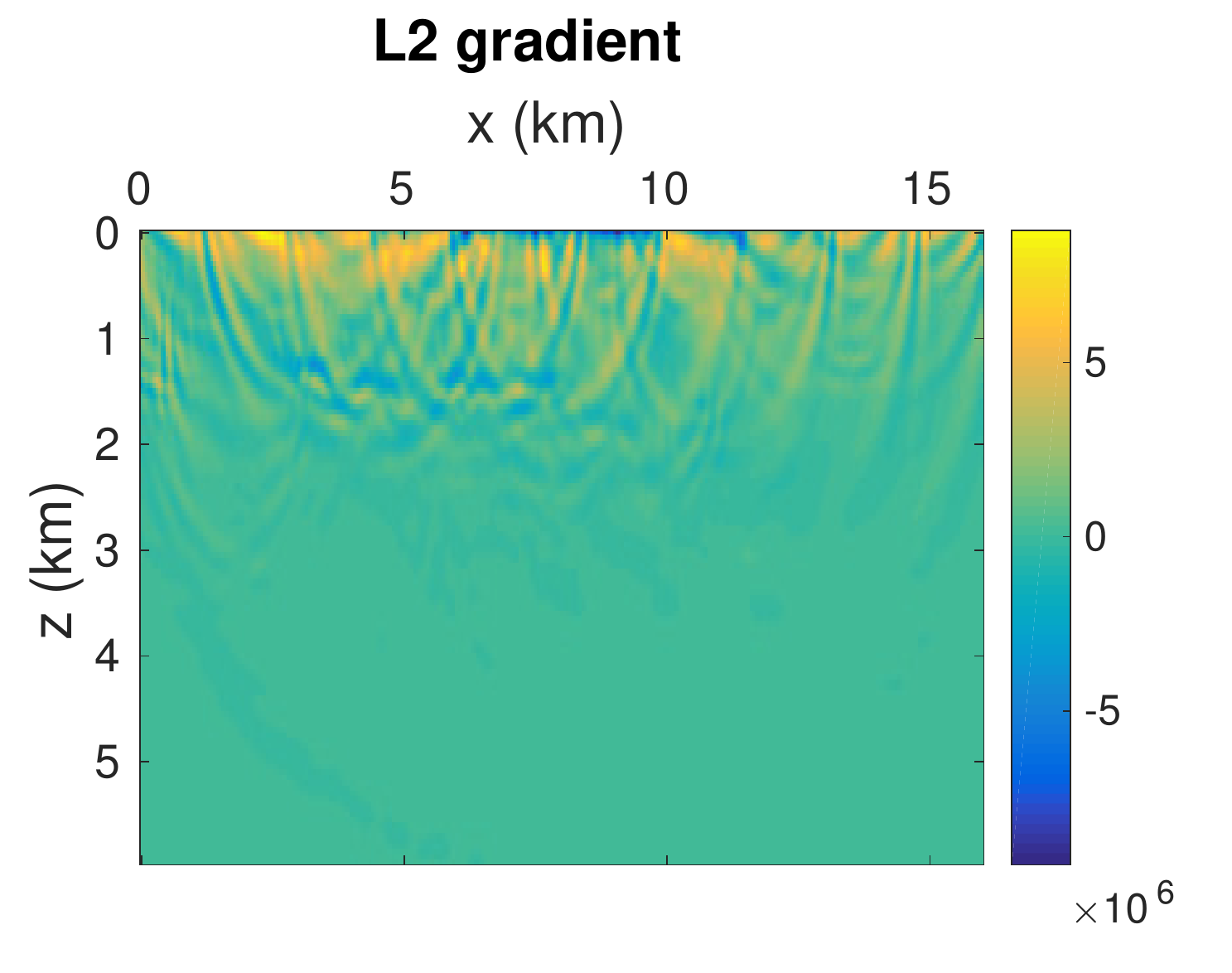}\label{fig:BP2_L2_grad}}
  \subfloat[]{\includegraphics[width=0.45\textwidth]{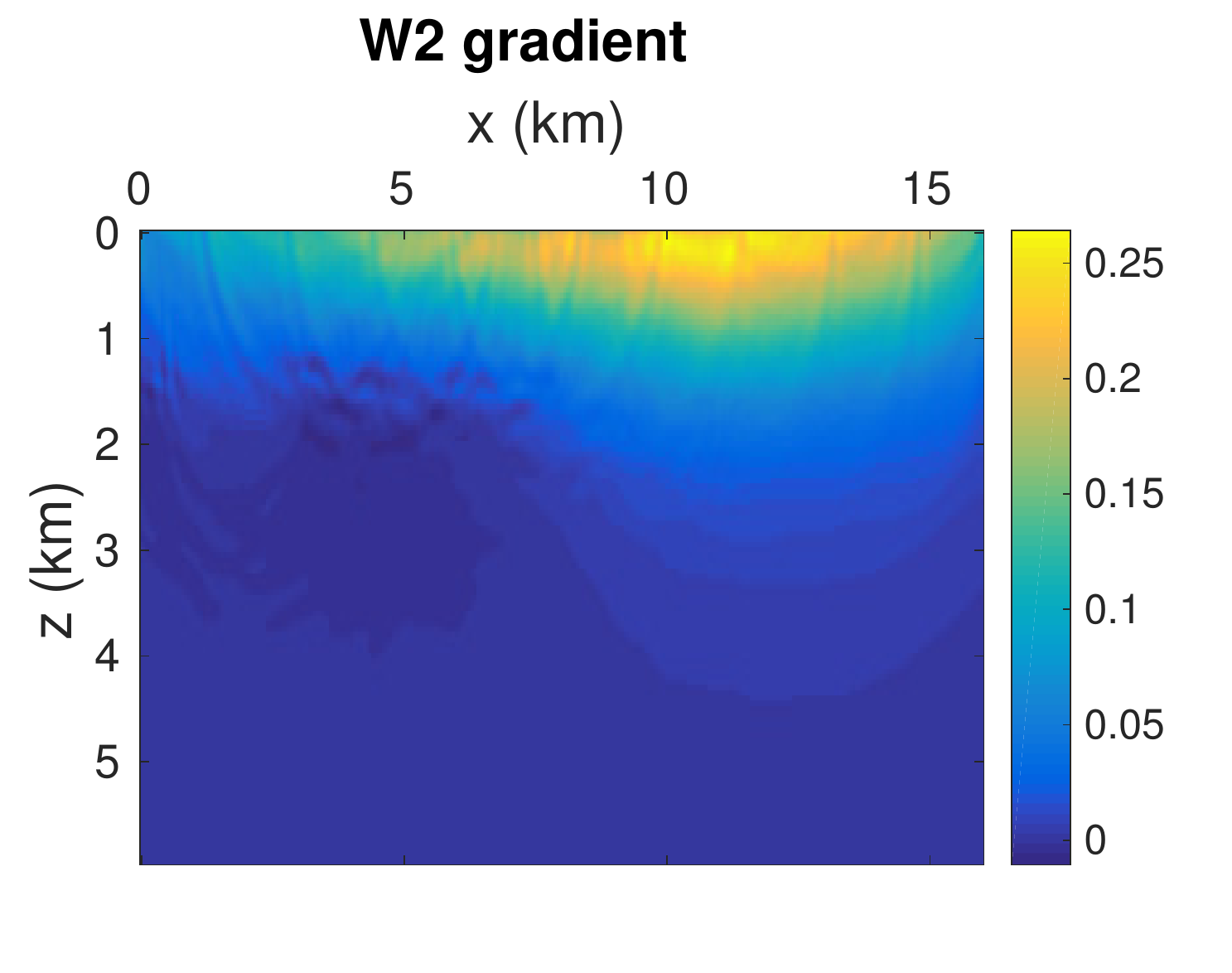}\label{fig:BP2_W2_grad}}
  \caption{The gradient in the first iteration of (A)~$L^2$ and (B)~trace-by-trace $W_2$ inversion for the second BP model}
  \label{fig:BP2_grad}
\end{figure}

\begin{figure}
\centering
 {\includegraphics[width=1.0\textwidth]{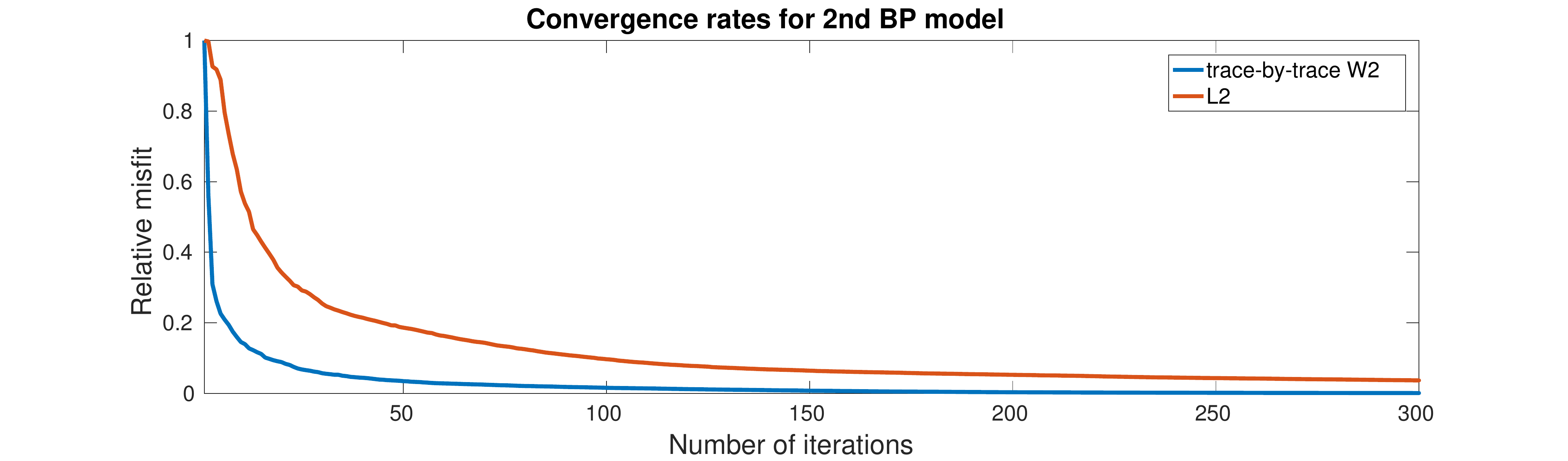}}
   \caption{The convergence curves for trace-by-trace $W_2$ and $L^2$ based inversion of the second BP model}
	\label{fig:BP2_conv}
\end{figure}

In our last experiment, we compare trace-by-trace $W_2$ and conventional $L^2$ as misfit functions for another modified BP 2004 model. Different from the previous experiment, the model is much larger as 6km deep and 16 km wide (Figure~\ref{fig:BP2_true}), similar to BP example in~\cite{W1_2D}. The inversion starts from an initial model with smoothed background without the salt (Figure~\ref{fig:BP2_v0}). We put 11 equally spaced sources on top at 250m depth in the water layer and 321 receivers on top at the same depth with 50m fixed acquisition. The discretization of the forward wave equation is 50m in the $x$ and $z$ directions and 50ms in time. The source is a Ricker wavelet with a peak frequency of 5Hz, and a band-pass filter is applied to keep the frequency components from 3 to 9Hz. The total acquisition time is 10s.

Again we compute the $W_2$ misfit trace by trace by solving the optimal transport problem in 1D as the Marmousi model and the Camembert model. Figure~\ref{fig:BP2_L2_grad} and Figure~\ref{fig:BP2_W2_grad} are the gradient in the first iteration of inversion using two misfit functions respectively. 

Starting from a smoothed initial model without the salt, in the first iteration $W_2$ inversion concentrates on the upper salt of the BP model (Figure~\ref{fig:BP2_W2_grad}). The darker area in the gradient matches salt part in the velocity model~(Figure~\ref{fig:BP2_true}). However, the gradient of $L^2$ is not very informative. Inversions are terminated after 300 l-BFGS iterations. Inversion with trace-by-trace $W_2$ misfit successfully constructed the shape of the salt bodies (Figure~\ref{fig:BP2_w2_1D}), while FWI with the conventional $L^2$ failed to recover boundaries of the salt bodies as shown by Figure~\ref{fig:BP2_L2}. The data residuals before and after trace-by-trace $W_2$ based FWI are presented in Figure~\ref{fig:BP2_res}. Trace-by-trace $W_2$ reduces the relative misfit to 0.1 in 20 iterations while $L^2$ converges slowly to a local minimum (Figure~\ref{fig:BP2_conv}).

\section{Discussion on two ways of using $W_2$}
The computational complexity of performing 1D optimal transport is extremely low compared with the cost of solving the \MA equation, which treats the synthetic and observed data as two objects and solves a 2D optimal transport problem.  From observation of the running time in our experiments, inversion with the trace-by-trace $W_2$ misfit requires less than 1.1 times the run-time required by inversion with the simple (and ineffective) $L^2$ misfit. 
Inversion using global $W_2$ comparison works for smaller scale models. It is more expensive since in each iteration we solve the \MA equation numerically to compute the misfit. The total inversion takes 3 to 4 times the run-time of the FWI with $L^2$ misfit in the experiments.

\begin{figure}
\centering
  \subfloat[]{\includegraphics[width=0.45\textwidth]{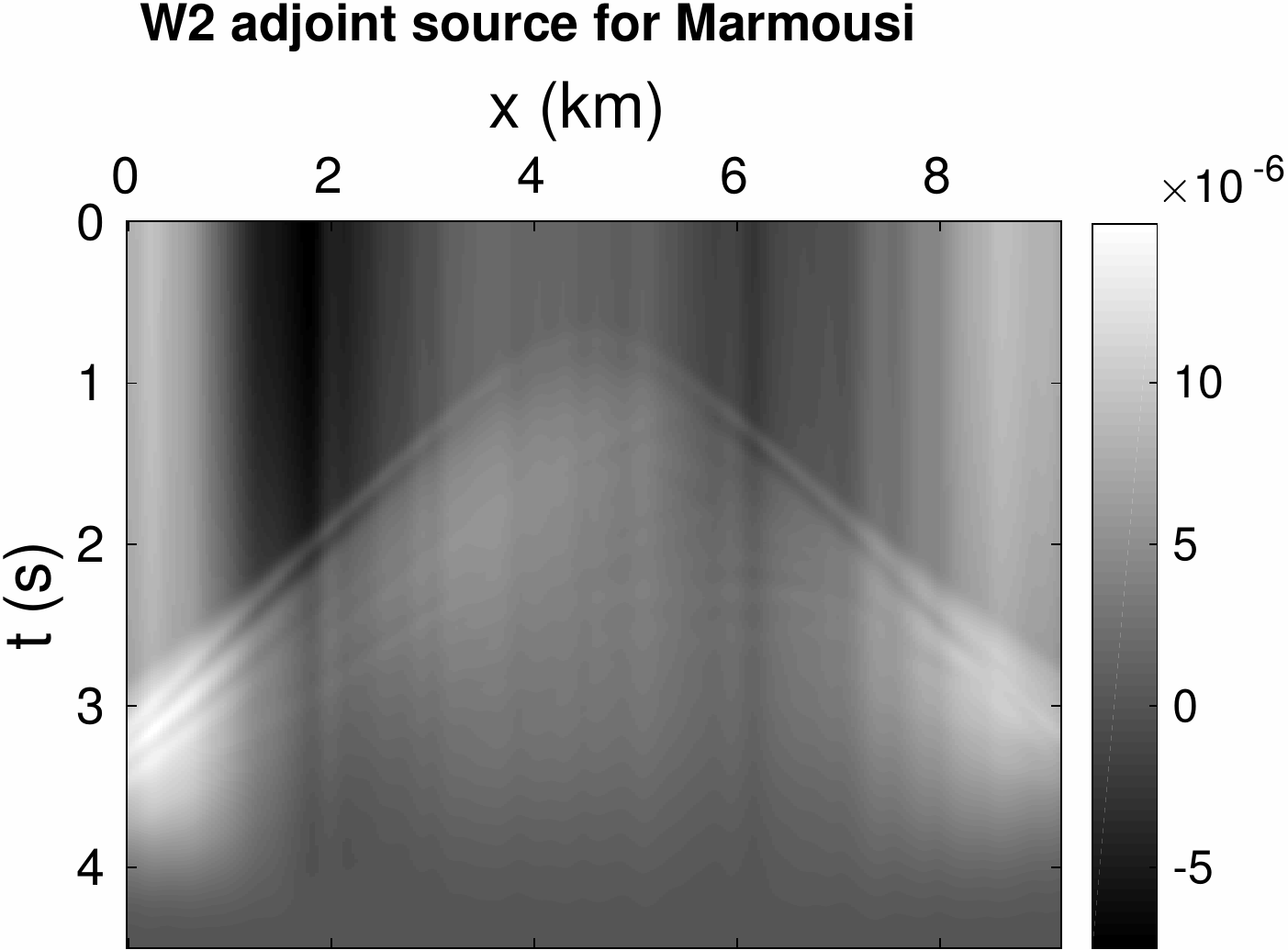}\label{fig:MARM2_DW2}}
  \subfloat[]{\includegraphics[width=0.45\textwidth]{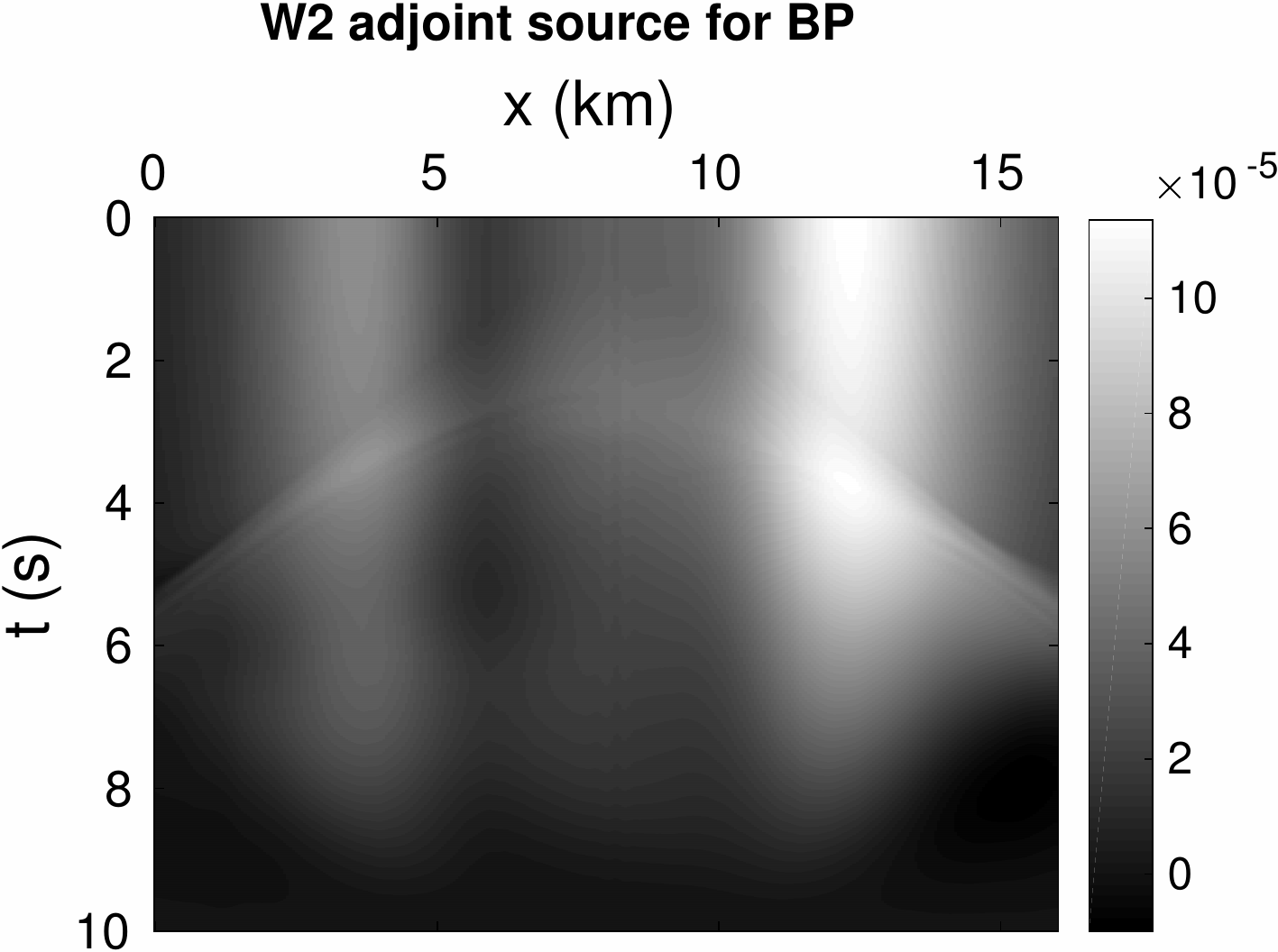}\label{fig:BP2_DW2}}
  \caption{The adjoint source in the first iteration of trace-by-trace $W_2$ (A) for the true Marmousi model and (B) for the second BP model}
  \label{fig:DW2}
\end{figure}

Figure~\ref{fig:cheese_dw2_1D} and Figure~\ref{fig:DW2} indicate one disadvantage of using $W_2$ trace by trace. The nonphysical variations in amplitude and non-uniform background of the adjoint source are caused by the fact that we rescale the data trace by trace in order to satisfy positivity and conservation of mass. On one hand, this may lead to non-uniform contributions of data misfits to the velocity update during the inversion process. Therefore, more careful treatment of the scaling in the trace-by-trace scheme may improve the convergence result being demonstrated in this study. On the other hand, in the experiments of the true Marmousi model and the second 2004 BP model, the gradients (Figure~\ref{fig:marm2_W2_grad} and Figure~\ref{fig:BP2_W2_grad}) do not have strong artifacts or nonphysical variation in the first iteration even if the corresponding adjoint sources (Figure~\ref{fig:DW2}) are irregular with strong horizontal variations. It will be interesting to study about the structure of the adjoint source in the success of FWI.


To compute the $W_2$ misfit globally we solve a 2D optimal transport problem based on \MA equation formulation. The numerical method for the \MA equation~\cite{FOFiltered} is proved to be convergent but requires the target profile $g$ to be Lipschitz continuous, and the discretization error is proportional to the Lipschitz constant of $g$. This means that for accurate results, enough data points are needed to effectively resolve steep gradients in the data; otherwise, the solver effectively regularizes the data before solving the \MA equation. This was evident in the example of the Marmousi model: the solver was much more robust to the scaled velocity benchmarks that provided better resolution of the fronts in the data. The trace-by-trace approach, on the other hand, can make use of exact formulas for 1D optimal transportation, which allows for accurate computations even when the data is highly non-smooth. 

The oscillatory artifacts in Figure~\ref{fig:cheese_w2_2D}, Figure~\ref{fig:marm_w2_2D} and Figure~\ref{fig:BP_w2_2D} likely originate from a combination of the numerical PDE solution discussed above and the insensitivity to noise of the $W_2$ measure. The fact that $W_2$ is insensitive to noise is good for noisy measurements, but not for artifacts in the synthetic data. This could, for example, be handled by TV regularization. We have chosen to present the raw results without pre or post processing. The trace-by-trace technique is also insensitive to noise, though to a lesser extent as there is only cancellation along one dimension. The 2D approach seems to have a slight edge for the Camembert model. The $L^2$ error between the converged velocity and the true model velocity with global $W_2$ is 25\% smaller than in the trace-by-trace case.


The trace-by-trace $W_2$ inversion results (Figure~\ref{fig:marm2_w2_1D} and Figure~\ref{fig:BP2_w2_1D}) are smooth. Starting from these velocities, we ran another 100 iterations of l-BFGS with $L^2$ norm as the misfit with the original source. However, we did not observe obvious improvements in the velocity resolution. In these experiments, FWI with trace-by-trace $W_2$ has reached the best resolution that $L^2$ can achieve at that frequency level. In the 6th experiment of the computation result, we start with a rough model built by 20 iterations of trace-by-trace $W_2$, and after the same number of iterations, the $W_2$ inversion result attained a lower model error than $L^2$ inversion. This demonstrates that the $W_2$ norm can not only build a good initial model, but also converge faster than the $L^2$ norm. One can also start with the trace-by-trace $W_2$ results in $L^2$ based FWI with higher frequency sources to deal with the cycle skipping issue.

\section{Conclusion}
We have developed a high-resolution full-waveform inversion (FWI) technique based on optimal transport and the quadratic Wasserstein metric ($W_2$). Here the $W_2$ misfit is coupled to efficient adjoint source computation for the optimization.
Our earlier work with $W_2$ was limited to a few degrees of freedom, but here we have presented successful inversion of the Marmousi, the 2004 BP, and the so-called Camembert models. This novel technique avoids cycle skipping as is demonstrated by numerical examples. 
The two-dimensional $W_2$ misfit is calculated by solving a relevant \MA equation and the latest version of the solver is outlined. We also show comparable results from trace-by-trace comparison with a $W_2$ misfit. This is as fast as the standard $L^2$ based FWI in terms of computation time, but more accurate, converges faster and also avoids cycle skipping. 

Our results clearly point to the quadratic Wasserstein metric as a potentially excellent choice for a misfit function in FWI. There are many possible directions for future improvements. 
In both the one and two-dimensional studies, the scaling or normalization of the signals play an important role. The linear normalization was by far the best for the large-scale inversion but does not satisfy the requirements of the theoretical result of convexity from shifts.  This should be further investigated and even better normalizations would be ideal.
Extending the one-dimensional trace-by-trace misfit to one-dimensional comparisons along additional directions is also possible. This has been successfully tried in other applications under the name of sliced Wasserstein distance.

\bibliographystyle{plain}
\bibliography{W2_app_ref}

\newpage
\appendix
\section{Derivation of Equation~\eqref{eqn:DW2_1D_2}}
We assume that $f(t)$ and $g(t)$ are continuous density functions in $[0,T_0]$, and $F(t) = \int_0^t f(\tau)d\tau$ and $G(t) = \int_0^t g(\tau)d\tau$. Now we perturb $f$ by an amount $\delta f$ and investigate the resulting change in~\eqref{myOT1D2} as a functional of $f$.
\begin{align}
W_2^2(f,g) + \delta W &= \int\limits_0^{T_0}|t-G^{-1}(F(t)+\delta F(t))|^2 (f(t)+\delta f(t)) dt\\
       			 &= \int\limits_0^{T_0}|t-G^{-1}(F(t)+\delta F(t))|^2 f(t) dt \\  \label{eqn:enq1}
       			 &\quad + \int\limits_0^{T_0}|t-G^{-1}(F(t))|^2 \delta f(t) dt + O((\delta f)^2).
\end{align}
Since $G$ is monotone increasing, so is $G^{-1}$. We have the following Taylor expansion of $G^{-1}$:
\begin{align}\label{eqn:G_inv}
G^{-1}(F(t)+\delta F(t)) &= G^{-1}(F(t)) + \frac{dG^{-1}(y)}{dy}\biggr\rvert_{y=F(t)}  \delta F(t) + O((\delta f)^2).
\end{align}
Substituting~\eqref{eqn:G_inv} back into~\eqref{eqn:enq1},  we obtain the first variation:
\begin{align}
\delta W &= \int\limits_0^{T_0} \left( \int_t^{T_0} -2(s-G^{-1}(F(s))\frac{dG^{-1}(y)}{dy}\biggr\rvert_{y=F(s)}  f(s) ds\right) \delta f(t) dt \\
&\quad + \int\limits_0^{T_0}|t-G^{-1}(F(t))|^2 \delta f(t) dt.
\end{align}

Thus the Fr\'{e}chet derivative of (29) with respect to $f$ is
\begin{equation}\label{eqn:1Dfrechet}
\frac{d W_2^2(f,g)}{df} = \int_t^{T_0}-2(s-G^{-1}(F(s))\frac{dG^{-1}(y)}{dy}\biggr\rvert_{y=F(s)}  f(s) ds+ |t-G^{-1}(F(t))|^2.
\end{equation}

In our numerical scheme, we discretize~\eqref{eqn:1Dfrechet} and derive~\eqref{eqn:DW2_1D_1} and \eqref{eqn:DW2_1D_2}.

\newpage
\end{document}